\documentclass[12pt,oneside,a4paper,onecolumn]{elsarticle}
\usepackage{times}
\usepackage{fullpage}

\usepackage{xcolor}
\usepackage{amsmath}
\usepackage{amssymb}
\usepackage{graphicx}
\usepackage{multicol}
\usepackage{caption}
\usepackage{cases}
\usepackage{subcaption}
\usepackage{tikz}
\usetikzlibrary{angles,quotes}
\usepackage{slashed}
\usetikzlibrary{patterns}
\usetikzlibrary{shapes.arrows}
\usepackage{pgfplots}
\usepackage{placeins}
\pgfplotsset{compat=newest}
\usepackage{mathtools} 
\usepackage{diagbox}
\usepackage{fixltx2e}
\usetikzlibrary{arrows.meta}

\newtheorem{rem}{Remark}

\usepackage{adjustbox}

\journal{Journal of \LaTeX\ Templates}

\begin{document}
\begin{frontmatter}

\title{An interface and geometry preserving phase-field method for fully Eulerian fluid-structure interaction }

\author[UBC]{Xiaoyu Mao}

\author[UBC]{Rajeev Jaiman\corref{cor1}}
\ead{rjaiman@mech.ubc.ca}
\cortext[cor1]{Corresponding author}
\address[UBC]{Department of Mechanical Engineering, University of British Columbia, Vancouver, Canada}




\begin{abstract}	
We present an interface and geometry preserving (IGP) method for the modeling of fully Eulerian fluid-structure interaction via phase-field formulation. While the hyperbolic tangent interface profile is preserved by the time-dependent mobility model, the proposed method maintains the geometry of the solid-fluid interface by reducing the volume-conserved mean curvature flow. To achieve the reduction in the curvature flow, we construct a gradient-minimizing velocity field (GMV) for the convection of the order parameter. The constructed velocity field enables the preservation of the solid velocity in the solid domain while extending the velocity in the normal direction throughout the diffuse interface region.  With this treatment, the GMV reduces the normal velocity difference of the level sets of the order parameter which alleviates the undesired thickening or thinning of the diffuse interface region due to the convection. During this process, the time-dependent mobility coefficient is substantially reduced and there is a lesser volume-conserved mean curvature flow. The GMV ensures that the diffuse interface region moves with the solid bulk such that the fluid-solid interface conforms to the geometry of the solid. Using the unified momentum equation and the phase-dependent interpolation, we integrate the IGP method into a fully Eulerian variational FSI solver based on the incompressible viscous fluid and the neo-Hookean solid. We first demonstrate the ability of the phase-field-based IGP method for the convection of circular and square interfaces with a prescribed velocity field. The variational FSI framework with the IGP method is then examined for the flow passing a fixed deformable block in a channel domain. Finally, the vibration of a plate attached behind a stationary cylinder subjected to incoming flow is employed to assess the fully Eulerian framework  for a large aspect ratio and sharp corners.
\end{abstract}


\end{frontmatter}


\section{Introduction}
Fluid-structure interaction (FSI) is a coupled highly-nonlinear multiphysics problem that can be found in various natural phenomena and industrial processes. Examples include from traditional aeroelasticity and flow-induced vibration problems in aerospace engineering \cite{shyy1999flapping,jaiman_jcp,jaiman_cmme,li2018novel}, marine/offshore \cite{jaiman2016partitioned,joshi20183d,joshi2019hybrid,kashyap2021}, biomedical \cite{peskin_IBM,peskin2002,griffith_annualreview_2020}, energy harvesting \cite{gurugubelli2015self} to the emerging fields of muscular hydrostat \cite{kier1985tongues,stavness2010byte} and soft robotics \cite{trivedi2008soft} and bio-inspired flying vehicles \cite{joshi2020variational}. 
The interface between the fluid and solid poses significant challenges in mathematical modeling and numerical simulation \cite{richter2017fluid,jaiman2022book}. Although various methods have been developed for handling the moving fluid-structure interface, solving large topological changes of solid while keeping the accuracy and stability of the solution remains to be challenging. During the contact and breaking-up of fluid-structure interfaces, the fluid-solid interfaces can undergo complex geometric changes which pose difficulty in the numerical implementation. The mutual dependence between the geometry and the underlying dynamics of interfaces is highly sensitive and error-prone in numerical simulations. Handling the discontinuity of physical properties along evolving interfaces needs careful considerations from both physical modeling and computational standpoints \cite{richter2017fluid,jaiman2022book}. Disparate kinematic descriptions of solids and fluids introduce intrinsic conflict and pose fundamental difficulties during the integration between the fluid and solid equations in a unified manner. Furthermore, multiphase FSI involves other well-known computational mechanics challenges with regard to satisfying conservation (e.g., mass and energy), handling high density and viscosity ratios, and turbulence effects \cite{jaiman2022book}. The motivation for this work primarily comes from bio-inspired locomotion \cite{kang2012dynamic} and flexible propeller blades \cite{young_jfs_2008,lampe2020partitioned} which require a robust and reasonably accurate fully Eulerian approach for the modeling of fluid-structure interaction.

The continuum hypothesis for the physical domains is typically assumed for large-scale numerical simulations of fluid-structure interaction. In the continuum mechanics formulation of fluid-structure interaction, the key challenges are associated with the conflict of dissimilar coordinate frames for the fluid and solid domains, the accurate treatment of the boundary conditions at the fluid-solid interface; and the design of stable and accurate discretizations for the coupled nonlinear PDEs  \cite{richter2017fluid,jaiman2022book}.
The kinematic description of continua is a fundamental consideration in the simulation of FSI problems, which can be classified into Lagrangian and Eulerian descriptions. From a computational point of view, the selection determines the relationship between the mesh and the deforming continuum. In the Lagrangian description, the grid points move consistently with the motion of material points. On the contrary, the grid is fixed spatially in the Eulerian description. This difference determines the accuracy and ease of the algorithm in handling large deformation of solid bodies under significant fluid forces
 and topological changes of fluid-solid interfaces. Due to the distinct constitutive relations of solids and fluids, they can be conveniently described in the Lagrangian and Eulerian frames, respectively.  In this work, we adopt a fully Eulerian description for both fluid and solid systems in a variational finite element framework \cite{dunne2006eulerian,richter2013fully,wick2013fully}.

Based on different kinematic descriptions for fluids and solids, various techniques treating the fluid-fluid and fluid-solid interfaces have been developed. One of the most accurate approaches is the arbitrary Lagrangian-Eulerian (ALE) method, in which interfaces are explicitly represented by conformal meshes \cite{hughes_ale,donea2017arbitrary}. The velocity and traction continuities can be accurately imposed as boundary conditions at the interface \cite{jaiman_ijnme,jaiman_jcp,jaiman_caf}. However, the conformal mesh may fail due to large motion and deformation, which requires special treatment such as remeshing strategies to avoid generating distorted elements. While the remeshing strategies are effective, they can be cumbersome in coding and software design. The Lagrangian-Eulerian approaches avoid these difficulties by allowing independent Lagrangian and Eulerian meshes for solids and fluids, respectively. For example, in the immersed boundary method (IBM) \cite{peskin1972flow,mittal_annualreview_2005}, the velocity and force of fluids and solids are projected back and forth between the Lagrangian grid and the Eulerian grid. In the front-tracking method \cite{unverdi1992front}, similar treatment is employed for the fluid-fluid interface. The fictitious domain method \cite{glowinski1994fictitious,baaijens2001fictitious} imposes the velocity continuity through a Lagrange multiplier. The extended finite element method (XFEM) \cite{sukumar2000extended,vaughan2007comparison} uses additional degrees of freedom and discontinuous shape functions for capturing the discontinuity of solution at the interface. In the fully Eulerian description, the governing equations for fluids and solids are unified as a one-field formulation, in which the velocity and stress continuities are naturally satisfied. The interfaces are captured by additional solution fields indicating the phases at Eulerian grid points. The interface capturing can be achieved through the volume of fluid, the level set or the phase-field method. The variation of the physical properties at interfaces can be explicitly modeled as functions of phase indicators \cite{mokbel2018phase}, or implicitly captured by the ghost fluid method  \cite{fedkiw1999non}. 

The fully Eulerian description for FSI problems has become an active area of research owing to its convenience in describing large motion and deformation of solids as well as topological changes of interfaces. This merit becomes prominent with the elevation of the geometric complexity in multiphase FSI problems. Some applications include contact dynamics, fracture and phase transfer of solids \cite{rycroft2020reference,wick2016coupling,zhu2021mixed}. In the fully Eulerian variational solver proposed in \cite{dunne2006eulerian}, the initial point set (IPS) method was used to track the incompressible neo-Hookean solid while the harmonic extension of the velocity in the fluid domain was used to avoid entanglement at fluid-solid interfaces. The constitutive relation of the solid was further generalized by \cite{richter2013fully} to St. Venant-Kirchhoff material. The decoupling of the momentum balance law with the structural displacement was accomplished in \cite{sun2014full}, where the stress was written as a function of velocity instead of displacement. The deformation of the solid was calculated by evolving the deformation gradient tensor while the interface was captured using the phase-field method. A thermodynamic consistent model based on the phase-field method was further developed in\cite{mokbel2018phase}, where the left Cauchy-Green tensor was employed to account for the evolution of the solid strain. In \cite{rycroft2020reference,kamrin2012reference}, the reference map technique was successfully demonstrated as an alternative to the IPS method.

However, accurate interface representation and evolution remain to be challenging in the interface capturing methods employed by the fully Eulerian formulation. Generally speaking, the interface capturing methods estimate the interface location while utilizing an implicit interface indicator in a fixed Eulerian grid. The indicator is almost a constant in the bulk of continua, while a rapid variation occurs across the interface. In the volume of fluid \cite{hirt1981volume}, level set \cite{malladi1995shape} and phase-field \cite{cahn1961spinodal,allen1979microscopic} methods, the interface indicator variables are namely volume fraction, level set function and order parameter, respectively. The interface can be reconstructed from the interface indicator variable, which forms the sharp interface approaches. Another approach is to represent the interface implicitly by the smooth transition of the interface capturing variable, which is classified as the diffuse interface approach \cite{anderson1998diffuse}. In the current work, we focus on the latter because they avoid the explicit interface reconstruction and ease the calculation of the normal and curvature of the interface.

In the diffuse interface approach, the transition layer between the phases has a finite thickness and employs definite profiles according to the diffuse interface model. In the level set method, the variation of the interface capturing variable across the interface is formulated as a signed distance function $d$. While in the phase-field method, the interface is associated with a smooth yet highly localized variation of the phase-field variable (i.e., order parameter), which takes a hyperbolic tangent profile $\tanh(d/(\sqrt{2}\varepsilon))$, where $\varepsilon$ controls the thickness of the diffuse interface region. When $\varepsilon$ is small so that the bulk of phases are distinct and the turning of the interface is well resolved, the diffuse interface description recovers to the sharp interface description \cite{elder2001sharp}. During the evolution of the interface, the interface capturing variable is convected by the flow field velocity. The relative normal velocity between the level sets of the interface capturing variable leads to an increase or decrease in the distance between the level sets, therefore distorting the signed distance function or the hyperbolic tangent profile. We refer to this as convective distortion \cite{mao2021variational}. In terms of the diffuse interface region, the convective distortion is reflected as the thickening or thinning effect \cite{jacqmin1999calculation}. The thickened diffuse interface region further elevates the inaccuracy. The thinned diffuse interface region brings all the level sets close to each other and forms large spatial gradients, which may cause numerical instability and convergence issues. 

To resolve the convective distortion, the reinitialization procedure is required in the level set method. Similarly, free energy minimization is employed to regularize the hyperbolic tangent profile and re-distance the diffuse interface region in the phase-field method. However, both the level set and phase-field methods can induce undesired displacement of the interface location, which is proportional to the local curvature \cite{hartmann2008differential,sun2007sharp}. When the convective distortion is severe, a significant reinitialization or free energy minimization process is required to restore the interface profile. The issue of undesired interface displacement becomes even worse at sharp corners of the solid where the curvature is not defined and tends to be very large in numerical simulation. Owing to the undesired displacement of the interface location, the geometry of the solid can be severely distorted during a simulation.
To summarize, there exists a conflict between the accurate interface representation and the evolution of a diffuse interface. If the interface is evolved exactly via the flow velocity, the diffuse interface representation can suffer from the convective distortion. On the other hand, if one wishes to recover the accurate diffuse interface representation, an undesired interface displacement will appear which will disturb the interface evolution. The conflict can be eased by directly modifying the reinitialization or the free energy minimization process. In \cite{hartmann2008differential}, the location of the interface is considered explicitly and the displacement caused by the reinitialization process is minimized. In \cite{sun2007sharp}, an anti-curvature term is suggested to nullify the curvature flow induced by the phase-field method. Recently in \cite{zhang2020efficient}, the authors introduced sub-iterations to correct the undesired interface displacement within the iteration of the reinitialization process. 

In the current work, instead of directly working on the reinitialization and free energy minimization processes, we preserve the interface geometry by minimizing the convective distortion. Inspired by the harmonic extension of the velocity in the fluid domain for the convection of IPS \cite{dunne2006eulerian}, we propose an auxiliary gradient-minimizing velocity field (GMV) for the convection of the order parameter in the phase-field method. The GMV preserves the solid velocity in the solid domain and further extends the solid velocity along the normal direction throughout the diffuse interface region. The stability of this process is enhanced by a diffusion term. Through this construction, the normal velocities of level sets of the order parameter are maintained the same inside the gradient-minimizing velocity field. As a result, the convective distortion decreases thereby reducing the need for the free energy minimization process \cite{mao2021variational}. This regularizes the hyperbolic tangent profile together with a lesser displacement of the interface. Hence the solid geometry is better preserved. In the context of the fully Eulerian FSI, the GMV keeps the velocity of the diffuse interface region close to the velocity of the solid bulk so that the diffuse interface region better conforms to the geometry of the solid. When the GMV is integrated with the Lagrange multiplier for the mass conservation \cite{brassel2011modified} and the time-dependent mobility coefficient for the interface profile preservation \cite{mao2021variational}, the resulting phase-field method better preserves the diffuse interface profile as well as the interface geometry. In the current paper, we refer to this as the interface and geometry preserving (IGP) phase-field method.

Besides the convection of the order parameter, the GMV is also used for the convection of the material coordinates, which tracks the solid in fixed Eulerian meshes. The dynamics of the interaction between the incompressible fluid and the incompressible neo-Hookean solid is solved through unified conservation equations via phase-dependent interpolation. The stabilized finite element technique is employed for the variational discretization while the generalized-$\alpha$ method is used for fully implicit time marching \cite{brooks1982streamline,chung1993,jaiman2022book}. We first test the effectiveness of the IGP method in preserving the interface profile and geometry in a prescribed velocity field. We examine the evolution of circular and square interfaces when convected by a velocity field that is extensional in the horizontal direction and compressional in the vertical direction. After investigating the IGP method alone, the IGP method is integrated into a variational fully Eulerian FSI formulation. The variational framework is examined by the case of flow passing a fixed deformable block in a channel domain. Finally, the vibration of a plate attached behind a stationary cylinder subjected to incoming channel flow, referred to as the cylinder-flexible plate problem, is considered to assess the solver for unsteady FSI involving solid with a large aspect ratio and sharp corners.

The organization of this paper is as follows: Section 2 presents the formulation for the IGP method and fully Eulerian FSI solver. Section 3 describes the variational discretization of the formulation. Section 4 presents the test cases, including the convection of circular and square interfaces in a prescribed velocity field, the channel flow passing a fixed deformable block and the cylinder-flexible plate problem. The conclusions are summarized in Section 5.

\section{Interface and geometry preserving method for fully Eulerian FSI}
In this section, we present the continuum formulation for the fully Eulerian FSI framework with the interface and geometry preserving phase-field method. We start with the mass and momentum conservation of the FSI problem. We next introduce the diffuse interface representation and evolution in the phase-field method and the gradient-minimizing velocity field for the convection of the order parameter. After that, we describe the material coordinates which track the solid in a fully Eulerian mesh for the calculation of the solid stress. Finally, we summarize all the equations and form the fully Eulerian FSI formulation with the IGP method.

\subsection{Conservation laws for the fully Eulerian FSI}
We start with the mass and momentum conservation of the FSI problem. In the current work, we restrict the material properties to the incompressible fluid and the incompressible Neo-Hookean solid. Consider a physical domain $\Omega\times]0,T[$ with spatial coordinates $\boldsymbol{x}$ and the temporal coordinate $t$. The domain is composed of the solid domain $\Omega^\mathrm{s}_t$ and the surrounding fluid domain $\Omega^\mathrm{f}_t$. The momentum balance law for the solid and the fluid can be written as:
\begin{align}\label{mom con}
	\rho^\mathrm{s}\left(\partial_{t}\boldsymbol{v}^\mathrm{s}+\boldsymbol{v}^\mathrm{s}\cdot\nabla\boldsymbol{v}^\mathrm{s}\right)=\nabla\cdot\boldsymbol{\sigma}^{\mathrm{s}}+\boldsymbol{b}^{\mathrm{s}},\quad 
	\rho^\mathrm{f}\left(\partial_{t}\boldsymbol{v}^\mathrm{f}+\boldsymbol{v}^\mathrm{f}\cdot\nabla\boldsymbol{v}^\mathrm{f}\right)=\nabla\cdot\boldsymbol{\sigma}^{\mathrm{f}}+\boldsymbol{b}^{\mathrm{f}},
\end{align}
where  $\boldsymbol{v}^\mathrm{s}$,  $\rho^{\mathrm{s}}$ , $\boldsymbol{\sigma}^{\mathrm{s}},\boldsymbol{b}^{\mathrm{s}}$ represent the velocity,  the density, the Cauchy stress and the body force of the solid respectively, and  $\boldsymbol{v}^\mathrm{f}$,  $\rho^{\mathrm{f}}$, $\boldsymbol{\sigma}^{\mathrm{f}},\boldsymbol{b}^{\mathrm{f}}$  represent the corresponding physical properties for the fluid.
With the incompressible condition in both the solid and the fluid, the mass conservation reduces to the continuity equations for both the solid and the fluid continua:
\begin{align}\label{mass con}
	\nabla\cdot \boldsymbol{v}^\mathrm{s}=0,\quad \nabla\cdot \boldsymbol{v}^\mathrm{f}=0.
\end{align} 

The mass and momentum balance equations can be integrated through a phase-dependent interpolation as follows:
\begin{align*}
	\alpha(\phi)\rho^\mathrm{s}\left(\partial_{t}\boldsymbol{v}^\mathrm{s}+\boldsymbol{v}^\mathrm{s}\cdot\nabla\boldsymbol{v}^\mathrm{s}\right)&+\left(1-\alpha(\phi)\right)\rho^\mathrm{f}\left(\partial_{t}\boldsymbol{v}^\mathrm{f}+\boldsymbol{v}^\mathrm{f}\cdot\nabla\boldsymbol{v}^\mathrm{f}\right)\\
	&=\nabla\cdot\left(\alpha(\phi)\boldsymbol{\sigma}^{\mathrm{s}}+\left(1-\alpha(\phi)\right)\boldsymbol{\sigma}^{\mathrm{f}}\right)+\alpha(\phi)\boldsymbol{b}^{\mathrm{s}}+\left(1-\alpha(\phi)\right)\boldsymbol{b}^{\mathrm{f}},\\
	\alpha(\phi) &\nabla\cdot \boldsymbol{v}^\mathrm{s}+\left(1-\alpha(\phi)\right)	\nabla\cdot \boldsymbol{v}^\mathrm{f}=0,
\end{align*}
where $\alpha(\phi)$ is a weight function satisfying $\alpha(\phi(\boldsymbol{x},t))=1, \boldsymbol{x}\in\Omega^{\mathrm{s}}_t ;$ $\alpha(\phi(\boldsymbol{x},t))=0, \boldsymbol{x}\in\Omega^{\mathrm{f}}_t$, and $\phi(\boldsymbol{x},t)$ being the order parameter in the phase-field method which indicates the local composition of the phases. A unified and continuous velocity field can be used to simplify the coupled fluid-solid equations as follows:
\begin{align}
	\rho(\phi)\left(\partial_{t}\boldsymbol{v}+\boldsymbol{v}\cdot\nabla\boldsymbol{v}\right)&=\nabla\cdot\boldsymbol{\sigma}(\phi)+\boldsymbol{b}(\phi),\\
		\nabla\cdot \boldsymbol{v}&=0,
\end{align}
where $\boldsymbol{v}$ is the unified velocity field subjected to density, stress and body force interpolation: $\rho(\phi) =	\alpha(\phi)\rho^\mathrm{s}+(1-\alpha(\phi))\rho^\mathrm{f}$, $\boldsymbol{\sigma}(\phi)=\alpha(\phi)\boldsymbol{\sigma}^{\mathrm{s}}+(1-\alpha(\phi))\boldsymbol{\sigma}^{\mathrm{f}}$ and $\boldsymbol{b}(\phi) =	\alpha(\phi)\boldsymbol{b}^\mathrm{s}+(1-\alpha(\phi))\boldsymbol{b}^\mathrm{f}$.

\subsection{IGP method and gradient-minimizing velocity field}
In this subsection, we first describe the diffuse interface representation and evolution via the phase-field method and provide a short review on the interface preserving phase-field method \cite{mao2021variational}. Furthermore, we point out the induced volume-conserved mean curvature flow which disturbs the geometry of the interface. We then introduce the gradient-minimizing velocity field for reducing the volume-conserved mean curvature flow. 
\subsubsection{Review of the interface preserving phase-field method}
In the phase-field method, an order parameter $\phi(\boldsymbol{x},t)$ is used to indicate the local composition of phases. In the current work, we set $\phi=1$ as an indicator for the solid phase, while $\phi=-1$ is used to track the fluid phase. The solid-fluid interface is represented by a region where $\phi$ transit smoothly from $\phi=1$ to $\phi=-1$. As a result of the free energy minimization $\mathcal{E}(\phi)=(F(\phi)+\varepsilon^2/2|\nabla\phi|^2)$, where $F(\phi)=1/4(\phi^2-1)^2$, the smooth transition evolves towards the hyperbolic tangent profile. The profile is given by $\phi=\tanh(d/\sqrt{2}\varepsilon)$, where $d$ is the signed distance function to the interface and $\varepsilon$ is the interface thickness parameter which controls the thickness of the transition region.

With the diffuse interface representation, the evolution of the interface now can be realized through the convection of the order parameter. However, when the level sets of the order parameter are convected in different normal velocities, the hyperbolic tangent profile will be distorted, and the thickness of the diffuse interface region will be changed. We characterize the thickness of the diffuse interface region with the distance between $\phi=\delta$ and $\phi=-\delta$. While the thinning of the diffuse interface region results in under-resolved high spatial gradients and further numerical instability, the thickening of the diffuse interface region leads to inaccuracy in the interface position. We refer to this phenomenon as convective distortion. To resolve this undesired artifact, we have developed a time-dependent mobility model. The model regularizes the hyperbolic tangent profile to re-distance the diffuse interface region according to the intensity of the convective distortion, thus preserving the hyperbolic tangent profile and the diffuse interface region \cite{mao2021variational}.

In the time-dependent mobility model, the convective distortion is quantified though the normal velocity gradient in the normal direction:
\begin{equation}
\frac{\partial v_n}{\partial n}=\frac{\nabla\phi\cdot\boldsymbol{v}\cdot\nabla\phi}{|\nabla\phi|^2}.
\end{equation} 
The convective form of the Allen-Cahn phase-field equation with a Lagrange multiplier for the mass conservation \cite{brassel2011modified} is given by :
\begin{equation}
	\frac{\partial \phi}{\partial t}+\boldsymbol{v}\cdot\nabla\phi=-\gamma(t)\left(\frac{\delta\mathcal{E}(\phi)}{\delta \phi} -\beta(t)\sqrt{F(\phi)}\right),
\end{equation}
where $\boldsymbol{v}$ is the convection velocity, $\gamma(t)$ is the time-dependent mobility model controls the intensity of the free energy minimization, and $\beta(t)=\frac{\int_{\Omega}F'(\phi)d\Omega}{\int_{\Omega}\sqrt{F(\phi)}d\Omega}$ is a Lagrange multiplier to ensure the mass conservation. As mentioned previously, the free energy minimization will evolve the interface profile towards the hyperbolic tangent profile.
The required free energy minimization to regularize the hyperbolic tangent profile against the convective distortion is given by:
\begin{equation}
	\gamma(t)=\frac{1}{\eta}\mathcal{F}\left(\left|\frac{\nabla\phi\cdot\nabla\boldsymbol{v}\cdot\nabla\phi}{|\nabla\phi|^2}\right|\right),\quad -\delta<\phi<\delta,
\end{equation}
where $\mathcal{F}(\cdot)$ is a RMS function and $\eta$ is a user-controlled parameter termed as the RMS interface distortion number  \cite{mao2021variational}. For a smaller value  $\eta$, the ratio between the free energy minimization  and the convective distortion is relatively larger and the interface distortion is reduced.  

However, the free energy minimization induces the undesired interface displacement. The velocity associated with the undesired interface displacement is proportional to the local curvature, which is given as:
\begin{equation}
	\boldsymbol{v}_{\kappa}=\gamma(t)\varepsilon^2(\kappa-\overline{\kappa})\boldsymbol{n},
\end{equation}
where $\kappa$ is the local curvature, $\overline{\kappa}$ is the average curvature along the interface and $\boldsymbol{n}$ is the normal vector of the interface. We refer to this as the volume-conserved mean curvature flow. The volume-conserved mean curvature flow will become significant in a fully Eulerian FSI when: (i) the solid moves at a relatively high speed, which induces significant convective distortion and requires large $\gamma(t)$ to regularize the hyperbolic tangent profile; (ii) the solid object involves sharp geometry features, where the curvature is not well defined and becomes a drastically large value of the curvature in numerical simulation.

\begin{rem}
While the desired traits of the phase-field method are maintained by employing the diffused interface description,  the mobility coefficient is adaptively reduced. We introduce a gradient-minimizing velocity field to reduce $\partial v_n/\partial n$, thus decreasing the mobility coefficient $\gamma(t)$ and the curvature flow velocity $\boldsymbol{v}_{\mathrm{\kappa}}$ during the interface evolution.
\end{rem}

\subsubsection{Gradient-minimizing velocity field for the phase-field-based IGP technique} 
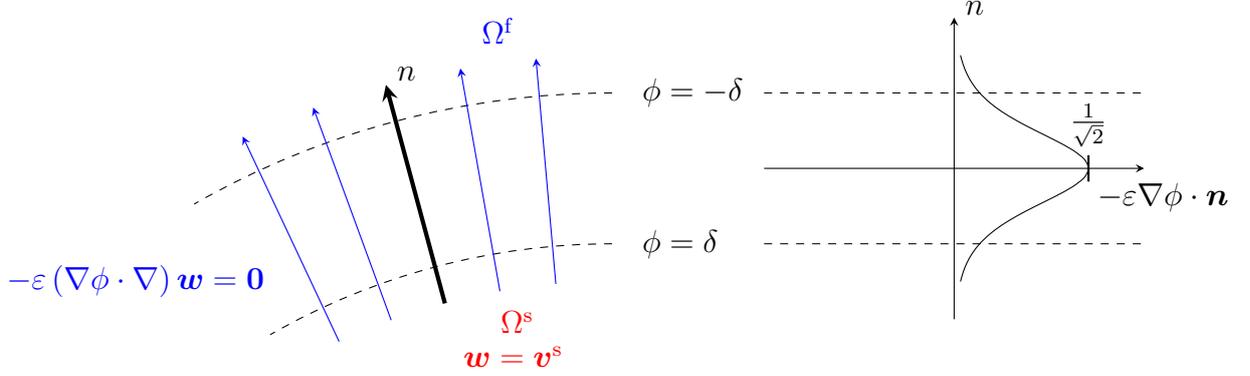
\begin{figure}[h]
	\centering
	\begin{tikzpicture}[scale=2.5]
		\node [left,red] at (0.8,-0.8) {$\boldsymbol{w}=\boldsymbol{v}^{\mathrm{s}}$};
		
		\draw [dashed](1,0.6) arc (90:120:4+0.4);
		\node [right] at (1.1,0.6) {$\phi=-\delta$};
		\draw [dashed](1,-0.2) arc (90:120:4-0.4);
		\node [below,red] at (0.5,-0.5) {$\Omega^{\mathrm{s}}$};
		\node [above,blue] at (0.4,0.8) {$\Omega^{\mathrm{f}}$};
		\node [right] at (1.1,-0.2) {$\phi=\delta$};
		
		\draw [-stealth,blue] (0.7037,-0.4129)--(0.5991,0.7825);
		\draw [-stealth,blue] (0.4096,-0.4517)--(0.2012,0.7301);
		\draw [ultra thick, -stealth] (0.1200,-0.5159)--(-0.1906,0.6433);
		\node [right] at (-0.1906,0.7) {$n$};
		\draw [-stealth,blue] (-0.1629,-0.6050)--(-0.5733,0.5226);
		\draw [-stealth,blue] (-0.4369,-0.7186)--(-0.9440,0.3690);

		\node [left,blue] at (-0.7713,-0.4) {$-\varepsilon\left(\nabla\phi\cdot\nabla\right)\boldsymbol{w}=\boldsymbol{0}$};

		\tikzset{shift={(2.8,0)}}
		\draw [-stealth](-1,0.2)--(1,0.2);
		\draw [-stealth](0,-0.6)--(0,1);
		
		\node[right] at (0,1.05) {$n$};
		\node[below] at (1.1,0.18) {$-\varepsilon\nabla\phi\cdot\boldsymbol{n}$};
		\draw [domain=-0.4:0.8, smooth, variable=\y]  plot ({(1-(tanh(((\y-0.2)/0.4*2.082)/(sqrt(2))))^2)/sqrt(2)}, {\y});
		\node[below] at (0.7071,0.6) {$\frac{1}{\sqrt{2}}$};
		\draw [dashed] (-1,-0.2)--(1,-0.2);
		\draw [dashed] (-1,0.6)--(1,0.6);
		\draw [ thick] (0.7071,0.13)--(0.7071,0.27);
	\end{tikzpicture}
	\caption{Illustration of gradient-minimizing velocity field. Propagation lines of $-\varepsilon\nabla\phi$ which propagates $\boldsymbol{w}$ in the normal direction throughout the diffuse interface region are shown as the blue arrows in the left figure, while the value of the propagation velocity in the normal direction $-\varepsilon\nabla\phi\cdot\boldsymbol{n}$ is shown in the right figure.}
	\label{normal-extension}
\end{figure}

The key proposition of the gradient-minimizing velocity field is to convect the level sets of the order parameter with approximately the same normal velocity in the normal direction. By employing such a velocity field, the distortion on the hyperbolic tangent profile or equivalently the thickening or thinning effect on the diffuse interface region can be reduced. Moreover,  there is a need for the diffuse interface region to conform to the solid bulk such that the normal diffuse interface velocity is tied with the velocity of the solid bulk. To achieve this constraint, we let the GMV follow the velocity field in the solid bulk. The solid velocity field along the periphery of the solid bulk is propagated in the normal direction throughout the diffuse interface region. Denoting the GMV as $\boldsymbol{w}$, the governing equation for the GMV is given by:
\begin{equation}
	\alpha(\phi)(\boldsymbol{w}-\boldsymbol{v})+(1-\alpha(\phi))\left(-\varepsilon\nabla\phi\cdot\nabla\right)\boldsymbol{w}=\boldsymbol{0},
\end{equation}
where $\alpha(\phi)$ is the weight function as mentioned previously. In the current work, we select $\alpha(\phi)=1/2(1+\phi)$. In evaluating $\alpha(\phi)$, we let $\alpha(\phi)=1$ when $\alpha(\phi)>1$ and $\alpha(\phi)=0$ when $\alpha(\phi)<0$ to ensure the stability of the solution.  The first term assigns the solid velocity to the GMV in the solid bulk, which we refer to as the Dirichlet term. The second term can be considered as a convection term, which propagates the velocity along the periphery of the solid bulk throughout the diffuse interface region. Since $\nabla\phi\sim O(1/\varepsilon)$, we premultiply $\varepsilon$ to ensure that the ratio between the convection term and the Dirichlet term is dependent on $\phi$ exclusively and independent of $\varepsilon$. Hence the governing equation for the GMV is consistent regardless of the diffuse interface thickness. The behavior of the governing equation is illustrated in Fig. \ref{normal-extension}.

To enhance the stability of the construction of the GMV, we add a diffusion term to ensure that $Pe=|v|h/2\nu\leq1$. We approximate the maximum magnitude of the propagation velocity $|v|_{\mathrm{max}}$ as $|v|_{\mathrm{max}}=|\left.-\varepsilon\partial\phi/\partial n\right|_{\phi=0}|=1/\sqrt{2}$, as shown in Fig. \ref{normal-extension}. The mesh size is approximated as $h=\varepsilon$, where the diffuse interface region is well resolved by four elements \cite{joshi2018positivity} . With these approximation, the diffusion coefficient can be selected as $\nu=\varepsilon/2\sqrt{2}$, and the governing equation for the GMV is finally given by:
\begin{equation}
	\alpha(\phi)(\boldsymbol{w}-\boldsymbol{v})+(1-\alpha(\phi))\left(\left(-\varepsilon\nabla\phi\cdot\nabla\right)\boldsymbol{w}+\frac{\varepsilon}{2\sqrt{2}}\Delta\boldsymbol{w}\right)=\boldsymbol{0}.
\end{equation}
The boundary condition is taken as $\nabla \boldsymbol{w}\cdot\boldsymbol{n}^{\Gamma}=0$, where $\boldsymbol{n}^{\Gamma}$ is the outer normal of the computational domain.

With the constructed GMV, the interface and geometry preserving phase-field method can be written as:
\begin{equation}
	\frac{\partial \phi}{\partial t}+\boldsymbol{w}\cdot\nabla\phi=-\gamma(t)\left(\frac{\delta\mathcal{E}(\phi)}{\delta \phi} -\beta(t)\sqrt{F(\phi)}\right),
\end{equation}
where 
\begin{equation}\label{dynamic mobility equation}
	\gamma(t)=\frac{1}{\eta}\mathcal{F}\left(\left|\frac{\nabla\phi\cdot\nabla\boldsymbol{w}\cdot\nabla\phi}{|\nabla\phi|^2}\right|\right),\quad -\delta<\phi<\delta.
\end{equation}
\begin{rem}
 The GMV reduces the convective distortion $\partial w_n	/\partial n=\nabla\phi\cdot\nabla\boldsymbol{w}\cdot\nabla\phi/|\nabla\phi|^2$ compared to that of flow field velocity $\partial v_n/\partial n$, which further decreases the dynamic mobility coefficient $\gamma(t)$ in Eq. (\ref{dynamic mobility equation}), and the resulting curvature flow $	\boldsymbol{v}_{\kappa}=\gamma(t)\varepsilon^2(\kappa-\overline{\kappa})\boldsymbol{n}$.
In the current paper, we select $\delta=0.9$ as the boundary of the diffuse interface region, within which $90\%$ of the variation of $\phi$ occurs.
\end{rem}
Owing to the reduction in the undesired convective distortion, the hyperbolic tangent profile in the diffuse interface region is preserved. Besides the convection of the order parameter, the GMV is utilized for the convection of the material coordinates to avoid the entanglement at the fluid-solid interface, which will be introduced in the next subsection.
\subsection{Material coordinates in the Eulerian description}

Following the convention of the continuum mechanics, a solid body $\mathcal{B}$ can be defined as a collection of particles. A one-to-one correspondence can be established between each particle of the body $P$ and its spatial coordinates in Euclidean space $\boldsymbol{x}\in\mathcal{R}$. The correspondence is referred to as a configuration of the body, which can be denoted as $\boldsymbol{\chi}:\boldsymbol{x}=\boldsymbol{\chi}(P)$, where  $P\in\mathcal{B}, \boldsymbol{x}\in\mathcal{R}$. To describe the deformation of the solid, we need to define a reference configuration. The reference configuration is usually selected as the initial configuration when the solid is undeformed and in a stress-free state. Mathematically, this can be denoted as:
\begin{equation} \label{reference configuration}
	\boldsymbol{X}(P)=\boldsymbol{\chi}_{\mathrm{ref}}(P,t=0)=\boldsymbol{x}(P,t=0).
\end{equation}
Note that Eq. (\ref{reference configuration}) establishes a one-to-one correspondence between $\boldsymbol{X}$ and $P$, which is invariant in time. In other words, the initial positions of the particles $\boldsymbol{X}$ can serve as unique labels for the solid particles. From this perspective, it can be considered as the material coordinates.

With the definition of the reference configuration, we can consider the deformation of a solid and the resulting restoring force. In the constitutive relation of the solid, stretch or compression of the line elements between any two particles causes restoring normal stress, while the change of the angle between two line elements causes restoring shear stress. Consequently, the solid particles take an ordered arrangement in any continuously deformed configurations without rupture \cite{wick2016coupling} or wrinkling \cite{leembruggen2021computational}. From the computational point of view, this facilitates the description of the solid in a Lagrangian frame of reference, where the computational nodes resolve the same particle without the issue of the entanglement of the computational grid. The current positions of the solid particles can be written as follows in the Lagrangian description:
\begin{equation}
	\boldsymbol{x}(\boldsymbol{X},t)=\boldsymbol{X}(P)+\boldsymbol{u}(\boldsymbol{X},t).
\end{equation}
Computationally, $\boldsymbol{X}(P)$ can play multiple roles for describing the motion of particles. It is the material coordinates of the particle $P$ and the position of the particle in the initial configuration. Since the same computational node is used for the particle in the Lagrangian description, $\boldsymbol{X}(P)$ is directly used as the coordinates of the computational node.  The displacement of the particle represented by $\boldsymbol{u}(\boldsymbol{X},t)$ is stored at the computational node located as $\boldsymbol{X}$.

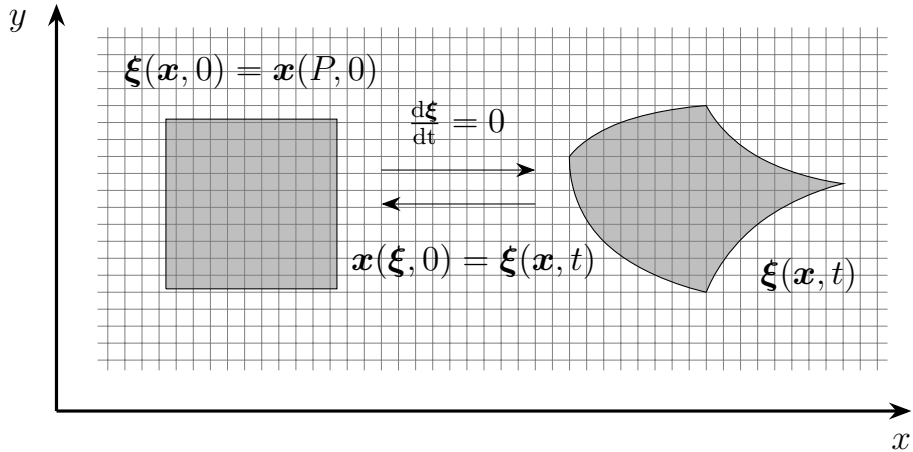
\begin{figure}[h]
	\centering
	\begin{tikzpicture}[scale=4.5,,every node/.style={scale=1.1}]
		\draw[-{Stealth[scale=1]},very thick] (0,0)--(2.5,0);
		\node [below ] at (2.47,-0.03) {$x$};
		\draw[-{Stealth[scale=1]},very thick] (0,0)--(0,1.2);
		\node [ left] at (-0.05,1.15) {$y$};
		
		\draw[step=0.05,gray,line width=0.009,fill opacity=0.7] (0.12,0.12) grid (2.43,1.13);

		\draw [fill=gray,fill opacity=0.5](0.32,0.36)--(0.82,0.36)--(0.82,0.86)--(0.32,0.86)--(0.32,0.36);
		\node at (0.57,1) {$\boldsymbol{\xi}(\boldsymbol{x},0)=\boldsymbol{x}(P,0)$};

		\draw [fill=gray,fill opacity=0.5,tension=1]  plot[smooth,tension=1] coordinates{(1.9,0.35)(2.05,0.55)(2.3,0.67)}--
			   plot[smooth] coordinates{(2.3,0.67)(2.05,0.75)(1.9,0.9)}--
			   plot[smooth] coordinates{(1.9,0.9)(1.65,0.85)(1.5,0.75)}--
			   plot[smooth] coordinates{(1.5,0.75)(1.6,0.5)(1.9,0.35)};

		\node at (2.2,0.4) {$\boldsymbol{\xi}(\boldsymbol{x},t)$};

		\draw [-{Stealth[scale=1.5]}] (0.95,0.71)--(1.4,0.71);
		\node [above] at(1.17,0.75) {$\mathrm{\frac{d \boldsymbol{\xi}}{dt}}=0$};
		\draw [-{Stealth[scale=1.5]}] (1.4,0.61)--(0.95,0.61);
		\node [below] at(1.22,0.53) {$\boldsymbol{x}(\boldsymbol{\xi},0)=\boldsymbol{\xi}(\boldsymbol{x},t)$};
	\end{tikzpicture}
	\caption{Illustration of the material coordinates for a deformable solid in the Eulerian frame. }
	\label{Material coodinates}
\end{figure}

 While the computational grid is well-behaved in the Lagrangian description for the solid due to the orderly arrangement of the solid particles, the surrounding mesh for the fluid domain can experience large stretch and fail when following the motion of the solid, especially when the solid is subjected to a large deformation or contact with each other. In such cases, an Eulerian description for the solid can be employed to circumvent this difficulty. In the Eulerian description for solid, the correspondence between the particles and the computational nodes are broken to allow large deformation of solids and topological changes of interfaces. The solid particles move through an Eulerian grid fixed in space. As a result, the coordinates of the computational nodes merely represent the spatial location of the node but no longer provide the initial position of the particles. Without the initial position of the solid particles, the relative displacement of the particles cannot be calculated due to the lack of reference. Hence the calculation of the strain and stress poses a difficulty. Therefore, we need an intermediate vector field to indicate the initial position of the particle located at the computational node. We denote the intermediate or supplementary vector field as $\boldsymbol{\xi}$. Since $\boldsymbol{\xi}$ gives the initial position of the particles, at $t=0$, we have:
 \begin{equation}
 	\boldsymbol{\xi}(\boldsymbol{x},t=0)=\boldsymbol{x}(P,t=0).
 \end{equation}
 For a solid particle, its initial position is time invariant. In other words, $\boldsymbol{\xi}$ can serve as unique labels to specify the solid particles, which essentially becomes a material coordinates in the Eulerian description. As markers of the solid particles, it should move along with the particles at the solid velocity in the Eulerian frame of reference. Therefore, the evolution of the material coordinates in the Eulerian description is given by:
 \begin{equation}\label{Eulerian evolution of material coordinates}
 	\frac{\partial\boldsymbol{\xi}(\boldsymbol{x},t)}{\partial t}+\boldsymbol{v}^{\mathrm{s}}\cdot\nabla\boldsymbol{\xi}(\boldsymbol{x},t)=\boldsymbol{0}.
 \end{equation}
 Furthermore, the initial position can be backtracked as:
 \begin{equation}
 	\boldsymbol{x}(\boldsymbol{\xi},0)=\boldsymbol{\xi}(\boldsymbol{x},t).
 \end{equation}
 The complete process is illustrated in Fig. \ref{Material coodinates}.

With the material coordinates, we can calculate the strain and stress of the deformed solid. The stretching or compressing of the solid relative to its reference configuration can be quantified as: $\boldsymbol{F}=\frac{\partial \boldsymbol{x}}{\partial \boldsymbol{\xi}}$, where $\boldsymbol{F}$ is referred to as the deformation gradient denser. Using $\frac{\partial \boldsymbol{\xi}}{\partial \boldsymbol{x}}\frac{\partial \boldsymbol{x}}{\partial \boldsymbol{\xi}}=\boldsymbol{I}$, one can write $\nabla\boldsymbol{\xi}=\frac{\partial\boldsymbol{\xi}}{\partial\boldsymbol{x}}=\boldsymbol{F}^{-1}$, where $\boldsymbol{\xi}$ at fixed Eulerian nodes represents the marker of the particle which moves to the current location, and the distance between the mesh nodes is the distances in the current configuration.  For the incompressible neo-Hookean solid, the constitutive relation is given by:
\begin{align}
	\boldsymbol{\sigma}^{\mathrm{s}}=-p^{\mathrm{s}}\boldsymbol{I}+\mu^{\mathrm{s}}(\boldsymbol{F}\boldsymbol{F}^{T}-\boldsymbol{I})=-p^{\mathrm{s}}\boldsymbol{I}+\mu^{\mathrm{s}}((\nabla\boldsymbol{\boldsymbol{\xi}})^{-1}(\nabla\boldsymbol{\xi})^{-T}-\boldsymbol{I}),
\end{align}
where $p^{\mathrm{s}}$ and $\mu^{\mathrm{s}}$ are the hydrostatic pressure and the shear modulus of the solid, respectively. This approach, which directly calculates the stress from the material coordinates, forms the reference map technique \cite{kamrin2012reference}. Another approach is to  substitute $ \boldsymbol{\xi}(\boldsymbol{x},t)=\boldsymbol{u}(\boldsymbol{\xi},t)-\boldsymbol{x}(\boldsymbol{\xi},0)$ into Eq. (\ref{Eulerian evolution of material coordinates}) and solve the displacement $\boldsymbol{u}$ explicitly, which recovers to the initial point set method \cite{dunne2006eulerian}. We select the former for the simplicity of the formulation.
For the completeness, we give the stress of the incompressible Newtonian fluid used in the current work as follows:
\begin{equation}
	\boldsymbol{\sigma}^{\mathrm{f}}=-p^{\mathrm{f}}\boldsymbol{I}+\mu^{\mathrm{f}} \left( \nabla \boldsymbol{v}^{\mathrm{f}} + (\nabla \boldsymbol{v}^{\mathrm{f}})^T\right),
\end{equation} 
where $p^{\mathrm{f}}$ is the hydrostatic pressure of the fluid and $\mu^{\mathrm{f}}$ is the dynamic viscosity of the fluid.

In the fully Eulerian FSI, the material coordinates $\boldsymbol{\xi}$ need to be convected inside the solid bulk and near the diffuse interface region to ensure the smoothness of the material coordinates. However, the unified velocity field near the diffuse interface is a composition of the fluid and the solid velocities. When the fluid velocity is involved, the material coordinates will be convected further downstream. This violates the physics of the solid and causes error and instability in the stress calculation. To avoid the entanglement of the material coordinates, we employ the GMV to convect the material coordinates, which is constructed purely based on the solid velocities:
\begin{align}
	\frac{\partial\boldsymbol{\xi}}{\partial t}+\boldsymbol{w}\cdot\nabla\boldsymbol{\xi}=\boldsymbol{0}.
\end{align}
\begin{rem}
It is worth pointing that the GMV plays a similar role with the harmonic velocity extension \cite{dunne2006eulerian} in our proposed formulation. Once the harmonic extension of velocity has been implemented in a solver, the IGP method can be easily coded by adding a convection term. 
\end{rem}
\subsection{Summary for the Eulerian FSI formulation with the IGP method}
Integrating above components, the fully Eulerian FSI formulation with the IGP method can be summarized as:
\begin{align}
	\left.
	\begin{aligned} \label{NS FULL}
		\rho(\phi)(\partial_t\boldsymbol{v}+\boldsymbol{v}\cdot\nabla\boldsymbol{v})&=\nabla\cdot\boldsymbol{\sigma}(\phi)+\boldsymbol{b}(\phi),\\
\nabla\cdot \boldsymbol{v}&=0,\\
\alpha(\phi)(\boldsymbol{w}-\boldsymbol{v})+&(1-\alpha(\phi))\left(\left(-\varepsilon\nabla\phi\cdot\nabla\right)\boldsymbol{w}+\frac{\varepsilon}{2\sqrt{2}}\Delta\boldsymbol{w}\right)=\boldsymbol{0},\\
\partial_t \phi+\boldsymbol{w}\cdot\nabla\phi&=-\gamma(t)\left(F(\phi)'-\varepsilon^2\Delta\phi-\beta(t)\sqrt{F(\phi)}\right),\\
\frac{\partial\boldsymbol{\xi}}{\partial t}+\boldsymbol{w}\cdot\nabla\boldsymbol{\xi}&=\boldsymbol{0},
	\end{aligned}\hspace{5pt}\right\}
\end{align}
with
\begin{align*}
	\rho (\phi) &=	\alpha(\phi)\rho^\mathrm{s}+(1-\alpha(\phi))\rho^\mathrm{f},\\ \boldsymbol{\sigma}(\phi)&=\alpha(\phi)\boldsymbol{\sigma}^{\mathrm{s}}+(1-\alpha(\phi))\boldsymbol{\sigma}^{\mathrm{f}},\\
	\boldsymbol{b} (\phi) &=	\alpha(\phi)\boldsymbol{b}^\mathrm{s}+(1-\alpha(\phi))\boldsymbol{b}^\mathrm{f},\\
	\alpha(\phi)&=1/2(1+\phi),\\
	\gamma(t)&=\frac{1}{\eta}\mathcal{F}\left(\left|\frac{\nabla\phi\cdot\nabla\boldsymbol{w}\cdot\nabla\phi}{|\nabla\phi|^2}\right|\right),\quad  -\delta<\phi<\delta,\\
	\boldsymbol{\sigma}^{\mathrm{s}}&=-p^{\mathrm{s}}\boldsymbol{I}+\mu^{\mathrm{s}}(\boldsymbol{F}\boldsymbol{F}^{T}-\boldsymbol{I}),\\
         \boldsymbol{\sigma}^{\mathrm{f}}&=-p^{\mathrm{f}}\boldsymbol{I}+\mu^{\mathrm{f}} \left( \nabla \boldsymbol{v}^{\mathrm{f}} + (\nabla \boldsymbol{v}^{\mathrm{f}})^T\right), \\
	\nabla\boldsymbol{\xi}&=\boldsymbol{F}^{-1}.
\end{align*}
Generally speaking, the GMV can be used in the diffuse interface method (where the interface is resolved by three to four elements) independent of the time-dependent mobility and the phase-field method to reduce the convective distortion. For example, the GMV should be able to reduce the convective distortion on the diffuse interface region in the level set method, thus reducing the frequency and iteration number of the reinitialization process, furthermore the displacement introduced by the reinitialization. The discretization of the continuum formulation will be introduced in the following section.

\section{Variational implementation of the Fully Eulerian FSI framework}
In this section, we present the variational implementation of our fully Eulerian FSI formulation. We start from the temporal discretization, which applies to each of the partitioned implicit solver where the temporal evolution is required. We then present the semi-discrete form of each component of the framework.  Specifically, we provide a detailed description on the linearization of the solid stress term in the momentum conservation equation. Finally, we present the linearized matrix form of the proposed variational framework.
\subsection{Temporal discretization}
We employ the generalized-$\alpha$ technique \cite{jansen2000generalized} for the temporal discretization which enables a user-controlled high frequency damping desirable for coarse discretizations in space and time. For a first-order system of variable $\varphi$, the generalized-$\alpha$ method is given by: 
\begin{align}
	\partial_t \varphi^{n+\alpha_m}&=f(\varphi^{n+\alpha}),\\
	\varphi^{n+1}&=\varphi^{n}+\Delta t \partial_t \varphi^n +\Delta t\varsigma (\partial_t \varphi^{n+1}-\partial_t \varphi^n),\\
	\partial_t \varphi^{n+\alpha_m} &= \partial_t \varphi^{n}+\alpha_m (\partial_t \varphi^{n+1}-\partial_t \varphi^n),\\
	\varphi^{n+\alpha}&=\varphi^n+\alpha(\varphi^{n+1}-\varphi^{n}),
\end{align} 
where $\alpha$, $\alpha_\mathrm{m}$ and $\varsigma$ are the generalized-$\alpha$ parameters which are dependent on the user-defined spectral radius $\rho_{\infty}$:
\begin{align}
	\alpha=\frac{1}{1+\rho_{\infty}},\ \alpha_m=\frac{1}{2}\left(\frac{3-\rho_{\infty}}{1+\rho_{\infty}}\right),\ \varsigma=\frac{1}{2}+\alpha_m-\alpha.
\end{align} 
The temporal evolution works in a predictor-multicorrector manner. In every nonlinear iteration, we first predict the solution at $n+1$, interpolate the solution to $n+\alpha$ and solve the first order system. After that, we correct the solution at $n+1$ according to the solution at $n+\alpha$. In the current work, the spectral radius is selected as $\rho_{\infty}=0$.

\subsection{Semi-discrete form of equations}
In this subsection, we present the semi-discrete form of all the components of the formulation, including the momentum and continuity equations, the gradient-minimizing velocity field, the Allen-Cahn equation and the evolution equation for the material coordinates.

\subsubsection{Momentum and mass conservation}
Suppose $\mathcal{S}^\mathrm{h}_{\boldsymbol{v}}$ and $\mathcal{S}^\mathrm{h}_{p}$  denote the space of trial solution such that:
\begin{align}
	\mathcal{S}^\mathrm{h}_{\boldsymbol{v}} &= \big\{ \boldsymbol{v}_\mathrm{h}\ |\ \boldsymbol{v}_\mathrm{h} \in (H^1(\Omega))^{d}, \boldsymbol{v}_\mathrm{h} = \boldsymbol{v}_{D}\ \mathrm{on}\ \Gamma_{D} \big\},\\
	\mathcal{S}^\mathrm{h}_{p} &= \big\{ p_\mathrm{h}\ |\ p_\mathrm{h} \in L^2(\Omega) \big\},
\end{align}
where $(H^1(\Omega))^{d}$ denotes the space of square-integrable $\mathbb{R}^{d}$-valued functions with square-integrable derivatives on $\Omega$, $L^2(\Omega)$ is the space of the scalar-valued functions that are square-integrable on $\Omega$ and $\Gamma_D$ represents the Dirichlet boundary for the velocity. Similarly, we define $\mathcal{V}^\mathrm{h}_{\boldsymbol{\psi}}$ and $\mathcal{V}^\mathrm{h}_{q}$ as the space of test functions such that:
\begin{align}
	\mathcal{V}^\mathrm{h}_{\boldsymbol{\psi}} &= \big\{ \boldsymbol{\psi}_\mathrm{h}\ |\ \boldsymbol{\psi}_\mathrm{h} \in (H^1(\Omega))^{d}, \boldsymbol{\psi}_\mathrm{h} = \boldsymbol{0}\ \mathrm{on}\ \Gamma_{D} \big\},\\
	\mathcal{V}^\mathrm{h}_{q} &= \big\{ q_\mathrm{h}\ |\ q_\mathrm{h} \in L^2(\Omega) \big\}.
\end{align}
The variational statement of the momentum and continuity equation can be written as: \\
find $\left[\boldsymbol{v}_\mathrm{h}(t^\mathrm{n+\alpha}),p_\mathrm{h}(t^\mathrm{n+1})\right]\in \mathcal{S}^\mathrm{h}_{\boldsymbol{v}} \times \mathcal{S}^\mathrm{h}_{p}$ such that $\forall \left[\boldsymbol{\psi}_\mathrm{h},q_\mathrm{h}\right]\in \mathcal{V}^\mathrm{h}_{\boldsymbol{\psi}} \times \mathcal{V}^\mathrm{h}_{q}$ for the 
momentum and continuity equations
\begin{align}
	&\int_{\Omega} \rho(\phi_\mathrm{h}^{n+\alpha}) ( \partial_t{\boldsymbol{v}}^{n+\alpha_m}_\mathrm
	{h} + {\boldsymbol{v}}_{\mathrm{h}} \cdot\nabla{\boldsymbol{v}}_{\mathrm{h}})\cdot\boldsymbol{\psi}_{\mathrm{h}} \mathrm{d\Omega} +
	\int_{\Omega} {\boldsymbol{\sigma}}_{\mathrm{h}}(\phi_\mathrm{h}^{n+\alpha})
	: \nabla\boldsymbol{\psi}_{\mathrm{h}} \mathrm{d\Omega} \nonumber \\
	+ &\displaystyle\sum_\mathrm{e=1}^\mathrm{n_{el}}\int_{\Omega
		^{\mathrm{e}}} \frac{\tau_\mathrm{m}}{\rho(\phi_\mathrm{h}^{n+\alpha})} (\rho(\phi_\mathrm{h}^{n+\alpha}
	){\boldsymbol{v}}_{\mathrm{h}}\cdot\nabla
	\boldsymbol{\psi}_{\mathrm{h}}+ \nabla q_{\mathrm{h}} )\cdot
	\boldsymbol{\mathcal{R}}_\mathrm{m} \mathrm
	{d\Omega^e} \nonumber \\
	+ &\int_{\Omega}q_{\mathrm{h}}(\nabla\cdot{\boldsymbol{v}}_{\mathrm{h}}) \mathrm{d\Omega} + \displaystyle\sum_\mathrm{e=1}^\mathrm{n_{el}}\int
	_{\Omega^{\mathrm{e}}} \nabla\cdot\boldsymbol{\psi}_{\mathrm{h}}\tau
	_\mathrm{c}\rho(\phi_\mathrm{h}^{n+\alpha}) \mathcal{R}_\mathrm
	{c} \mathrm{d\Omega^e}\nonumber\\
	= & \int_{\Omega}\boldsymbol{b}(\phi^{n+\alpha}_\mathrm{h})\cdot\boldsymbol{\psi}_{\mathrm{h}}\mathrm{d\Omega}+\int_{\Gamma
		_{H}} \boldsymbol{h}\cdot\boldsymbol{\psi}_{\mathrm{h}}
	\mathrm{d\Gamma}, \label{PG_NS}
\end{align}
where $\boldsymbol{\mathcal{R}}_\mathrm{m}$ and $\mathcal{R}_\mathrm{c}$  denote the element-wise residuals for the momentum and continuity equations, respectively.
In Eq.~(\ref{PG_NS}), the terms in the first line represent the Galerkin projection of the momentum equation in the test function space $\boldsymbol{\psi}_\mathrm{h}$ and the second line comprises of the Petrov-Galerkin stabilization term for the momentum equation. The third line denotes the Galerkin projection and stabilization terms for the continuity equation. The terms in the last line are the Galerkin projection of the body force and the Neumann boundary condition.  The stabilization parameters $\tau_\mathrm{m}$ and $\tau_\mathrm{c}$  in Eqs.~(\ref{PG_NS})are given by \cite{shakib1991new, brooks1982streamline}:
\begin{align}
	\tau_\mathrm{m} &= \left( \bigg( \frac{2}{\Delta t}\bigg)^2 + \boldsymbol{v}_\mathrm{h}\cdot\boldsymbol{G}\boldsymbol{v}_\mathrm{h} + C_I \bigg( \frac{\mu(\phi)}{\rho(\phi)}\bigg)^2 \boldsymbol{G}:\boldsymbol{G} \right)^{-1/2},\qquad \tau_\mathrm{c} = \frac{1}{\mathrm{tr}(\boldsymbol{G})\tau_\mathrm{m}},
\end{align}
where $C_I$ is a constant derived from the element-wise inverse estimates \cite{harari1992c}, $\boldsymbol{G}$ is the element contravariant metric tensor and $\mathrm{tr}(\boldsymbol{G})$ is the trace of the contravariant metric tensor. This stabilization in the variational form circumvents the Babu$\mathrm{\check{s}}$ka-Brezzi condition that is required to be satisfied by any standard mixed Galerkin method \cite{Johnson}.  

\subsubsection{Gradient-minimizing velocity field}
For the gradient-minimizing velocity field, we define the space of trial solution  $\mathcal{S}_{\boldsymbol{w}}^{\mathrm{h}}$ and test functions $\mathcal{V}_{\boldsymbol{w}}^{\mathrm{h}}$ as:
\begin{align}
	\mathcal{S}^\mathrm{h}_{\boldsymbol{w}} &= \big\{ \boldsymbol{w}_\mathrm{h}\ |\ \boldsymbol{w}_\mathrm{h} \in (H^1(\Omega))^{d}\big\},\\
	\mathcal{V}^\mathrm{h}_{\boldsymbol{\psi}} &= \big\{ \boldsymbol{\psi}_\mathrm{h}\ |\ \boldsymbol{\psi}_\mathrm{h} \in (H^1(\Omega))^{d} \big\}.
\end{align}
 The variational statement  of the equation for constructing the gradient-minimizing velocity field can be written as: find $\boldsymbol{w}_{\mathrm{h}}(t^{\mathrm{n+\alpha}})\in\mathcal{S}^{\mathrm{h}}_{\boldsymbol{w}}$ such that $\forall \boldsymbol{\psi}_{\mathrm{h}}\in\mathcal{V}^{\mathrm{h}}_{\boldsymbol{\psi}}$
\begin{align}
	\int_{\Omega}\alpha(\phi_\mathrm{h}^{n+\alpha})(\boldsymbol{w}_\mathrm{h}-\boldsymbol{v}^{n+\alpha}_\mathrm{h})\cdot\boldsymbol{\psi}_\mathrm{h} \mathrm{d\Omega}&+\int_{\Omega}(1-\alpha(\phi_\mathrm{h}^{n+\alpha}))\left(\left(-\varepsilon\nabla\phi^{n+\alpha}_\mathrm{h}\cdot\nabla\right)\boldsymbol{w}_\mathrm{h}\right)\cdot\boldsymbol{\psi}_\mathrm{h} \mathrm{d\Omega}\nonumber\\&+\int_{\Omega}(1-\alpha(\phi_\mathrm{h}^{n+\alpha}))\frac{\varepsilon}{2\sqrt{2}}\nabla\boldsymbol{w}_\mathrm{h}:\nabla\boldsymbol{\psi}_\mathrm{h}\mathrm{d\Omega}=0.
\end{align}
Note that the generalized-$\alpha$ method does not apply to this elliptic equation. Instead, the elliptic equation for the GMV is solved via Newton-Raphson sub-iterations until convergence within every nonlinear iteration. The constructed GMV is subsequently passed to the Allen-Cahn equation and the evolution equation for the material coordinates.

\subsubsection{Allen-Cahn equation}
Suppose $\mathcal{S}^\mathrm{h}_{\phi}$ and $\mathcal{V}^\mathrm{h}_{\psi}$ denote the space of trial solution and test functions such that:
\begin{align}
	\mathcal{S}^\mathrm{h}_{\phi} &= \big\{ \phi_\mathrm{h}\ |\ \phi_\mathrm{h} \in H^1(\Omega), \phi_\mathrm{h} = \phi_{D}\ \mathrm{on}\ \Gamma^\mathrm{\phi}_{D} \big\},\\
	\mathcal{V}^\mathrm{h}_{\psi} &= \big\{ \psi_\mathrm{h}\ |\ \psi_\mathrm{h} \in H^1(\Omega), \psi_\mathrm{h} = 0\ \mathrm{on}\ \Gamma^\mathrm{\phi}_{D} \big\},
\end{align}
where $\Gamma^{\phi}_{D}$ denotes the Dirichlet boundary for the order parameter.
The variational statement of the Allen-Cahn equations can be written as: find $\phi_\mathrm{h}(t^\mathrm{n+\alpha})\in\mathcal{S}^\mathrm{h}_{\phi}$ such that $\forall  \psi_\mathrm{h}\in \mathcal{V}^\mathrm{h}_{\psi}$ for the Allen-Cahn equation:
\begin{align}
	\label{PPV_AC}
	&\int_{\Omega}\bigg( \psi_{\mathrm{h}}\partial_t{\phi}_{\mathrm{h}}^{n+\alpha_m} +
	\psi_{\mathrm{h}}\big(\boldsymbol{w}_\mathrm{h}^{n+\alpha}\cdot\nabla\phi_{\mathrm{h}}\big) + \gamma(t^{n+\alpha}) \big(\nabla
	\psi_{\mathrm{h}}\cdot(\hat{k}\nabla\phi_{\mathrm{h}} ) + \psi_{\mathrm{h}}\hat{s}\phi
	_{\mathrm{h}} - \psi_{\mathrm{h}}\hat{f}\big) \bigg) \mathrm{d}\Omega\nonumber\\
	+& \displaystyle\sum_\mathrm{e=1}^\mathrm{n_{el}}\int_{\Omega
		^{\mathrm{e}}}\bigg( \Big(\boldsymbol{w}_\mathrm{h}^{n+\alpha}\cdot\nabla \psi_{\mathrm{h}}
	\Big)\tau_{\phi}\Big( \partial_t{\phi}_{\mathrm{h}}^{n+\alpha_m} + \boldsymbol{w}_\mathrm{h}^{n+\alpha}\cdot
	\nabla\phi_{\mathrm{h}} -\gamma\big(t^{n+\alpha}\big)\big(\nabla\cdot(\hat{k}\nabla\phi_{\mathrm{h}}) -
	\hat{s}\phi_{\mathrm{h}} +\hat{f}\big) \Big) \bigg) \mathrm{d}\Omega^{\mathrm{e}}=0,
\end{align}
where $\boldsymbol{w}$, $\hat{k}$, $\hat{s}$ and $\hat{f}$ are the GMV, the modified diffusion coefficient, modified reaction coefficient and modified source respectively which are defined in \cite{joshi2018positivity}. The $\gamma(t^{n+\alpha})$ is the time-dependent mobility model, the implementation details of which can be found in \cite{mao2021variational}.
In Eq.~(\ref{PPV_AC}), the first line is the Galerkin projection of the transient, convection, diffusion, reaction and source terms, the second line represents the Petrov-Galerkin stabilization terms where
the stabilization parameters $\tau_\phi$ are given by \cite{shakib1991new, brooks1982streamline}:

\begin{align}
	\tau_{\phi} &= \left( \bigg(\frac{2}{\Delta t} \bigg)^2 + \boldsymbol{w}_\mathrm{h}\cdot\boldsymbol{G}\boldsymbol{w}_\mathrm{h} + 9\hat{k}^2 \boldsymbol{G}:\boldsymbol{G} + \hat{s}^2 \right)^{-1/2}.
\end{align}

\subsubsection{Material coordinates in the Eulerian description}
Similarly, we define the space of trial solution $\mathcal{S}_{\boldsymbol{\xi}}^{\mathrm{h}}$ and test functions $\mathcal{V}_{\boldsymbol{\psi}}^{\mathrm{h}}$ as:
\begin{align}
	\mathcal{S}^\mathrm{h}_{\boldsymbol{\xi}} &= \big\{ \boldsymbol{\xi}_\mathrm{h}\ |\ \boldsymbol{\xi}_\mathrm{h} \in (H^1(\Omega))^{d}\big\},\\
	\mathcal{V}^\mathrm{h}_{\boldsymbol{\psi}} &= \big\{ \boldsymbol{\psi}_\mathrm{h}\ |\ \boldsymbol{\psi}_\mathrm{h} \in (H^1(\Omega))^d \big\}.
\end{align}

The variational statement of the material coordinates can be written as:
find $\boldsymbol{\xi}_{\mathrm{h}}(t^{\mathrm{n+\alpha}})\in\mathcal{S}^{\mathrm{h}}_{\boldsymbol{\xi}}$ such that $\forall \boldsymbol{\psi}_{\mathrm{h}}\in\mathcal{S}^{\mathrm{h}}_{\boldsymbol{\psi}}$
\begin{align}
	&\int_{\Omega}\left(\partial_t\boldsymbol{\xi}_{\mathrm{h}}^{n+\alpha_m}+\boldsymbol{w}_{\mathrm{h}}\cdot\nabla\boldsymbol{\xi}_{\mathrm{h}}\right)\cdot\boldsymbol{\psi}_{\mathrm{h}}\mathrm{d\Omega}\\
	&+\sum_{\mathrm{e=1}}^{n_{\mathrm{el}}}\int_{\Omega^{\mathrm{e}}}\tau_{\boldsymbol{\xi}}\left(\boldsymbol{w}_\mathrm{h}\cdot\nabla\boldsymbol{\xi}_{\mathrm{h}}\right)\cdot\boldsymbol{\mathcal{R}}_{\boldsymbol{\xi}} \mathrm{d}\Omega^{\mathrm{e}}={0},
\end{align}
where $\boldsymbol{\mathcal{R}}_{\boldsymbol{\xi}}$  and $\tau_{\boldsymbol{\xi}}$ denote the element-wise residuals for the evolution equation of the material coordinates and the stabilization parameters, respectively. The first line is the Galerkin projection, the second line represents the Petrov-Galerkin stabilization term. The stabilization parameter is selected as $\tau_{\boldsymbol{\xi}}=\left((2/\Delta t)^2+\boldsymbol{w}_{\mathrm{h}}\cdot\boldsymbol{G}\boldsymbol{w}_{\mathrm{h}}\right)$.

\subsection{Linearization of the stress term for solid}
We employ Newton-Raphson iterations to find the solution of the momentum conservation equation. Therefore, the directional derivative of the solid stress term with respect to the velocity is needed. Similar to \cite{sun2014full}, we write the stress term as a function of the velocity utilizing the evolution of the left Cauchy-Green tensor. As a result, the Jacobian of the stress term with respect to the velocity can be easily derived and implemented numerically. Furthermore, the momentum conservation equation and the evolution equation for the material coordinates can be fully decoupled. According to the generalized-$\alpha$ discretization, the left Cauchy-Green tensor at $t^{\mathrm{n+\alpha}}$ can be derived as: 
\begin{align}
	\boldsymbol{B}^{n+\alpha}&=\boldsymbol{B}^{n}+\alpha (\boldsymbol{B}^{n+1}-\boldsymbol{B}^{n}),\\
	&=\boldsymbol{B}^{n}+\alpha \Delta t(\partial_t \boldsymbol{B}^n+\varsigma (\partial_t\boldsymbol{B}^{n+1}-\partial_t\boldsymbol{B}^n)),\\
	&=\boldsymbol{B}^{n}+\alpha \Delta t(\partial_t\boldsymbol{B}^n+\frac{\varsigma}{\alpha_m} (\partial_t\boldsymbol{B}^{n+\alpha_m}-\partial_t\boldsymbol{B}^n)),\\
	&=\boldsymbol{B}^{n}+\alpha \Delta t\left(\left(1-\frac{\varsigma}{\alpha_m}\right)\partial_t\boldsymbol{B}^n+\frac{\varsigma}{\alpha_m} \partial_t\boldsymbol{B}^{n+\alpha_m}\right).
\end{align}
The evolution equation of the left Cauchy-Green tensor is given by:
\begin{align}
	\partial_t\boldsymbol{B}^{n+\alpha_m}=-(\boldsymbol{w}^{n+\alpha}\cdot\nabla)\boldsymbol{B}^{n+\alpha}+\nabla\boldsymbol{w}^{n+\alpha}\cdot\boldsymbol{B}^{n+\alpha}+\boldsymbol{B}^{n+\alpha}\cdot(\nabla\boldsymbol{w}^{n+\alpha})^T.
\end{align}
The Jacobian matrix in the solid domain can be calculated as:
\begin{align}
	\frac{\delta\boldsymbol{B}^{n+\alpha}}{\delta \boldsymbol{v}^{n+\alpha}}=	\frac{\delta\boldsymbol{B}^{n+\alpha}}{\delta \boldsymbol{w}^{n+\alpha}}=\frac{\alpha\varsigma\Delta t}{\alpha_m}\left(-(\boldsymbol{N}\cdot\nabla)\boldsymbol{B}^{n+\alpha}+\nabla\boldsymbol{N}\cdot\boldsymbol{B}^{n+\alpha}+\boldsymbol{B}^{n+\alpha}\cdot\left(\nabla\boldsymbol{N}\right)^T\right),
\end{align} 
where $\boldsymbol{N}$ is a vector composed of the shape function.
Notice that $\boldsymbol{B}$ is not explicitly available in the current formulation. Therefore, we need to construct $\boldsymbol{B}^n$, $\boldsymbol{B}^{n+\alpha}$ and $\partial_t\boldsymbol{B}^n$ from known $\boldsymbol{\xi}^n$ and $\boldsymbol{\xi}^{n+\alpha}$.  This is realized by the backward projection technique. Taking $\boldsymbol{B}^n$ as an example, The nodal value of $\boldsymbol{\xi}^n$ is used to interpolate the $\nabla\boldsymbol{\xi}^n$ at the quadrature points, and $L^2$-projection is used to project the value on the quadrature points back to the nodes \cite{jaiman2016partitioned}. For node $p$, we have:
\begin{align}
	\nabla\boldsymbol{\xi}^n\large|_p=\frac{\sum_e\int_{\Omega^e}N_p \nabla\boldsymbol{\xi}^n d \Omega^{e}}{\sum_e\int_{\Omega^e}N_pd\Omega^e}.
\end{align}
Then, $\boldsymbol{B}^n$ can be calculated as $\boldsymbol{B}^n=(\nabla\boldsymbol{\xi}^n)^{-1}(\nabla\boldsymbol{\xi}^n)^{-T}$. $\boldsymbol{B}^{n+\alpha}$ can be calculated using $\boldsymbol{\xi}^{n+\alpha}$ in a similar manner. With $\boldsymbol{B}^n$ and $\boldsymbol{B}^{n+\alpha}$ readily available,  $\partial_t \boldsymbol{B}^n$ is initialized as zero and updated in the following time steps according to the generalized-$\alpha$ method:
\begin{align}
	\partial_t\boldsymbol{B}^{n+\alpha_m}=\left(1-\frac{\alpha_m}{\varsigma}\right)\partial_t\boldsymbol{B}^n+\frac{\alpha_m}{\varsigma\Delta t \alpha}\left(\boldsymbol{B}^{n+\alpha}-\boldsymbol{B}^n\right).
\end{align}
To avoid possible confusions caused by the tensor notation, the component-wise formulation for the residual and Jacobian matrix of the stress term is shown in \ref{Ap2}.
With the residual and Jacobian matrix of the solid stress, that of the stress in the unified momentum conservation equation is given through phase-dependent interpolation:
\begin{align}
	\boldsymbol{\sigma}&=\alpha(\phi)\boldsymbol{\sigma}^{\mathrm{s}}+(1-\alpha(\phi))\boldsymbol{\sigma}^{\mathrm{f}}\\
	\frac{\delta\boldsymbol{\sigma}}{\delta\boldsymbol{v}}&=\alpha(\phi)\frac{\delta\boldsymbol{\sigma}^{\mathrm{s}}}{\delta\boldsymbol{v}}+(1-\alpha(\phi))\frac{\delta\boldsymbol{\sigma}^{\mathrm{f}}}{\delta\boldsymbol{v}}
\end{align}

\subsection{Implementation details}
In this subsection, we present the implementation details of our variational framework. The fully Eulerian FSI framework is decoupled and solved in a  partitioned-block iterative manner which leads to flexibility and ease in its implementation to the existing variational solvers. The root finding process of each block employs the Newton-Raphson method, which can be expressed in terms of the solution increments of the velocity, the pressure, the order parameter and the material coordinates ($\Delta \boldsymbol{u},\Delta p,\Delta \phi$ and $\Delta \boldsymbol{\xi}$ respectively).  
We start with the increment of the velocity and pressure: 
\begin{align} \label{LS_NS}
	\begin{bmatrix}
		\boldsymbol{K}_{\Omega}& & \boldsymbol{G}_{\Omega}\\ \noalign{\vspace{4pt}}
		-\boldsymbol{G}^T_{\Omega}& &\boldsymbol{C}_{\Omega}
	\end{bmatrix}
	\begin{Bmatrix}
		\Delta\boldsymbol{v}\\ \noalign{\vspace{4pt}}
		\Delta p
	\end{Bmatrix}
	&=
	-\begin{Bmatrix}
		\overline{\boldsymbol{\mathcal{R}}}_\mathrm{m} \\ \noalign{\vspace{4pt}}
		\overline{\mathcal{R}}_\mathrm{c}
	\end{Bmatrix}
\end{align}
where $\boldsymbol{K}_{\Omega}$ is the stiffness matrix of the momentum equation consisting of transient, convection, viscous and Petrov-Galerkin stabilization terms, $\boldsymbol{G}_{\Omega}$ is the gradient operator, $\boldsymbol{G}^T_{\Omega}$ is the divergence operator for the continuity equation and $\boldsymbol{C}_{\Omega}$ is the stabilization term for cross-coupling of pressure terms. $\overline{\boldsymbol{\mathcal{R}}}_\mathrm{m}$ and $\overline{\mathcal{R}}_\mathrm{c}$ represent the weighted residuals of the variational forms of the momentum and continuity equation. The updated velocity is used to construct the gradient-minimizing velocity field. Noting that the governing equation for GMV is an elliptic PDE, the generalized-$\alpha$ method is not applied. 
We solve the elliptic equation of GMV for multiple sub-iterations within the nonlinear iteration till convergence is reached:
\begin{align}
	\begin{bmatrix}
		\boldsymbol{K}_{\boldsymbol{w}}
	\end{bmatrix}
	\begin{Bmatrix}
		\Delta\boldsymbol{w}
	\end{Bmatrix}
	&=
	-\begin{Bmatrix}
		\overline{\boldsymbol{\mathcal{R}}}(\boldsymbol{w})
	\end{Bmatrix} \label{LS_AC}
\end{align}
where $\boldsymbol{K}_{\boldsymbol{w}}$ is the stiffness matrix for the construction equation of GMV and $	\overline{\boldsymbol{\mathcal{R}}}(\boldsymbol{w})$ is the weighted residual of the governing equation of GMV. After that, the constructed GMV is passed to the Allen-Cahn equation and the evolution equation of the material coordinates:
\begin{align}
	\begin{bmatrix}
		\boldsymbol{K}_{AC}
	\end{bmatrix}
	\begin{Bmatrix}
		\Delta\phi
	\end{Bmatrix}
	&=
	-\begin{Bmatrix}
		\overline{\mathcal{R}}(\phi)
	\end{Bmatrix} \label{LS_AC}\\
	\begin{bmatrix}
		\boldsymbol{K}_{MC}
	\end{bmatrix}
	\begin{Bmatrix}
		\Delta\boldsymbol{\xi}
	\end{Bmatrix}
	&=
	-\begin{Bmatrix}
		\overline{\boldsymbol{\mathcal{R}}}(\boldsymbol{\xi})
	\end{Bmatrix} \label{LS_MC}
\end{align}
where $\boldsymbol{K}_{\mathrm{AC}}$ and $\boldsymbol{K}_{MC}$ are the stiffness matrices of the Allen-Cahn equation and the evolution equation of the material coordinates, $\overline{\mathcal{R}(\phi)}$ and $	\overline{\boldsymbol{\mathcal{R}}}(\boldsymbol{\xi})$ are the weighted residuals of the Allen-Cahn equation and the governing equation for the material coordinates. This finishes one nonlinear iteration.
The nonlinear iteration stops when the ratio between the $L^2$ norm of the increment and the current value is less than $5\times 10^{-4}$ for all the blocks, or the nonlinear iteration number exceeds the maximum iteration number, which is set as four in the current work. 

\section{Test cases}
In this section, we examine the effect of the IGP method for cases of increasing complexity. We first demonstrate the interface and geometry preserving effect of the IGP method in cases of convection of circular and square interfaces in a prescribed velocity field. After clarifying the interface and geometry preserving effect, we incorporate the FSI dynamics in the case of channel flow passing a fixed deformable block, which causes a static deformation of the solid. Finally, we consider the cylinder-flexible plate problem to showcase the IGP method in the FSI case with oscillatory structure motion, large aspect ratio and sharp corner. As a complement, a benchmarking of our fully Eulerian FSI solver in the case of deformation of a solid block driven by a cavity flow is presented in \ref{cavity block}.

\subsection{Convection of square and circular interfaces in a prescribed velocity field}
In this subsection, we demonstrate the interface and geometry preserving effect of the IGP method with a prescribed velocity field. The evolution of circular and square interfaces is considered because they cover the geometry of planar interface, curved interface and sharp corner, which can serve as building blocks for more complex geometries. The velocity field is prescribed such that the interface is thickened in the $X$-direction and thinned in the $Y$-direction.

\begin{figure}[h]
	\begin{minipage}[b]{0.5\textwidth}
		\centering
		\includegraphics[scale=0.6,trim=0 0 0 0,clip]{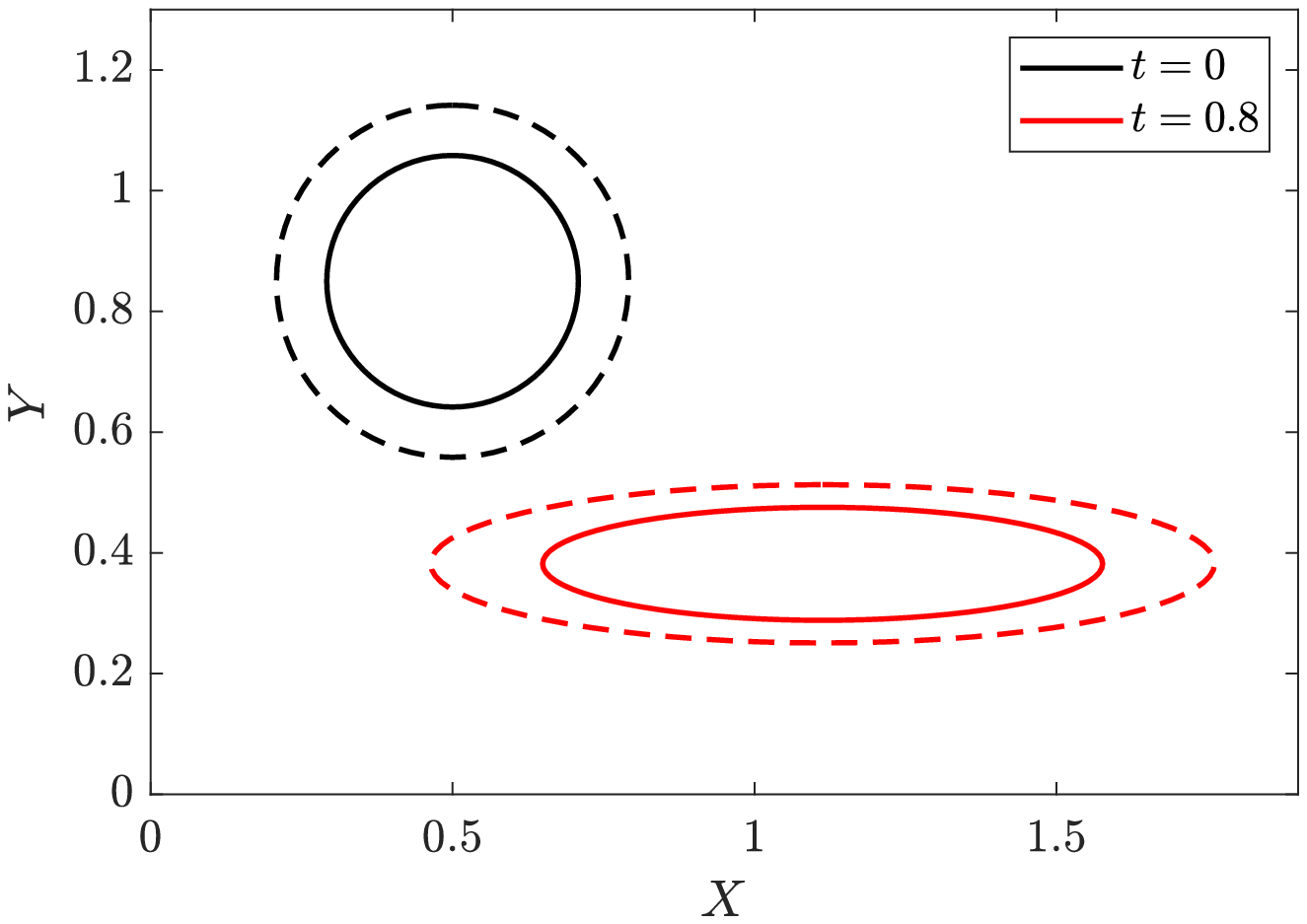}
		\caption*{(a)}
	\end{minipage}
	\begin{minipage}[b]{0.5\textwidth}
		\centering
		\includegraphics[scale=0.6,trim=0 0 0 0,clip]{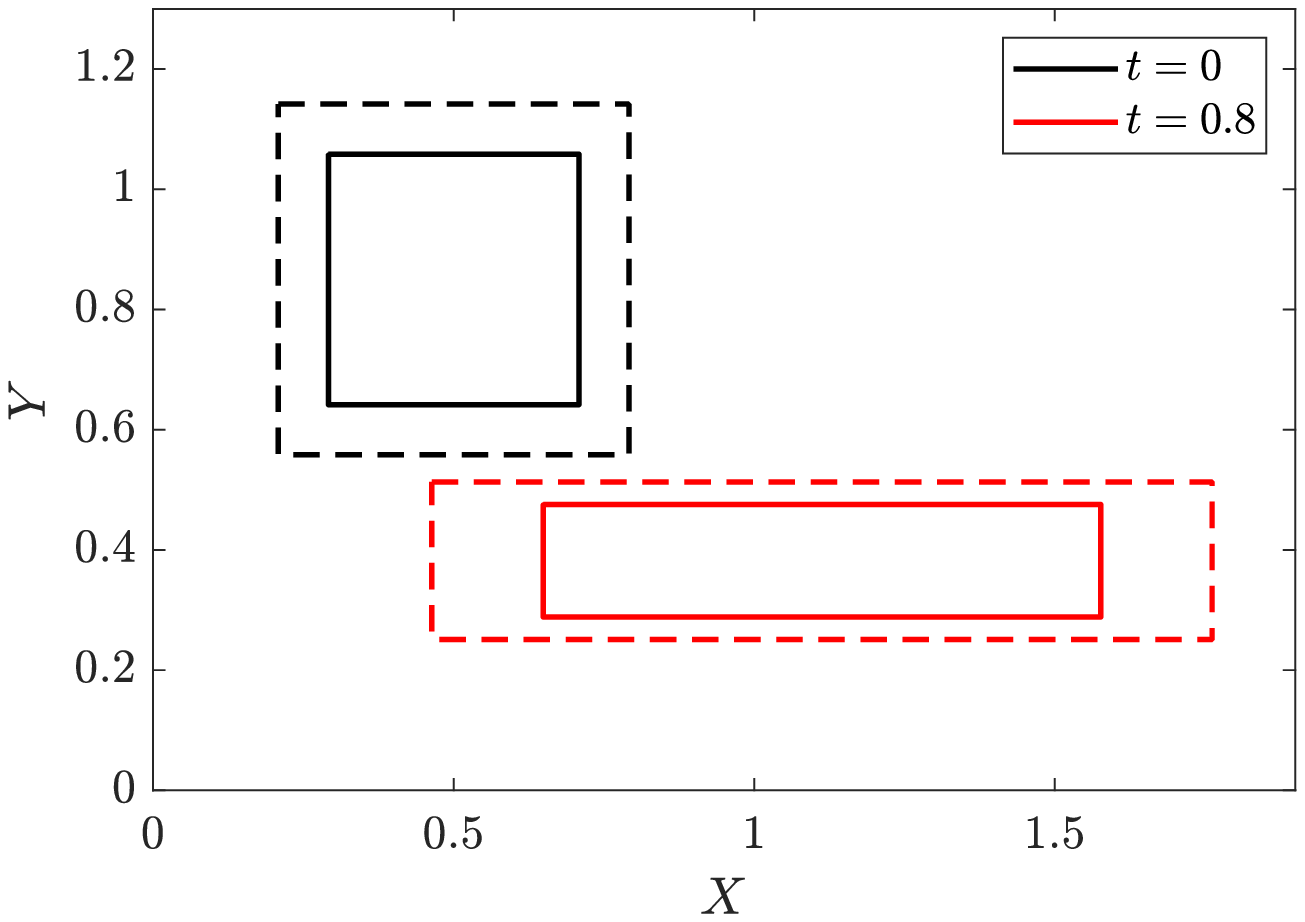}
		\caption*{(b)}
	\end{minipage}
	\caption{The interface distortion of a circular and square interface in the velocity field $v_x=x,v_y=-y$ till $t=0.8$. The diffuse interface region is shown as the space between $\phi=0.9$ (solid line) and $\phi=-0.9$ (dashed line).}
	\label{UV_ana}
\end{figure}

The computational domain is defined as $[0,2]\times[0,1.5]$. We first examine a circular interface, the order parameter field of which is initialized as:
\begin{equation}
	\phi(x,y)=\tanh\left(\frac{R-\sqrt{(x-x_c)^2+(y-y_c)^2}}{\sqrt{2}\varepsilon}\right),
\end{equation}
where $R=0.25$ and $(x_c,y_c)=(0.5,0.85)$ are the radius and the center of the circular interface given by $\phi=0$. The diffuse interface thickness parameter is selected as $\varepsilon=1/50$. The circular interface is convected in an incompressible velocity field $v_x=x,v_y=-y$.  In this velocity field, the location of the interface can be calculated analytically by solving ordinary differential equations $d\boldsymbol{x}/dt=\boldsymbol{v}$. In the $X$-direction, the velocity gradient $\partial v_x/\partial x=1>0$, which means that the relative velocity of the level sets of the order parameter is positive. Therefore, the level sets will move far away from each other and the diffuse interface region is thickened in the $X$-direction. Similarly, due to the negative velocity gradient in the $Y$-direction $\partial v_y/\partial y=-1<0$, the diffuse interface region is thinned in the $Y$-direction. To clearly show the resulting interface distortion due to the convection, we use $\phi=\delta$ and $\phi=-\delta$ to illustrate the diffuse interface region, where $\delta=0.9$ is selected in the current and following cases. Considering the interface at $t=0.8$, the diffuse interface region at the initial and final time instances are plotted in Fig. \ref{UV_ana} (a). With the same set up, the interface distortion of a square interface whose diagonal points of $\phi=0$ are located at $(0.25,0.6)$ and $(0.75,1.1)$ is shown in Fig. \ref{UV_ana} (b).

\begin{figure}[h]
	
	\begin{minipage}[b]{0.5\textwidth}
		\centering
		\includegraphics[scale=0.6,trim=0 0 0 0,clip]{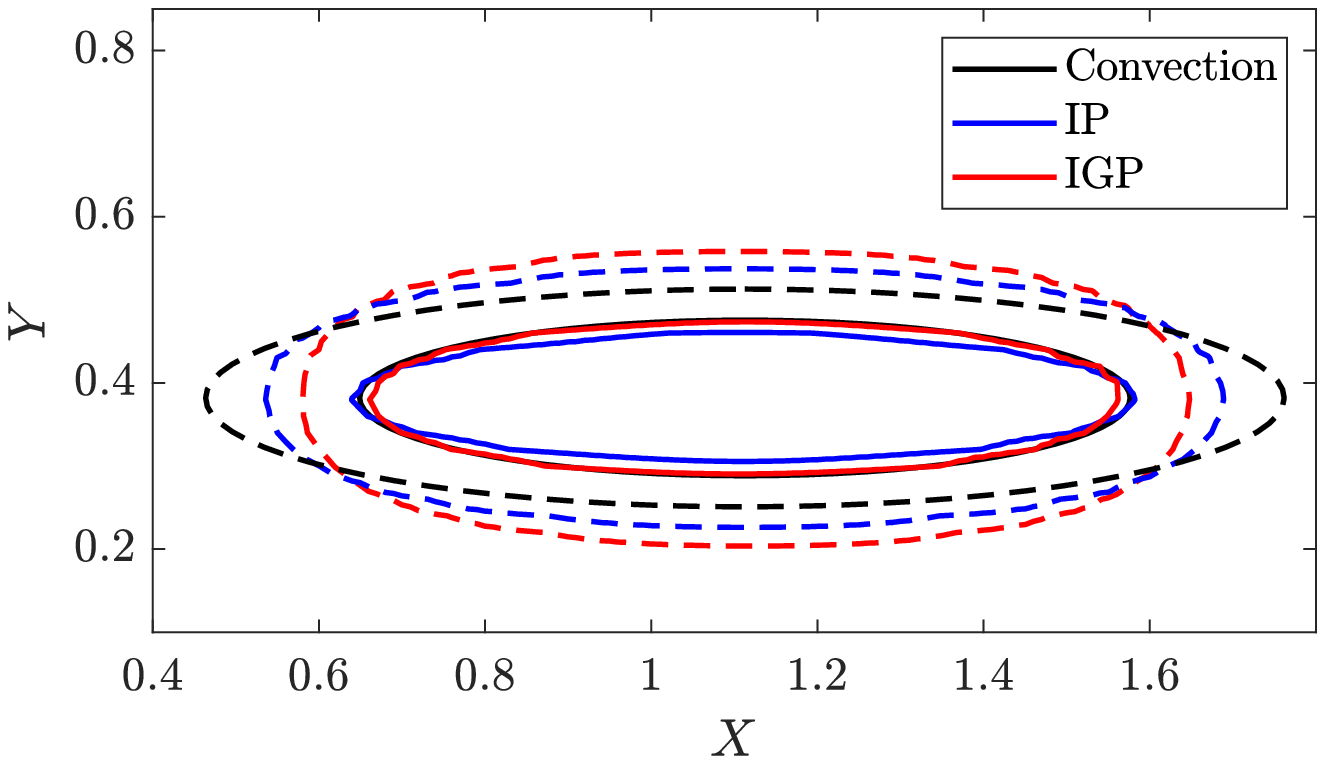}
		\caption*{\hspace{0.8cm}(a)}
	\end{minipage}
	\begin{minipage}[b]{0.5\textwidth}
		\centering
		\includegraphics[scale=0.6,trim=0 0 0 0,clip]{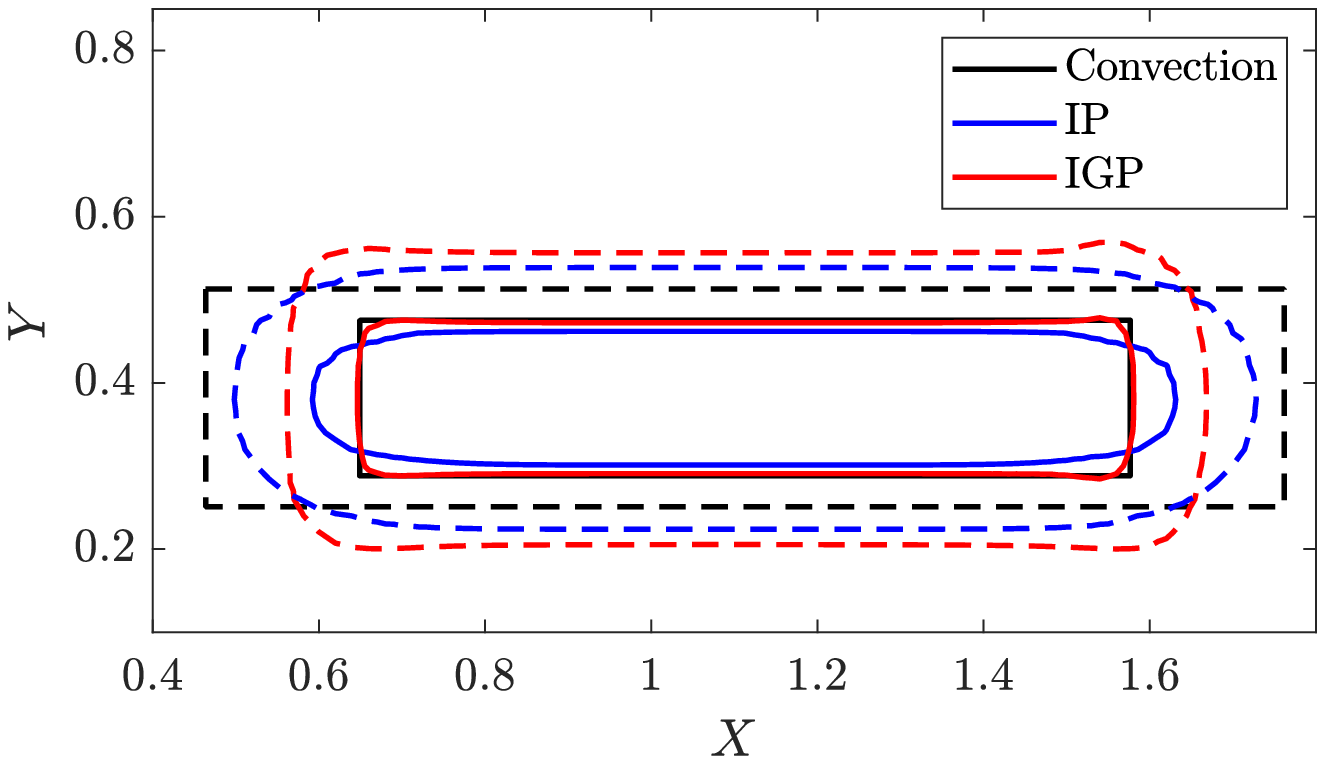}
		\caption*{\hspace{0.8cm}(b)}
	\end{minipage}

	\begin{minipage}[b]{0.5\textwidth}
		\centering
		\includegraphics[scale=0.6,trim=0 0 0 0,clip]{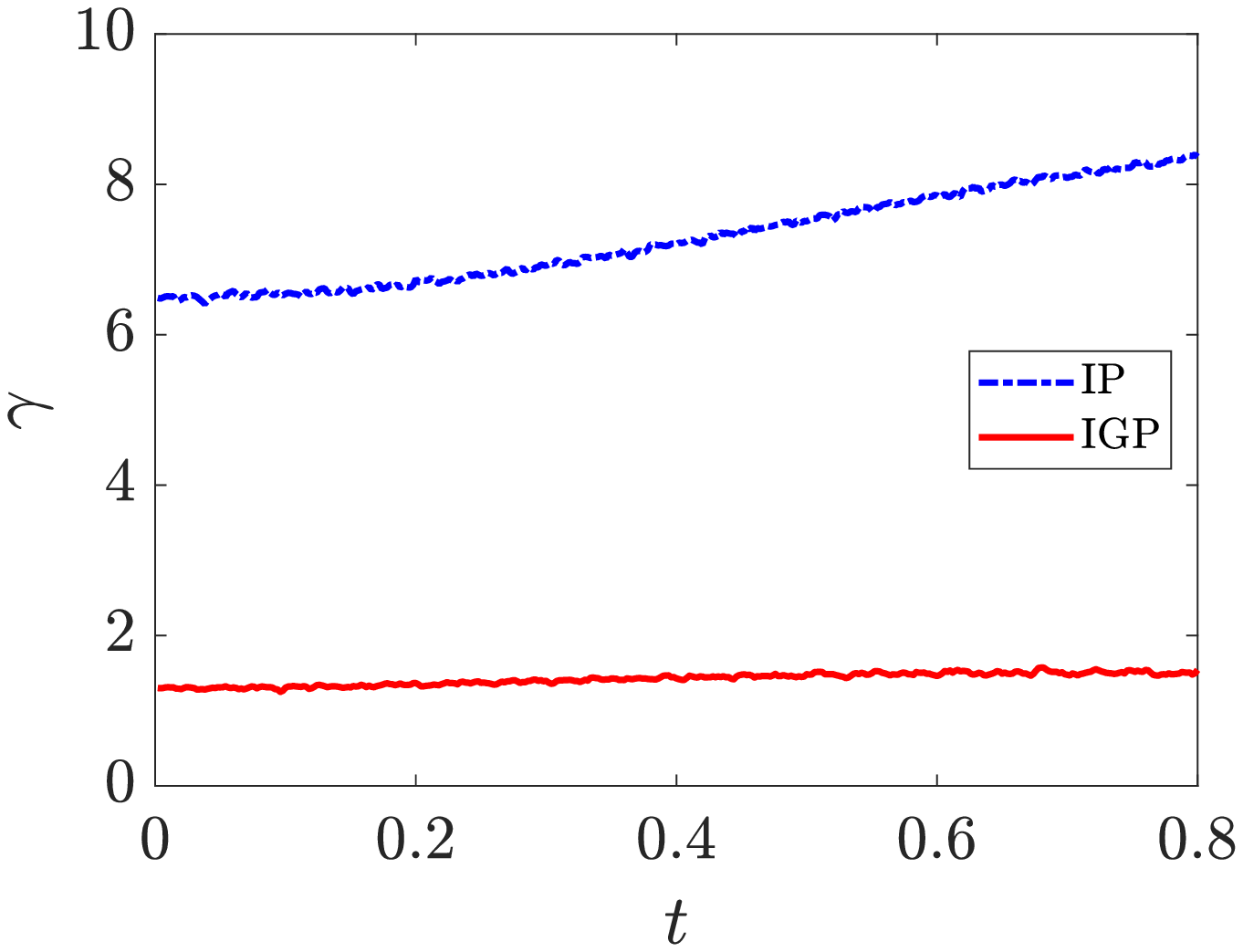}
		\caption*{\hspace{1cm}(c)}
	\end{minipage}
	\begin{minipage}[b]{0.5\textwidth}
		\centering
		\includegraphics[scale=0.6,trim=0 0 0 0,clip]{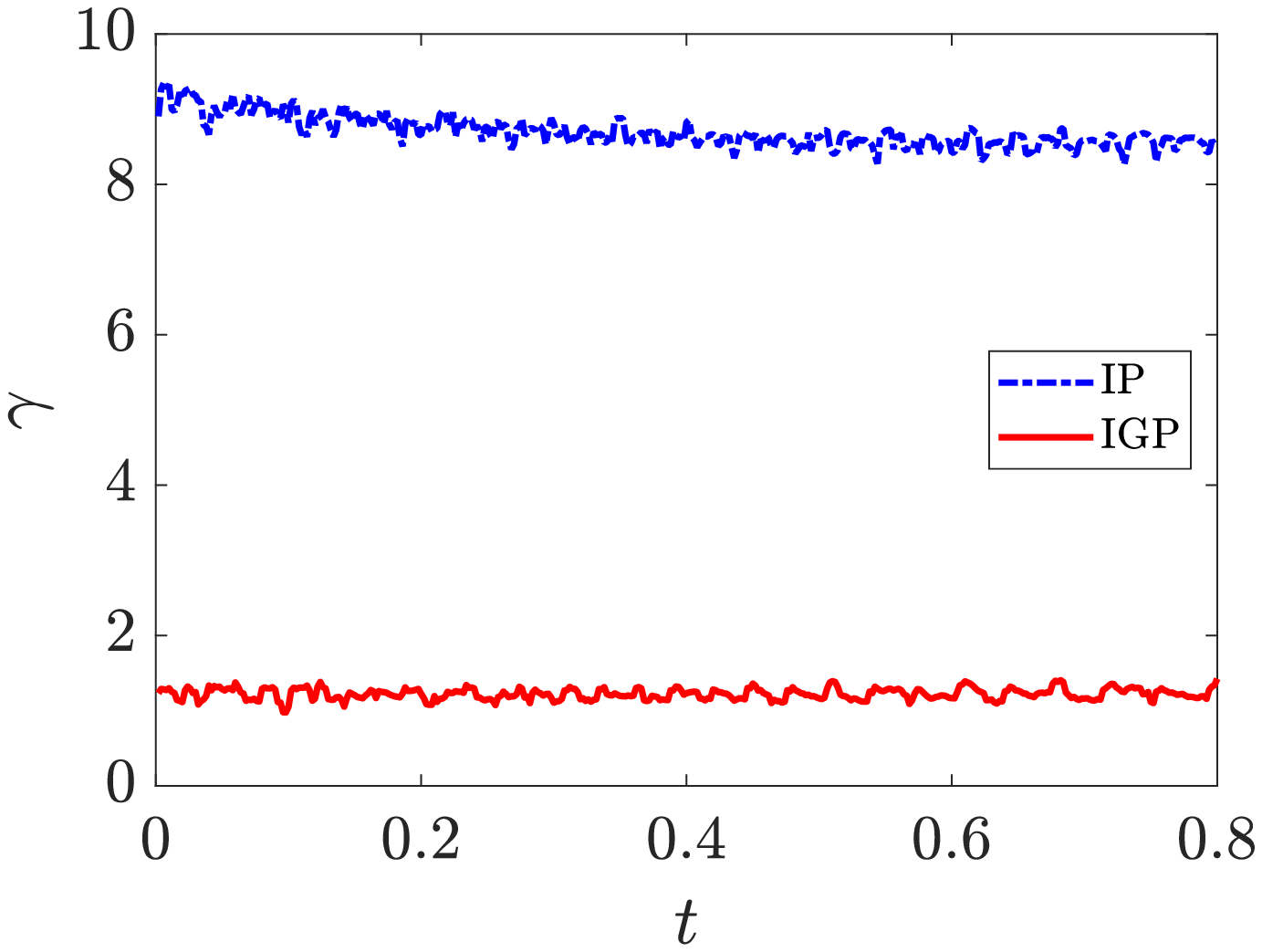}
		\caption*{\hspace{1cm}(d)}
	\end{minipage}
	
	\caption{Diffuse interface region at $t=0.8$ given by solely convection, the IP method and the IGP method in the convection of circular and square interfaces, and the corresponding time history of the mobility coefficient for quantifying the volume-conserved mean curvature flow. }
	\label{UV_num}
\end{figure}

We solve this problem numerically with the IP method and the IGP method respectively. The computational domain is discretized with a structured triangle mesh, the size of which is taken as $h=1/50$. The RMS interface distortion parameter is selected as $\eta=0.1$. The time step is taken as $\Delta t=0.002$. The diffuse interface region calculated through the IP method and the IGP method for the circular and square interfaces are shown in Fig. \ref{UV_num} (a) and (b). The time history of the mobility coefficient is shown in Fig. \ref{UV_num} (c) and (d) respectively.

As shown in Fig. \ref{UV_num} (a) and (b), in the analytical solution which only considers the convection, the diffuse interface region becomes thickened and thinned in $X$- and $Y$-directions, respectively. In contrast, both the IP and IGP methods can maintain the thickness of the diffuse interface region for both the circular and square interfaces. While the geometry preservation effect of the IGP method is not obvious in the circular interface case, it can be seen clearly in the square interface case. Because in the latter case, the sharp corner leads to a significant difference between the local curvature and the average curvature along the interface, thus larger volume-conserved mean curvature flow. Qualitatively, we can compare the reduction of the volume-conserved mean curvature flow by comparing the mobility coefficient ($\boldsymbol{v}_{\kappa}=\gamma(t)\varepsilon^2(\kappa-\overline{\kappa})\boldsymbol{n}$). As seen in Fig. \ref{UV_num} (c) and (d), a reduction in the mobility coefficient at around factors of 5 and 9 are attained through the IGP method.


\subsection{Channel flow passing a fixed deformable block}
\begin{figure}[h]
	\begin{minipage}[b]{0.48\textwidth}
	\centering
	\begin{tikzpicture}[scale=8]
		\draw (0,0) rectangle (0.8,0.4);
		\node at (0.55,0.2) {$\Omega^f$};
		\draw [fill=gray,fill opacity=0.5] (0.2,0) rectangle (0.3,0.2);

		\node at (0.25,0.1) {$\Omega^s$};
		\draw[scale=1, domain=0:0.4, smooth, variable=\y]  plot ({-2*\y*(\y-0.4)}, {\y});
		\draw[-stealth,thick] (0,0.2)--(0.08,0.2);
		\draw[-stealth,thick] (0,0.1)--(0.06,0.1);
		\draw[-stealth,thick] (0,0.3)--(0.06,0.3);
		\filldraw (0.25,0.2) circle (0.008);
		\node at (0.25,0.33) {$\mathrm{Marker}$};
		\node at (0.25,0.27) {$\mathrm{point}$};
	\end{tikzpicture}
\caption*{(a)}
\end{minipage}
\begin{minipage}[b]{0.48\textwidth}
	\centering
	\includegraphics[scale=0.13,trim=0 0 0 0,clip]{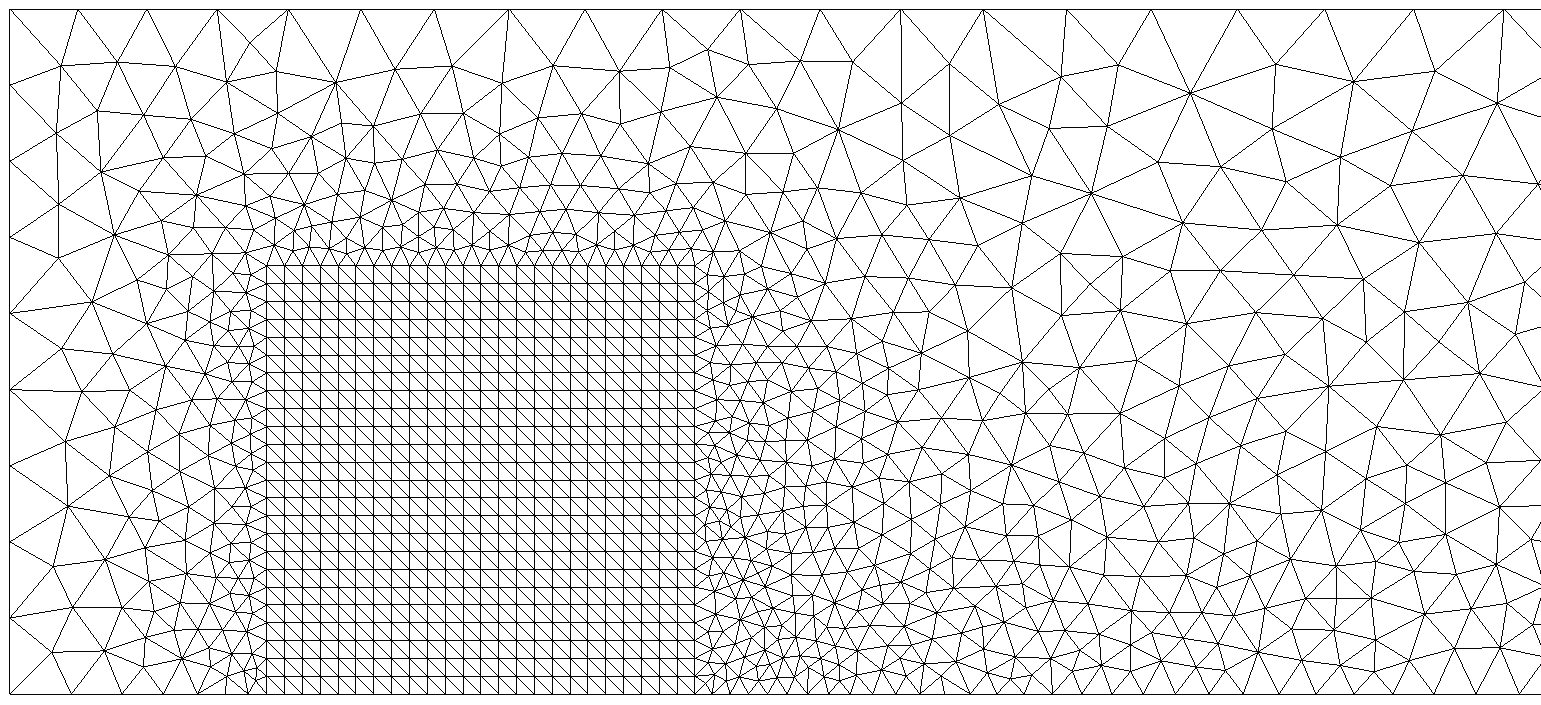}
	\caption*{(b)}
\end{minipage}
	\caption{Flow passing a fixed deformable block in a channel domain: (a) schematic diagram showing the computational domain, (b) structured finite element mesh covering the block and the surrounding unstructured mesh.}
	\label{CPB}
\end{figure}

In this subsection, we incorporate FSI dynamics with the case of channel flow passing a fixed deformable block, which causes static deformation of the solid. The computational domain $[0,2.5]\times[0,0.4]$ is considered. The off-diagonal coordinates of the rectangular block are located at $(0.2,0)$ and $(0.3,0.2)$. The bottom of the block is fixed at the wall. The order parameter field representing the block is initialized by calculating the signed distance to the left, top and right boundary of the block $d$ (bottom boundary is not included because there is no fluid-solid interface). Then $d$ is composed by a hyperbolic tangent function, which result in $\phi(\boldsymbol{x},0)=\tanh(d/\sqrt{2}\varepsilon)$.  An inflow condition is prescribed at the left boundary, whose $X$-component velocity at the inlet is given by a parabolic function with ramping in time:
\begin{align}
	v_{x}(0,y,t)=
	\begin{cases} 0.5 \left(1-\cos(2\pi t)\right)y(H-y)/(0.5H)^2, \quad\quad t\leqslant0.5,\\
		y(H-y)/(0.5H)^2,\hspace{4.05cm}t>0.5,
	\end{cases}
\end{align}
where $H=0.4$ represents the height of the computational domain. The no-slip boundary condition is applied on the top and bottom boundary. The outflow condition is applied on the right boundary. The densities of the solid and the fluid are chosen as $\rho^{\mathrm{s}}=1000$, $\rho^{\mathrm{f}}=1000$,  the dynamic viscosity of the fluid is $\mu^{\mathrm{f}}=1$, the shear modulus of the solid is $\mu^{\mathrm{s}}=2\times10^6$. A marker is placed at the mid point of the top boundary of the block to monitor its deformation, which is initialized as $\boldsymbol{\xi}=(0.25,0.2)$ . The problem setup is illustrated in Fig. \ref{CPB} (a).

\subsubsection{Comparison between the IP method and the IGP method}
\begin{figure}[h]
	
	\begin{minipage}[b]{0.48\textwidth}
		\centering
		\includegraphics[scale=0.6,trim=0 0 0 0,clip]{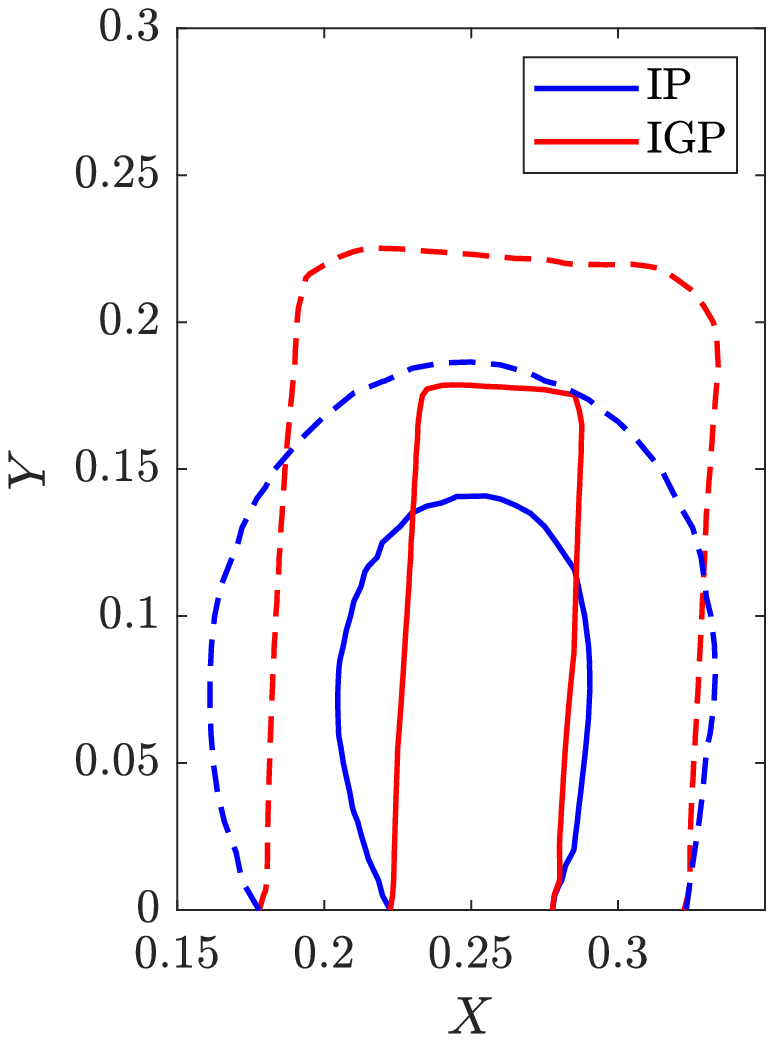}
		\caption*{\hspace{0.8cm}(a)}
	\end{minipage}
	\begin{minipage}[b]{0.48\textwidth}
		\centering
		\includegraphics[scale=0.6,trim=0 0 0 0,clip]{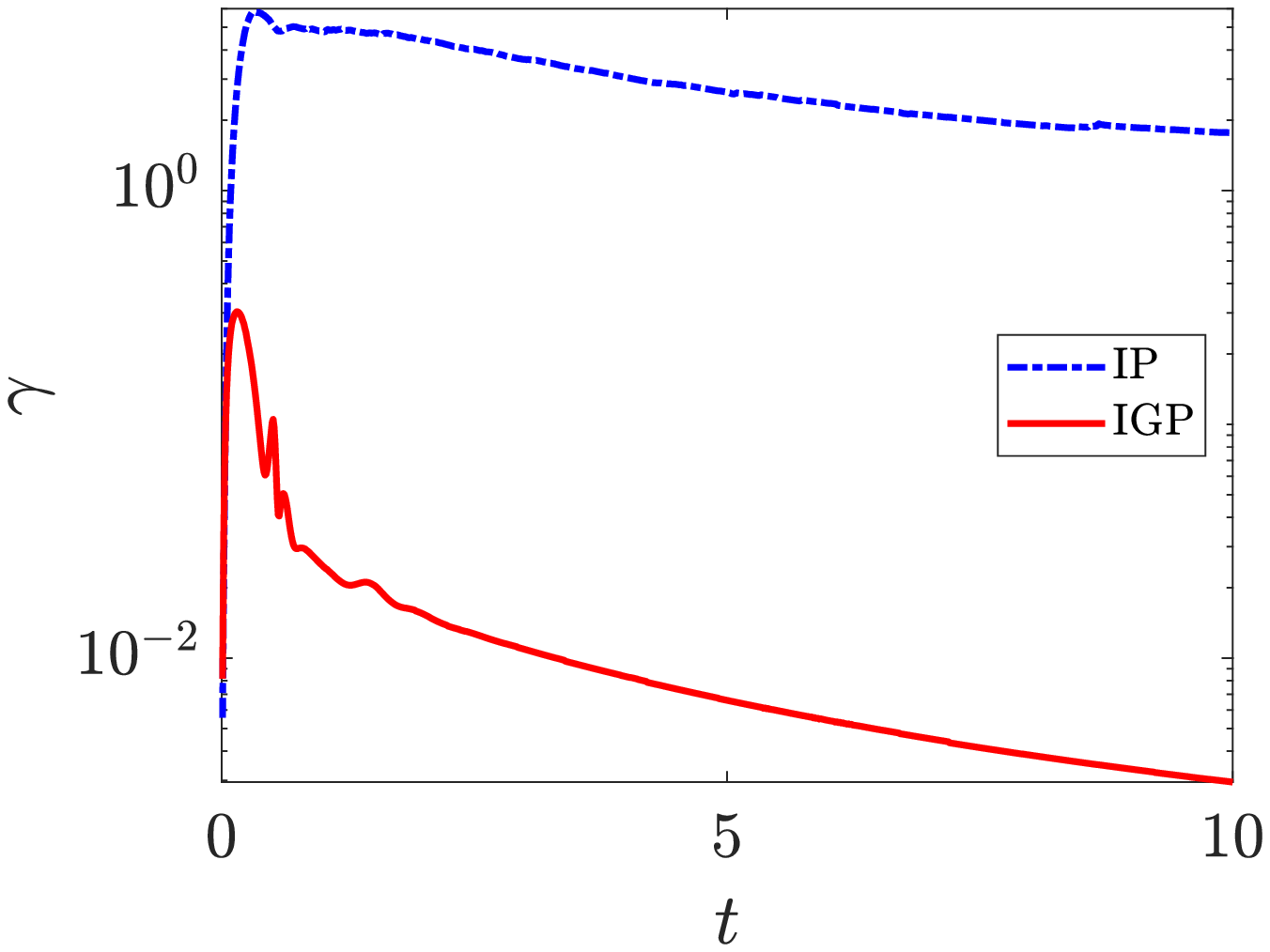}
		\caption*{\hspace{1.8cm}(b)}
	\end{minipage}
	\caption{Flow passing a fixed deformable block in a channel: (a) diffuse interface region at $t=10$ given by the IP method and the IGP method, (b) time history of the mobility coefficient.}
	\label{CPBcom}
\end{figure}

To begin with, we perform a systematic comparison between the IP method and the IGP method in a coarse mesh. The computational domain near the block $[0.15,0.4]\times[0,0.25]$ is discretized with structured triangle mesh at size $h=0.01$, while the rest of the computational domain is discretized with an unstructured triangle mesh. The mesh is illustrated in Fig. \ref{CPB} (b). The diffuse interface thickness parameter is selected as $\varepsilon/h=1$, where the diffuse interface region is resolved with four elements in the normal direction. The RMS interface distortion parameter is selected as $\eta=0.1$. The time step is selected as $\Delta t=0.01$. We compare the solution from the IP method and the IGP method, where the convection velocity for the order parameter is changed from $\boldsymbol{v}$ to $\boldsymbol{w}$ while the rest are identical. Specifically, $\boldsymbol{w}$ is still used for the convection of the material coordinates for both methods to avoid entanglement of $\boldsymbol{\xi}$. While the diffuse interface region at $t=10$ is shown in Fig. \ref{CPBcom} (a), the time history of the mobility coefficient is depicted in Fig. \ref{CPBcom} (b). As observed, the IGP method preserves the geometry of the block substantially. The mobility coefficient, which is proportional to the curvature flow, is reduced at around two orders of magnitude.

\subsubsection{Convergence of the diffuse interface model}
We next turn out attention to the convergence of the interface modeling. In the phase-field diffuse interface model, the diffuse interface profile should be close to the hyperbolic tangent profile so that the transition of physical properties across the diffuse interface region is uniform. Moreover, the volume-conserved mean curvature flow should be small enough so that the geometry of the interface is not severely disturbed. We first consider a group of interface parameters $\eta_0$ and $\varepsilon_0$ for the flow field velocity $\boldsymbol{v}$. The interface distortion is controlled by $\eta_0$. The volume-conserved mean curvature flow is given by:
\begin{align}
	\boldsymbol{v}_{\kappa0}(\boldsymbol{v})=\frac{1}{\eta_0}\mathcal{F}\left(\left|\frac{\partial v_n}{\partial n}\right|_{\varepsilon_0}\right)\varepsilon_0^2(\kappa-\overline{\kappa})\boldsymbol{n}.
\end{align}
Now we consider the convergence of the interface modeling for interface parameters $\eta_1=(1/2) \eta_0,\varepsilon_1=(1/2) \varepsilon_0$.
The convergence of the interface profile is relatively straightforward. The interface distortion is directly proportional to $\eta$. For example $\eta_1=(1/2)\eta_0<\eta_0$, the interface distortion is decreased at $\eta_1$. For the volume-conserved mean curvature flow, the current velocity can be derived as:
\begin{align}\label{reduce curvature flow vel}
	\boldsymbol{v}_{\kappa1}(\boldsymbol{v})=\frac{1}{\eta_1}\mathcal{F}\left(\left|\frac{\partial v_n}{\partial n}\right|_{\varepsilon_1}\right)\varepsilon_1^2(\kappa-\overline{\kappa})\boldsymbol{n}.
\end{align}
Note that in FSI problems, the physical properties vary rapidly across the diffuse interface region, which leads to a large velocity gradient. As we reduce the interface thickness by $\varepsilon_1=(1/2)\varepsilon_0$, the transition happens at a shorter distance. The resulting velocity change happens in a shorter distance as well, which leads to an increased velocity gradient. We approximate this increment through:
\begin{equation}
	\left|\frac{\partial v_n}{\partial n}\right|_{\varepsilon_1}\approx 2\left|\frac{\partial v_n}{\partial n}\right|_{\varepsilon_0}.
\end{equation}
Substituting $\eta_1=(1/2)\eta_0$, $\varepsilon_1=(1/2)\varepsilon_0$ and $\left|\partial v_n/\partial n\right|_{\varepsilon_1}\approx 2\left|\partial v_n/\partial n\right|_{\varepsilon_0}$ into Eq. (\ref{reduce curvature flow vel}), we have:
\begin{equation}
	\boldsymbol{v}_{\kappa1}(\boldsymbol{v})\approx\boldsymbol{v}_{\kappa0}(\boldsymbol{v}).
\end{equation}
Now we consider the gradient-minimizing velocity field $\boldsymbol{w}$. Since $\boldsymbol{w}$ is a constructed velocity field based on the solid velocity, which is independent of the transition at the fluid-solid interface, we can expect:
\begin{equation}
		\left|\frac{\partial w_n}{\partial n}\right|_{\varepsilon_1}\approx \left|\frac{\partial w_n}{\partial n}\right|_{\varepsilon_0}.
\end{equation}
Substituting $\eta_1=(1/2)\eta_0$, $\varepsilon_1=(1/2)\varepsilon_0$ and $\left|\partial w_n/\partial n\right|_{\varepsilon_1}\approx \left|\partial w_n/\partial n\right|_{\varepsilon_0}$, we have:
\begin{equation}
		\boldsymbol{v}_{\kappa1}(\boldsymbol{w})\approx(1/2)\boldsymbol{v}_{\kappa0}(\boldsymbol{w})\leq\boldsymbol{v}_{\kappa0}(\boldsymbol{w}),
\end{equation}
where the reduction of the volume-conserved mean curvature flow can be realized. Together with the interface profile convergence ensured by $\eta_1\leq\eta_0$, the interface modeling convergence is investigated via the refinement approach that scales down $\eta$ and $\varepsilon$ simultaneously with the same factor. Although in practical FSI simulations, the reduction in the volume-conserved mean curvature flow may be less than the approximation from this analysis, the reason of which can include the change of dynamics during the refinement, an improved model convergence can be achieved by the IGP method compared to the IP method.

\begin{figure}[h]
	
	\begin{minipage}[b]{0.48\textwidth}
		\centering
		\includegraphics[scale=0.6,trim=0 0 0 0,clip]{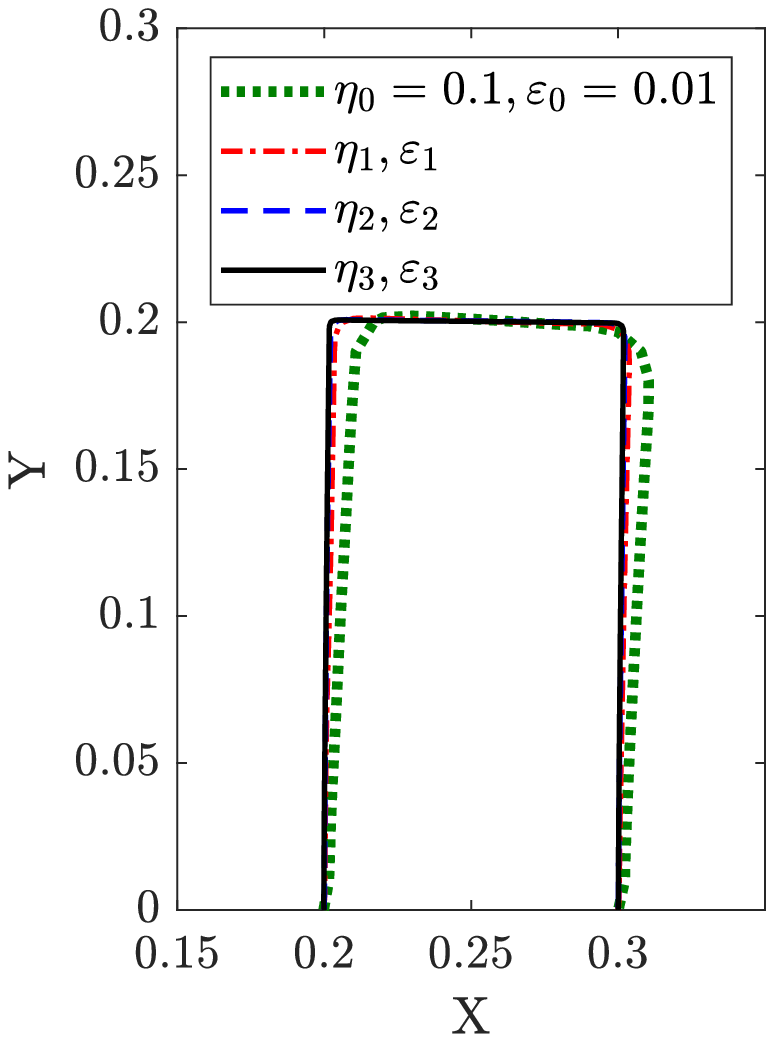}
		\caption*{\hspace{0.8cm}(a)}
	\end{minipage}
	\begin{minipage}[b]{0.48\textwidth}
		\centering
		\includegraphics[scale=0.6,trim=0 0 0 0,clip]{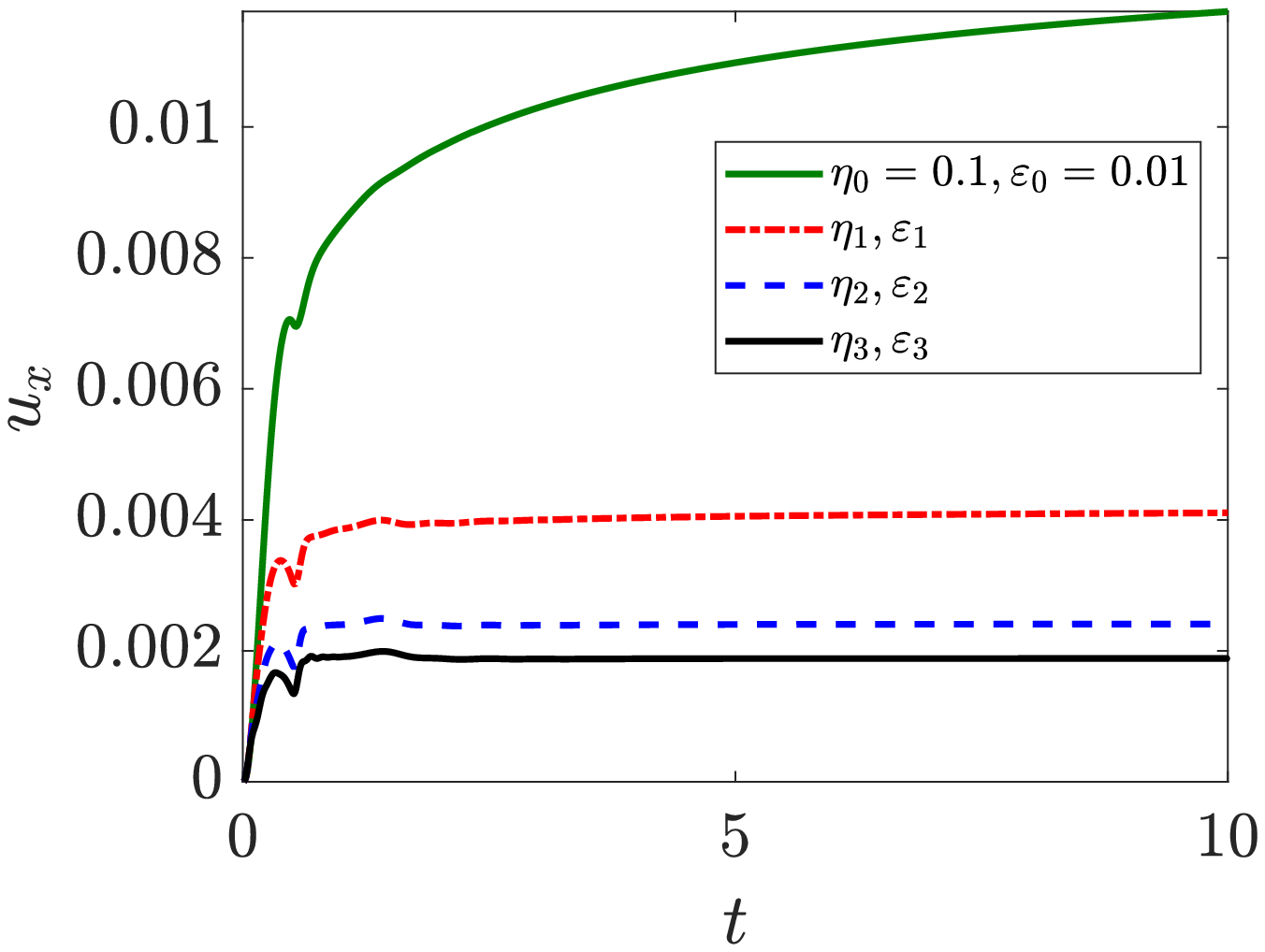}
		\caption*{\hspace{1.8cm}(b)}
	\end{minipage}
	\caption{Convergence of the diffuse interface modeling for the channel flow with a fixed deformable block. The subscript denotes the number of times of the bisection ($\eta_n=(1/2)^n\eta_0,\varepsilon_n=(1/2)^n\varepsilon_0$) : (a) interface position represented by $\phi=0$, (b) time history of $X$-displacement of the marker point.}
	\label{CPB_conv}
\end{figure}

\begin{figure}[h]
		\centering
		\includegraphics[scale=0.5,trim=1 1 1 1,clip]{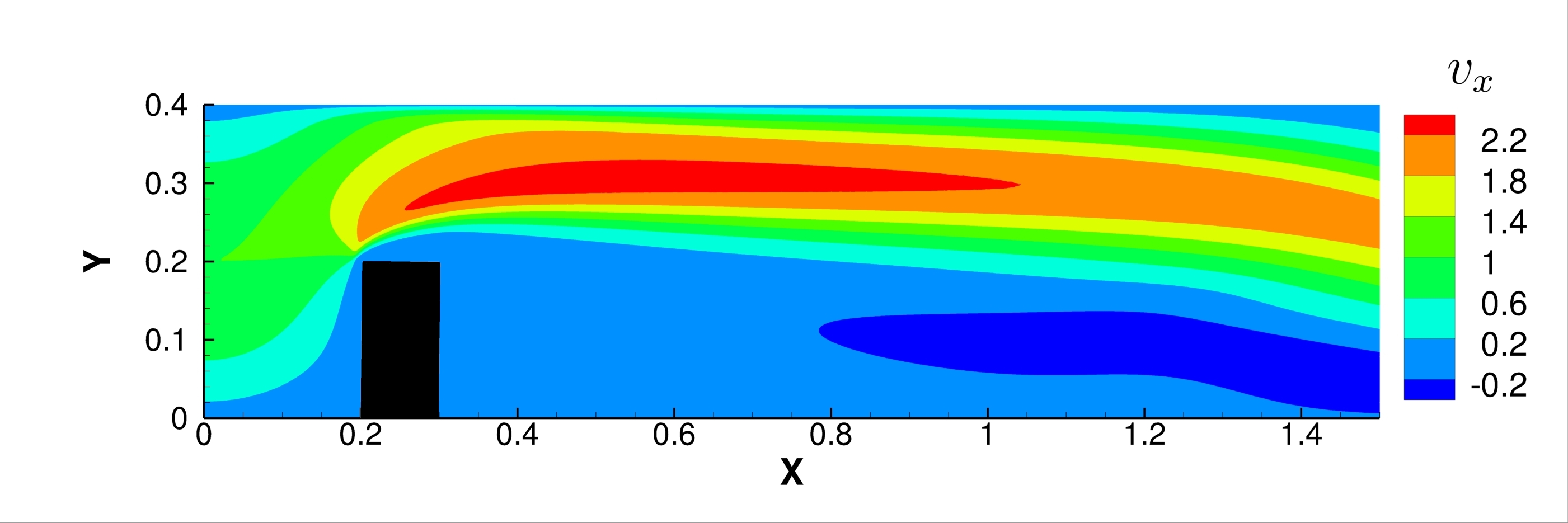}
	\caption{2D channel flow with a fixed deformable block: Demonstration of the flow field $v_x$ and the deformed solid depicted by $\phi>0$ at $t=10$}
	\label{CPB_field}
\end{figure}

As discussed above, in the current convergence study, we bisect $\eta$ and $\varepsilon$ simultaneously from $\eta_0=0.01$, $\varepsilon_0=0.1$ to $\eta_3=0.00125$, $\varepsilon_3=0.0125$, where the subscript denotes the number of times of the bisection. The mesh size is subdivided accordingly to keep $\varepsilon/h=1$. The time step is taken as $\Delta t=0.01$ and the solution at $t=10$ is analyzed. The resulting convergence of the interface position represented by $\phi=0$ is shown in Fig. \ref{CPB_conv} (a). The convergence of the $X$-displacement of the marker points is shown in Fig. \ref{CPB_conv} (b). As observed, with the reduction of $\epsilon$ and $\eta$, the solid bulk is modeled more accurately, which leads to higher stiffness and less deformation. The horizontal $X$-displacement of the marker point converges to the true solution. The block depicted by $\phi>0$ and the flow field velocity $v_x$ at $t=10$ in the finest case is shown in Fig. \ref{CPB_field}. In the contour, we can observe that the flow is blocked by the solid and flows through the top gap, which forms a high-velocity region. The negative velocity behind the block shows the formation of a recirculation zone.

\subsubsection{Demonstration on soft solid}
\begin{figure}[h]
	
	\begin{minipage}[b]{0.48\textwidth}
		\centering
		\includegraphics[scale=0.6,trim=0 0 0 0,clip]{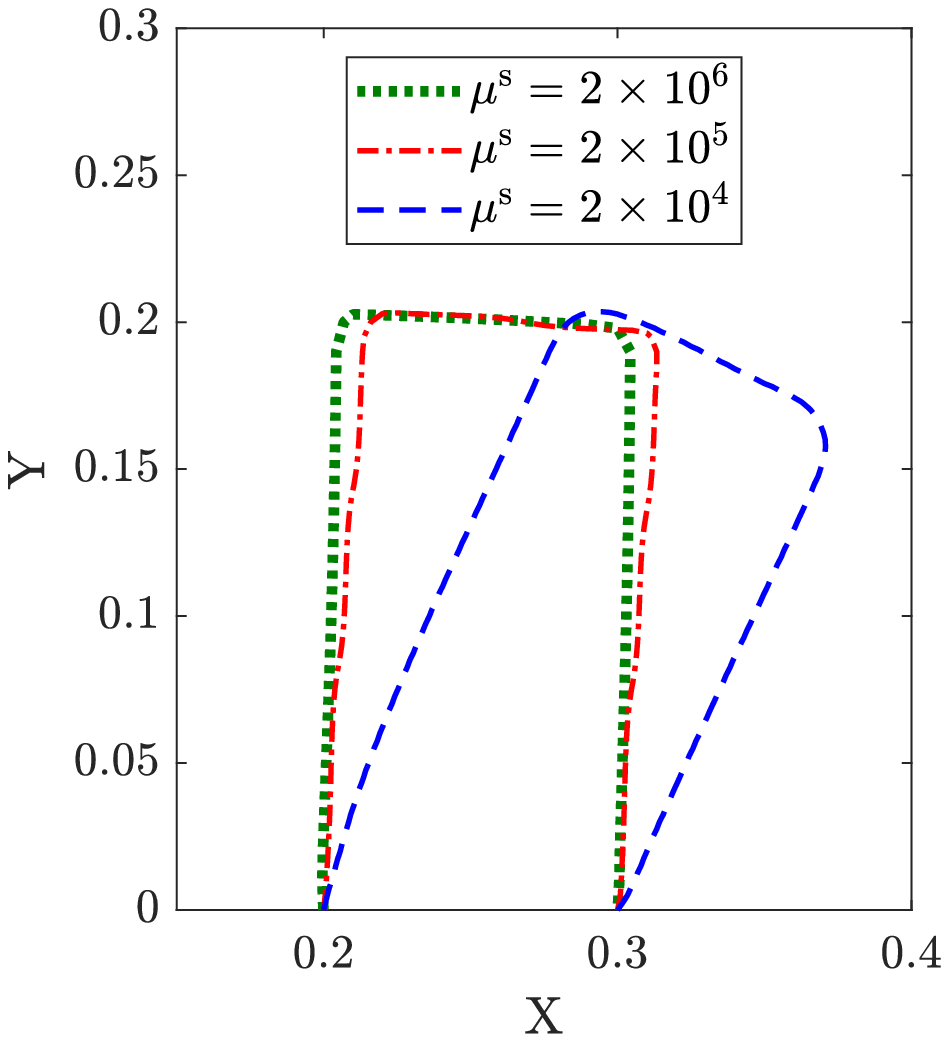}
		\caption*{\hspace{0.8cm}(a)}
	\end{minipage}
	\begin{minipage}[b]{0.48\textwidth}
		\centering
		\includegraphics[scale=0.6,trim=0 0 0 0,clip]{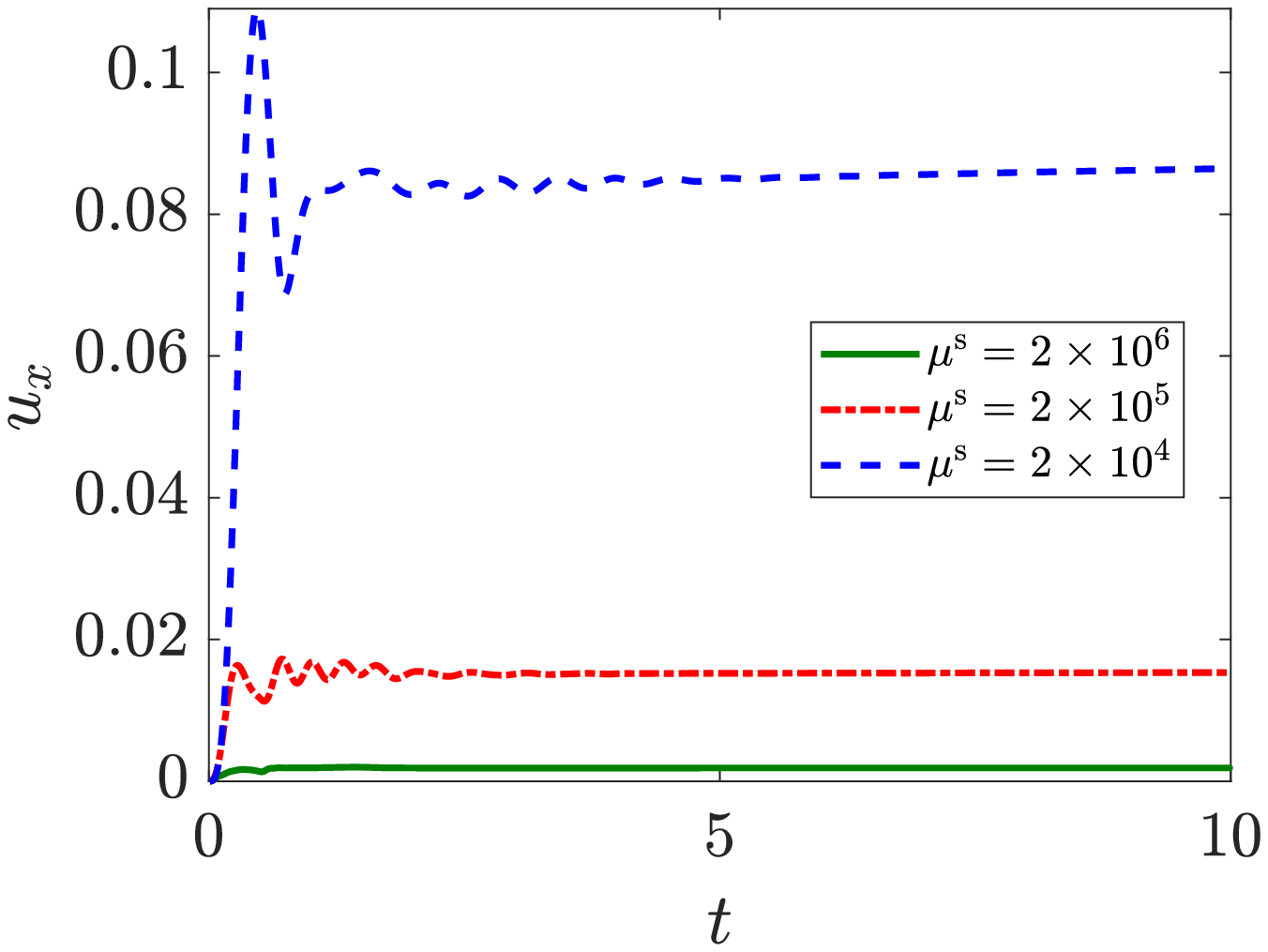}
		\caption*{\hspace{1.8cm}(b)}
	\end{minipage}
	\caption{Deformation of the block with different share modulus in a channel flow: (a) interface position represented by $\phi=0$, (b) $X$-displacement of the marker point}
	\label{CPB_soft}
\end{figure}

Next, we demonstrate the flow passing a fixed deformable block case on softer solids, which induces larger deformation. With the same problem setup, discretization and interface parameters in the finest case of the diffuse interface model convergence study, we decrease the shear modulus to $\mu^{\mathrm{s}}=2\times 10^{5}$ and $\mu^{\mathrm{s}}=2\times 10^{4}$ respectively.  The problems can be solved without much difficulty. The interface position given by $\phi=0$ and the time history of the $X$-displacement of the marker point is shown in Fig. \ref{CPB_soft} (a) and (b) respectively.

\subsection{Cylinder-flexible plate problem}
\begin{figure}[h]
 	\begin{minipage}[b]{0.48\textwidth}
	\centering
	\trimbox{0 45 0 0}{
		\begin{tikzpicture}[scale=1.1,every node/.style={scale=1}]
			
			\begin{axis}[ 
				ymin=0,
				ymax= 0.41,
				xmax=1,
				xmin=0,
				xticklabel=\empty,
				yticklabel=\empty,
				ytick style={draw=none},
				xtick style={draw=none},
				axis line style = {draw=none},
				minor tick num=0,
				axis lines = middle,
				label style = {at={(ticklabel cs:1.1)}},
				axis equal
				]
				
				\draw (axis cs: 0,0)--(axis cs: 1,0)--(axis cs: 1,0.41)--(axis cs: 0,0.41)--(axis cs: 0,0);
				\draw[scale=1, domain=0:0.41, smooth, variable=\y]  plot ({-2*\y*(\y-0.41)}, {\y});
				\draw (axis cs: 0.2,0.2) circle (0.05);
				\draw [fill=gray,fill opacity=0.5] (axis cs: 0.249,0.19)--(axis cs: 0.6,0.19)--(axis cs: 0.6,0.21)--(axis cs: 0.249,0.21);
				\draw [-stealth] (axis cs: 0,0.1) --(axis cs:0.062,0.1);
				\draw [-stealth] (axis cs: 0,0.2) --(axis cs:0.084,0.2);
				\draw [-stealth] (axis cs: 0,0.3) --(axis cs:0.062,0.3);
				\node at (axis cs: 0.15,0.35) {$\Omega^{\mathrm{f}}$};
				\draw (axis cs:0.425,0.2)--(axis cs:0.44,0.25);
				\node [above] at (axis cs:0.45,0.25) {$\Omega^{\mathrm{s}}$};
				\filldraw (0.6,0.2) circle (0.004);
				\node [above] at (axis cs:0.75,0.2) {$\mathrm{Marker}$};
				\node [below] at (axis cs:0.75,0.2) {$\mathrm{point}$};
		\end{axis}
	\end{tikzpicture}}
	\caption*{\hspace{30pt}(a)}
	\end{minipage}
\hfill
	\begin{minipage}[b]{0.48\textwidth}
	\centering
	\includegraphics[scale=0.105,trim= 0 0 0 0,clip]{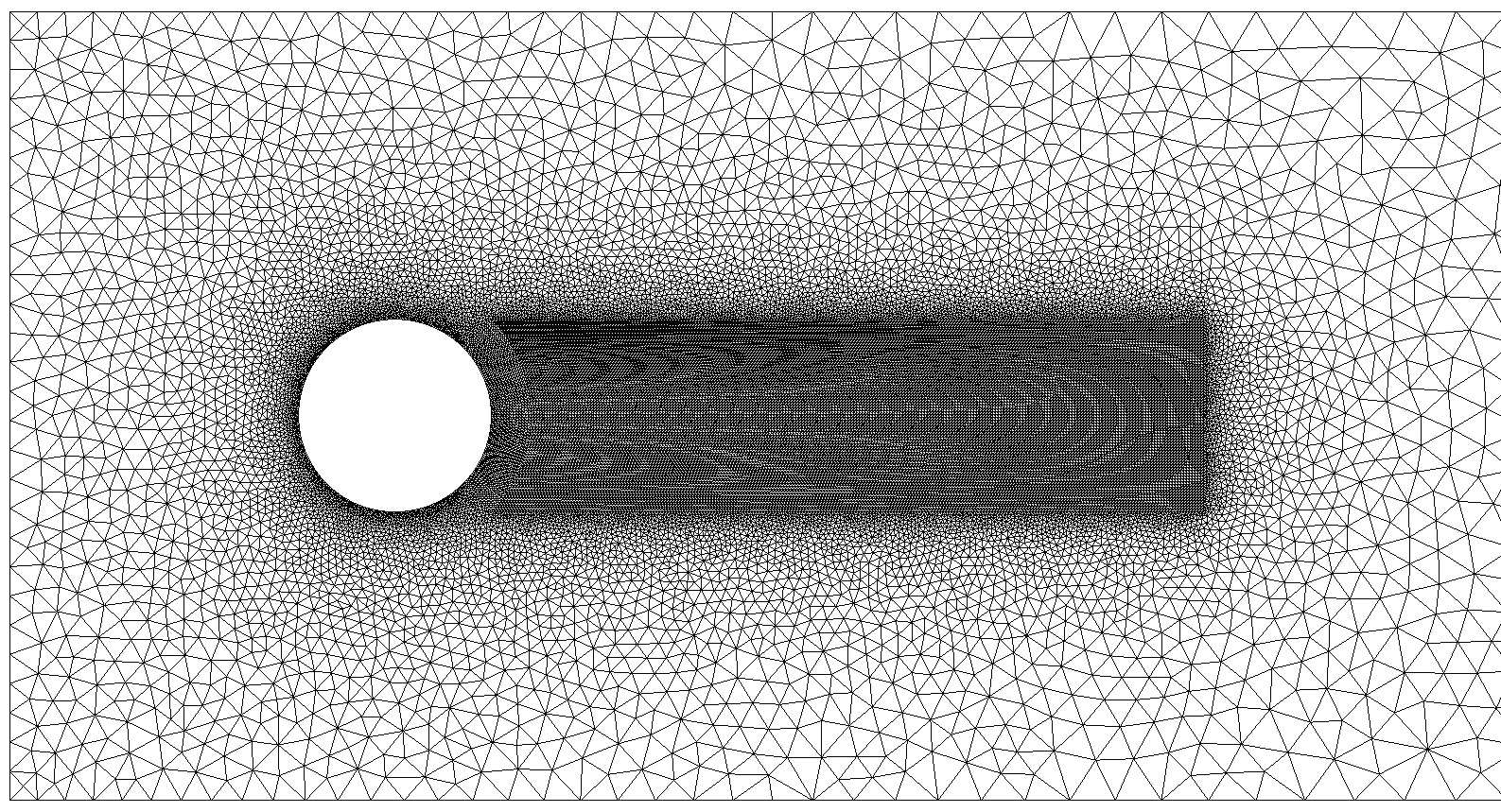}
	\caption*{\hspace{30pt}(b)}
	\end{minipage}

	\caption{Cylinder-flexible plate problem: (a) schematic diagram showing the computational domain, (b) structured finite element mesh covering the area where the plate moves through and the surrounding unstructured mesh. }
	\label{cpi}
\end{figure}

For further assessment, we consider a cylinder-flexible plate problem, which results in a periodic transverse oscillation of the plate. The computational domain $[0,2.5]\times[0,0.41]$ is considered for the case. The center of the rigid cylinder of radius $r=0.05$ is located at $(0.2,0.2)$. The flat plate of thickness $T=0.02$ and length $L=0.35$ is attached in the downstream direction of the rigid cylinder, the right lower end of which is located at $(0.6,0.19)$. The order parameter field representing the plate is initialized by calculating the signed distance of each node to the top, right and bottom boundary $d$. Then $d$ is composed by a hyperbolic tangent function, which results in  $\phi(\boldsymbol{x},0)=\tanh(d/\sqrt{2}\varepsilon)$. The steady flow inflows from the left boundary, whose $X$-component velocity at the inlet is given by a parabolic function with ramping in time:
\begin{align}
	v_{x}(0,y,t)=
	\begin{cases} 0.5 \left(1-\cos(2\pi t)\right)3.0y(H-y)/(0.5H)^2, \quad\quad t\leqslant0.5,\\
		3.0y(H-y)/(0.5H)^2,\hspace{4.05cm}t>0.5,
\end{cases}
\end{align}
where $H=0.41$ represents the height of the computational domain. The no-slip boundary condition is applied on the top and bottom boundary. The outflow condition is applied on the right boundary. The densities of the solid and the fluid are chosen as $\rho^{\mathrm{s}}=1000$ and $\rho^{\mathrm{f}}=1000$, respectively.  The dynamic viscosity of the fluid is taken as $\mu^{\mathrm{f}}=1$. The shear modulus of the solid is selected as  $\mu^{\mathrm{s}}=2\times10^{6}$. The displacement at the mid point of the end of the plate is used to characterize the vibration amplitude. The problem setup is illustrated in Fig. \ref{cpi} (a).

\subsubsection{Comparison between the IP method and the IGP method}
We perform a comparison between the IP method and the IGP on a coarse mesh. The computational domain is discretized with a unstructured triangle mesh, while a structured refined mesh is employed where the flexible plate passes through during the vibration. The mesh is illustrated in Fig. (\ref{cpi}) (b). The interface thickness parameter is selected as $\varepsilon=0.001\times\sqrt{2}$, where the factor of $\sqrt{2}$ is multiplied for the convenience of the convergence study later. To further reduce the computational cost, the resolution at the interface is selected as around three elements across the diffuse interface region, which leads to $h=\varepsilon\times4/3$ in the refined region. The RMS interface distortion parameter is selected as $\eta=0.1\times\sqrt{2}$. The time step is selected as $\Delta t=0.001$.

\begin{figure}[h]
	
	\begin{minipage}[b]{0.48\textwidth}
		\centering
		\includegraphics[scale=0.6,trim=0 0 0 0,clip]{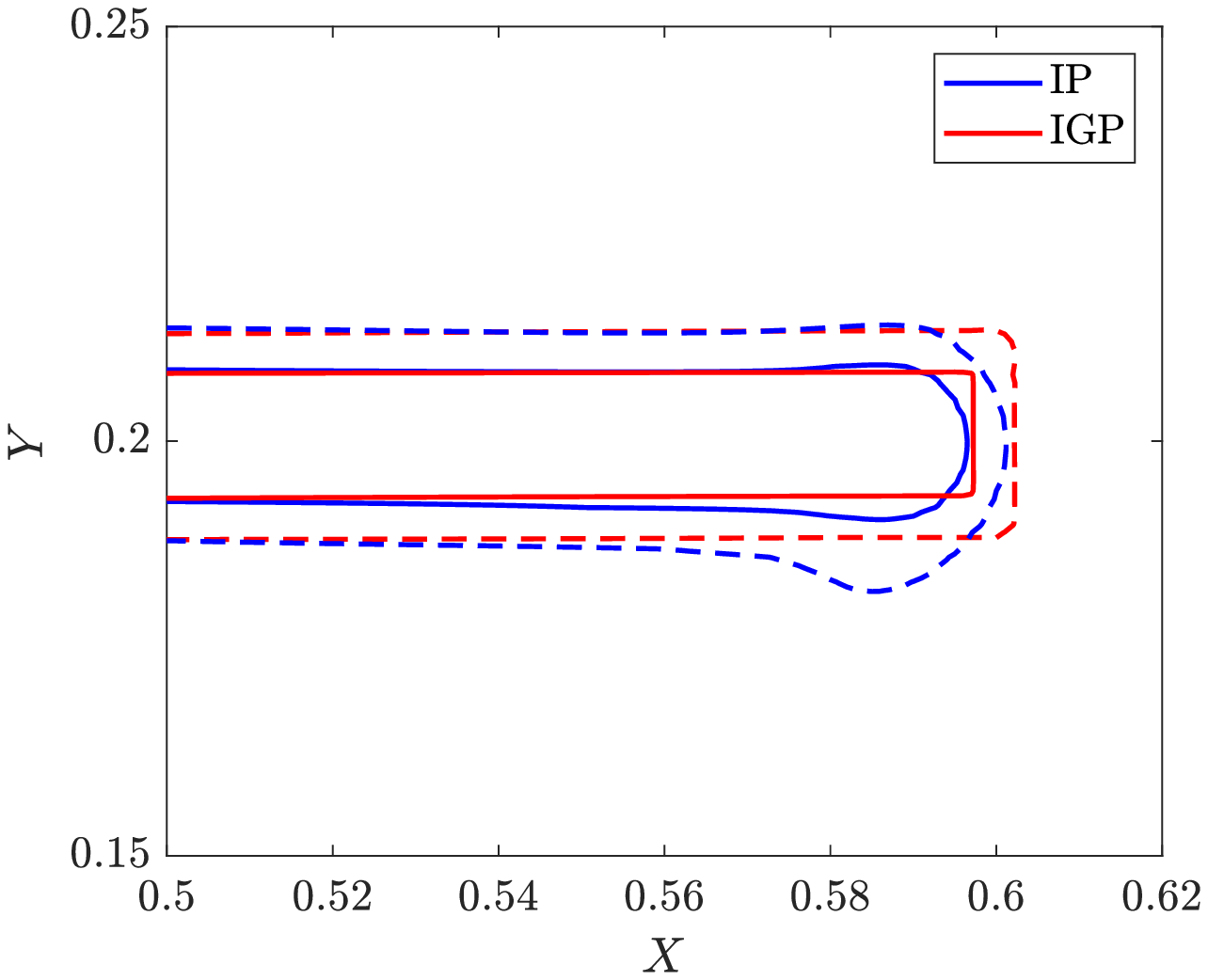}
		\caption*{\hspace{0.8cm}(a)}
	\end{minipage}
	\begin{minipage}[b]{0.48\textwidth}
		\centering
		\includegraphics[scale=0.6,trim=0 0 0 0,clip]{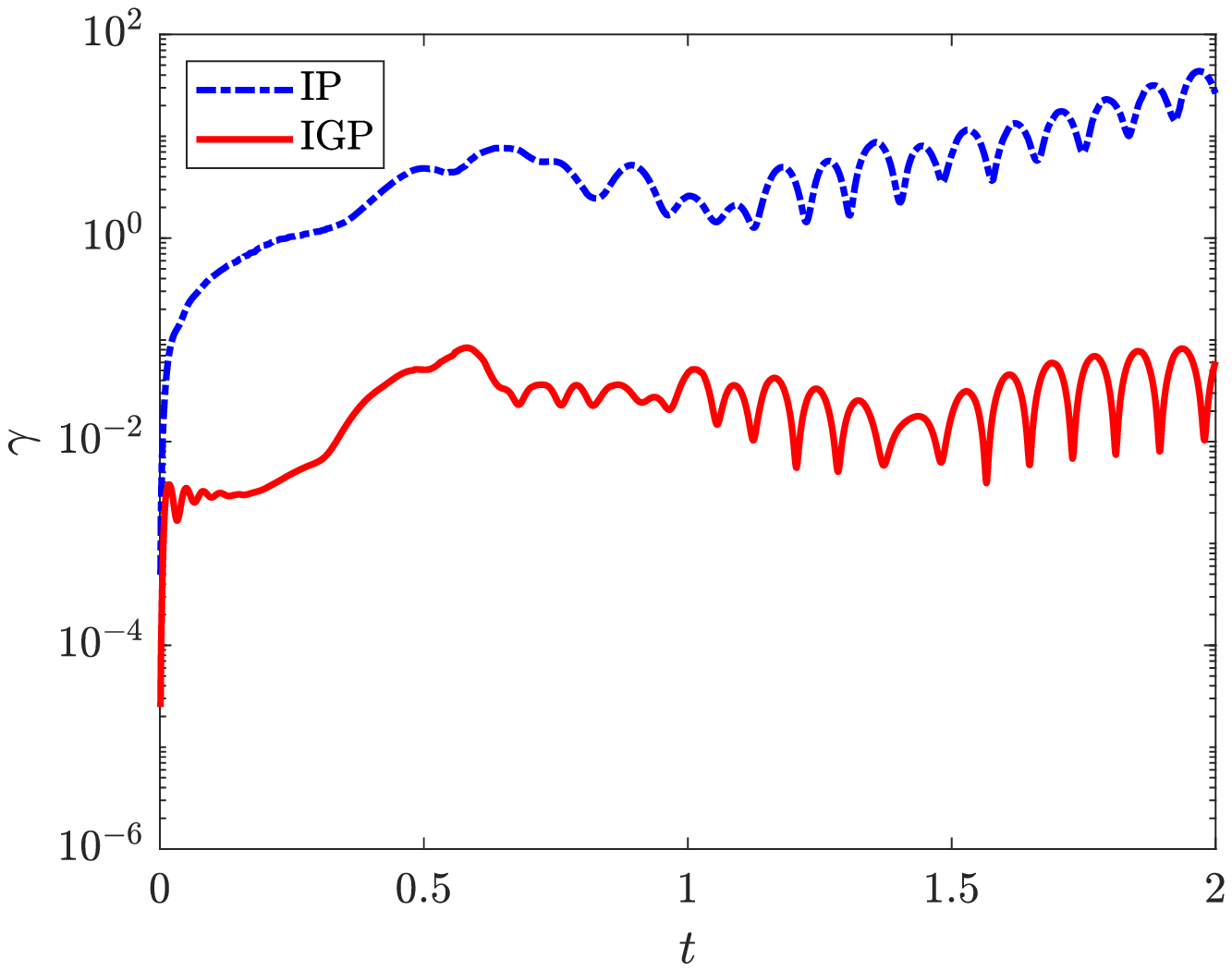}
		\caption*{\hspace{1.3cm}(b)}
	\end{minipage}
	\caption{Cylinder-plate problem: (a) diffuse interface region at $t=2$ given by the IP method and the IGP method, (b) time history of the mobility coefficient.}
	\label{Trkcom}
\end{figure}

We first focus on the analysis of the solution from $t=0$ to $t=2$. The comparison of the diffuse interface region from the IP method and the IGP method at $t=2$ is shown in Fig. \ref{Trkcom} (a). The time history of the mobility coefficient is shown in Fig. \ref{Trkcom} (b). Similar to the previous cases, we can see that the geometry of the plate is well preserved in the IGP method, and the mobility coefficient is reduced by around two orders of magnitude.

\begin{figure}[h]
	
	\begin{minipage}[b]{0.48\textwidth}
		\centering
		\includegraphics[scale=0.5,trim=0 0 0 0,clip]{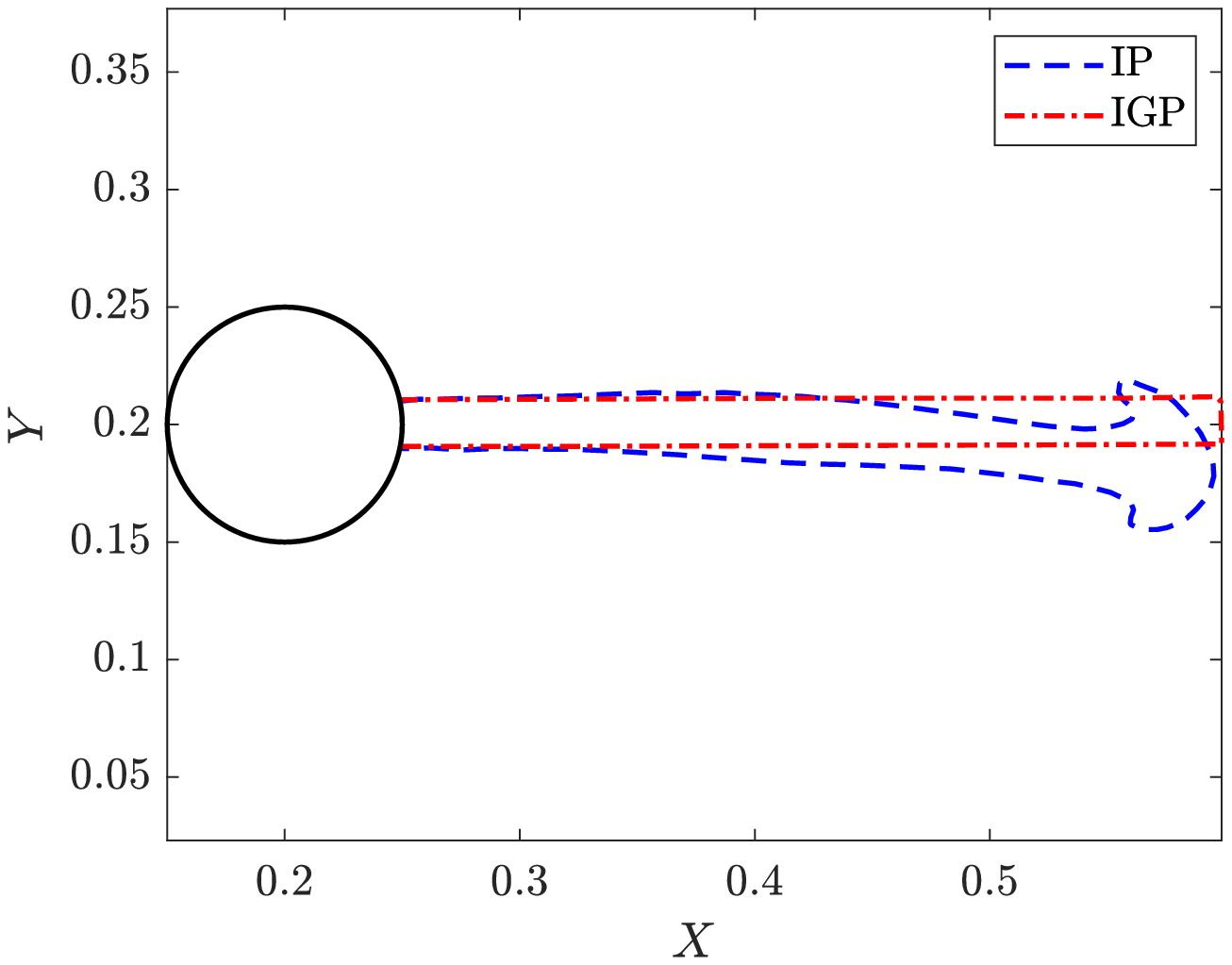}
		\caption*{\hspace{0.8cm}(a)}
	\end{minipage}
	\begin{minipage}[b]{0.48\textwidth}
		\centering
		\includegraphics[scale=0.5,trim=0 0 0 0,clip]{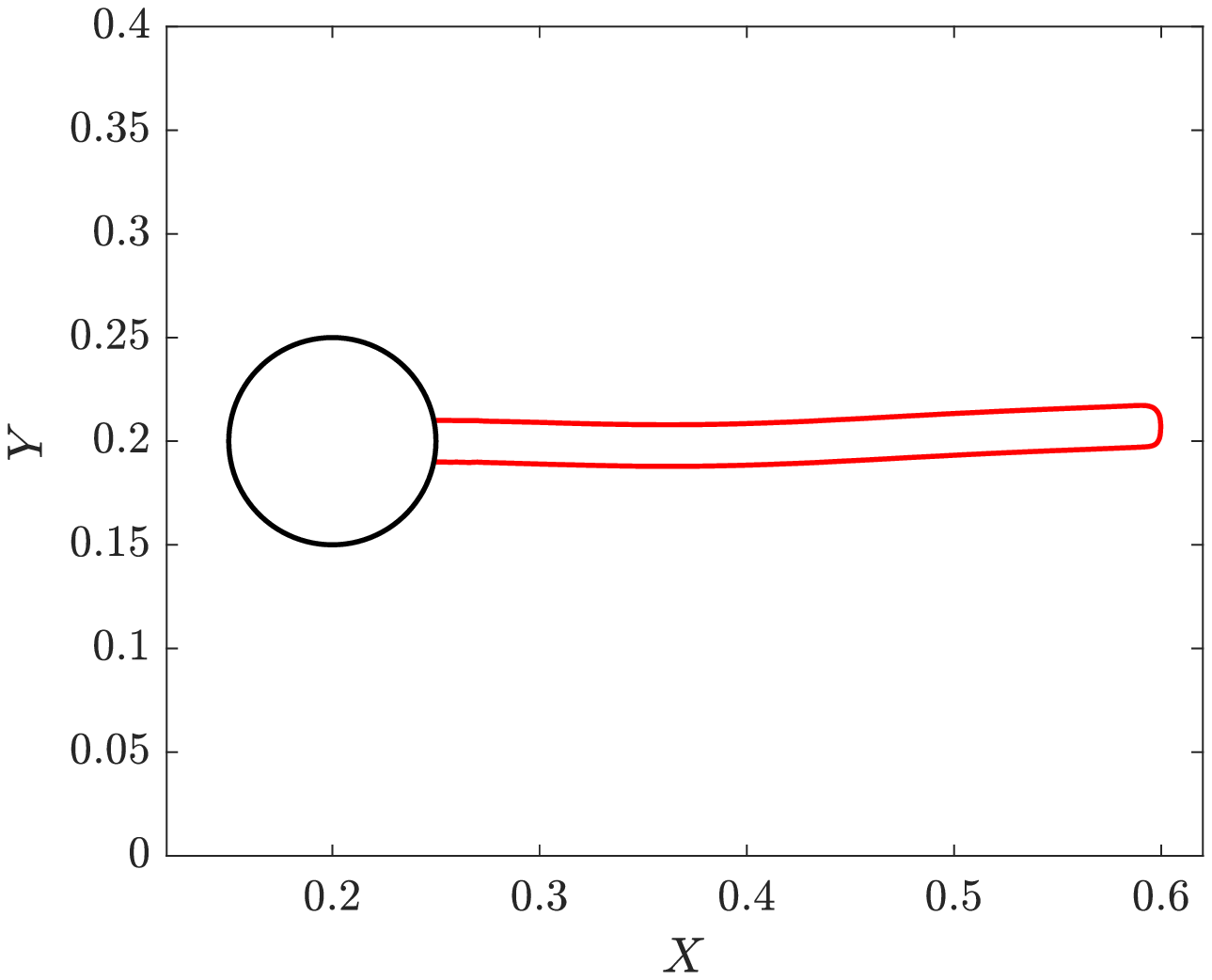}
		\caption*{\hspace{0.8cm}(b)}
	\end{minipage}

	\caption{Cylinder-flexible plate problem: (a) interface position $\phi=0$ given by the IP method and the IGP method at $t=2.3$, (b) interface position $\phi=0$ given by the IGP method at $t=6$.}
	\label{TrkAT}
\end{figure}

\begin{figure}[h]
	\begin{minipage}[b]{0.48\textwidth}
	\centering
	\includegraphics[scale=0.5,trim=0 0 0 0,clip]{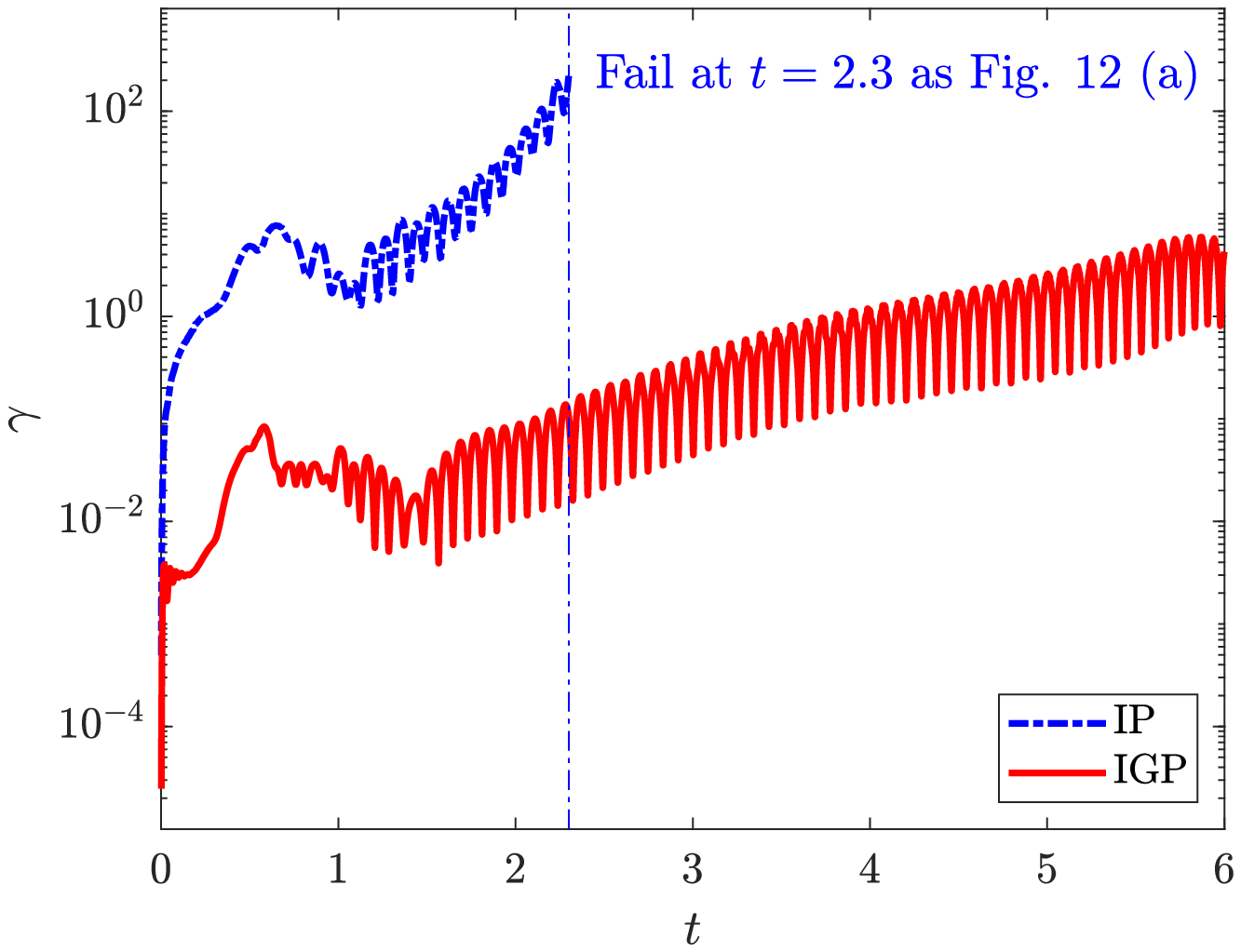}
	\caption*{\hspace{0.8cm}(a)}
\end{minipage}
\hspace{0.1cm}
\begin{minipage}[b]{0.48\textwidth}
	\centering
	\includegraphics[scale=0.5,trim=0 0 0 0,clip]{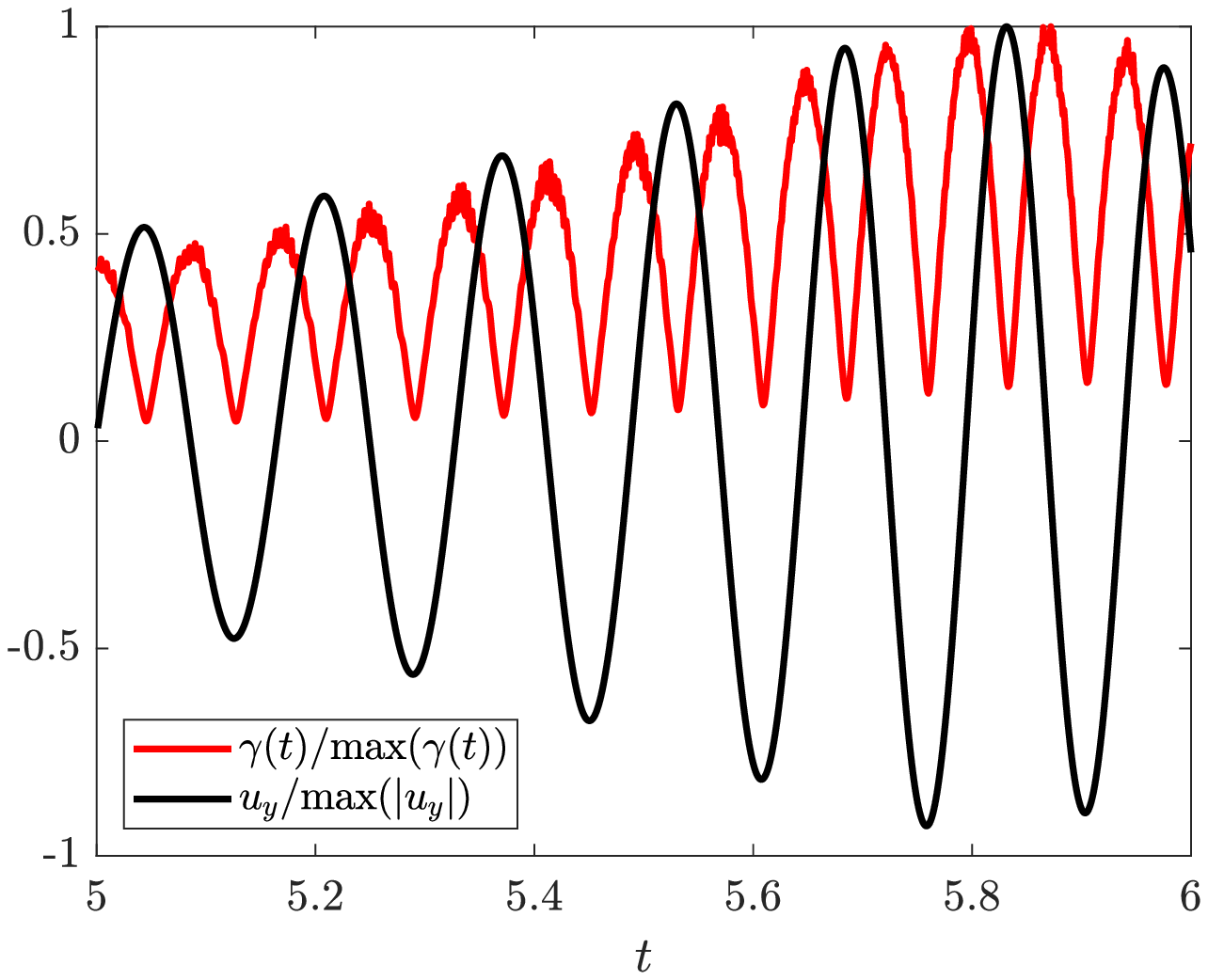}
	\caption*{\hspace{0.8cm}(b)}
\end{minipage}
	\caption{Cylinder-flexible plate problem: (a) comparison of the mobility coefficient given by the IP method and the IGP method, (b) time history of the normalized mobility and the $Y$-displacement.}
\label{Trk_Mob}
\end{figure}

In the simulation of the IP method after $t=2$, the error quickly accumulates because the diffuse interface region no longer conforms to the solid bulk tracked by $\boldsymbol{\xi}$. As a result, the interface geometry becomes erroneous and the simulation fails.  For example, we show the interface position given by the IP method at $t=2.3$ in Fig. \ref{TrkAT} (a). On the other hand, the IGP method preserves the interface geometry throughout the simulation. The interface position $\phi=0$ at $t=6$ is shown in Fig. \ref{TrkAT} (b).
Finally, we present the complete time history of the mobility coefficient from the IGP method in Fig. \ref{Trk_Mob} (a). As observed, the mobility coefficient is reduced significantly in the IGP method compared to the IP method. Notably, a temporal oscillation can be observed clearly in the time history of the mobility coefficient. To further examine this oscillation, we plot the normalized mobility coefficient and the normalized $Y$-displacement of the plate together to analyze their phases in \ref{Trk_Mob} (b). As observed, when the plate arrives at the neutral position $u_y=0$, where the speed of the plate reaches its maximum, the mobility coefficient reaches its maximum. When the transverse $Y$-displacement reaches its maximum or minimum value, where the velocity of the plate is close to zero, the mobility coefficient gets close to zero as well. This observation is consistent with our statement that when the solid bulk moves at a relatively high speed, a significant convective distortion will be introduced, thus requiring larger mobility to preserve the diffuse interface from being distorted.

\subsubsection{Convergence of the diffuse interface model}
\begin{figure}[h]
	
	\centering
	\includegraphics[scale=0.6,trim= 0 0 0 0,clip]{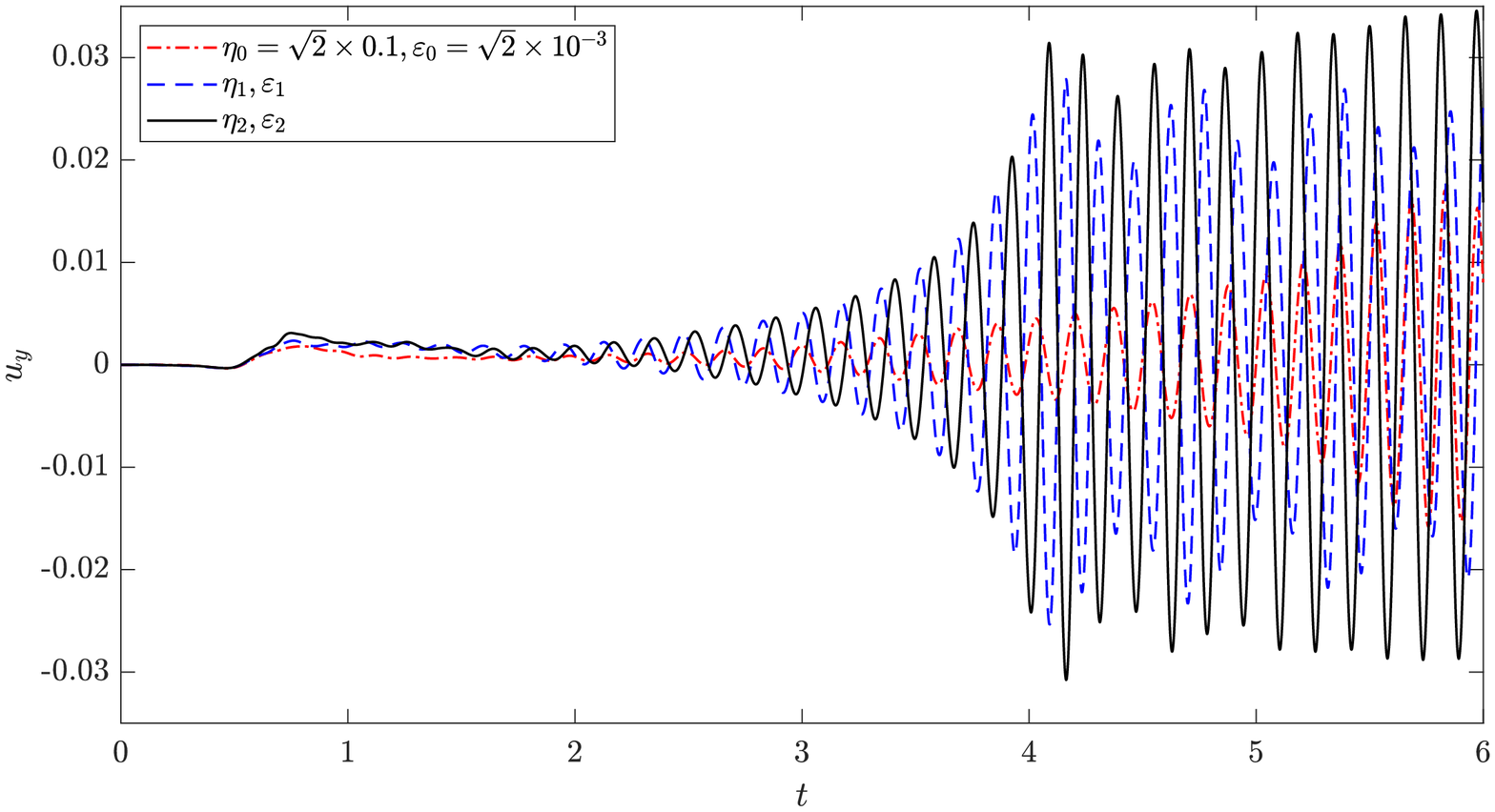}
	\caption*{\hspace{30pt}(a)}
	\caption{Convergence of the diffuse interface model in the cylinder-flexible plate problem. The subscript denotes the number of times of scaling down by the factor of $1/\sqrt{2}$. ($\eta_n=(1/\sqrt{2})^n\eta_0,\varepsilon_n=(1/\sqrt{2})^n\varepsilon_0$). The mesh resolution is kept as three elements across the diffuse interface region by taking $h=4/3\ \varepsilon$.   } 
	\label{tkcv}
\end{figure}

\begin{figure}[h]
	
	\begin{minipage}[b]{0.5\textwidth}
		\centering
		\includegraphics[scale=0.5,trim= 2 2 2 2,clip]{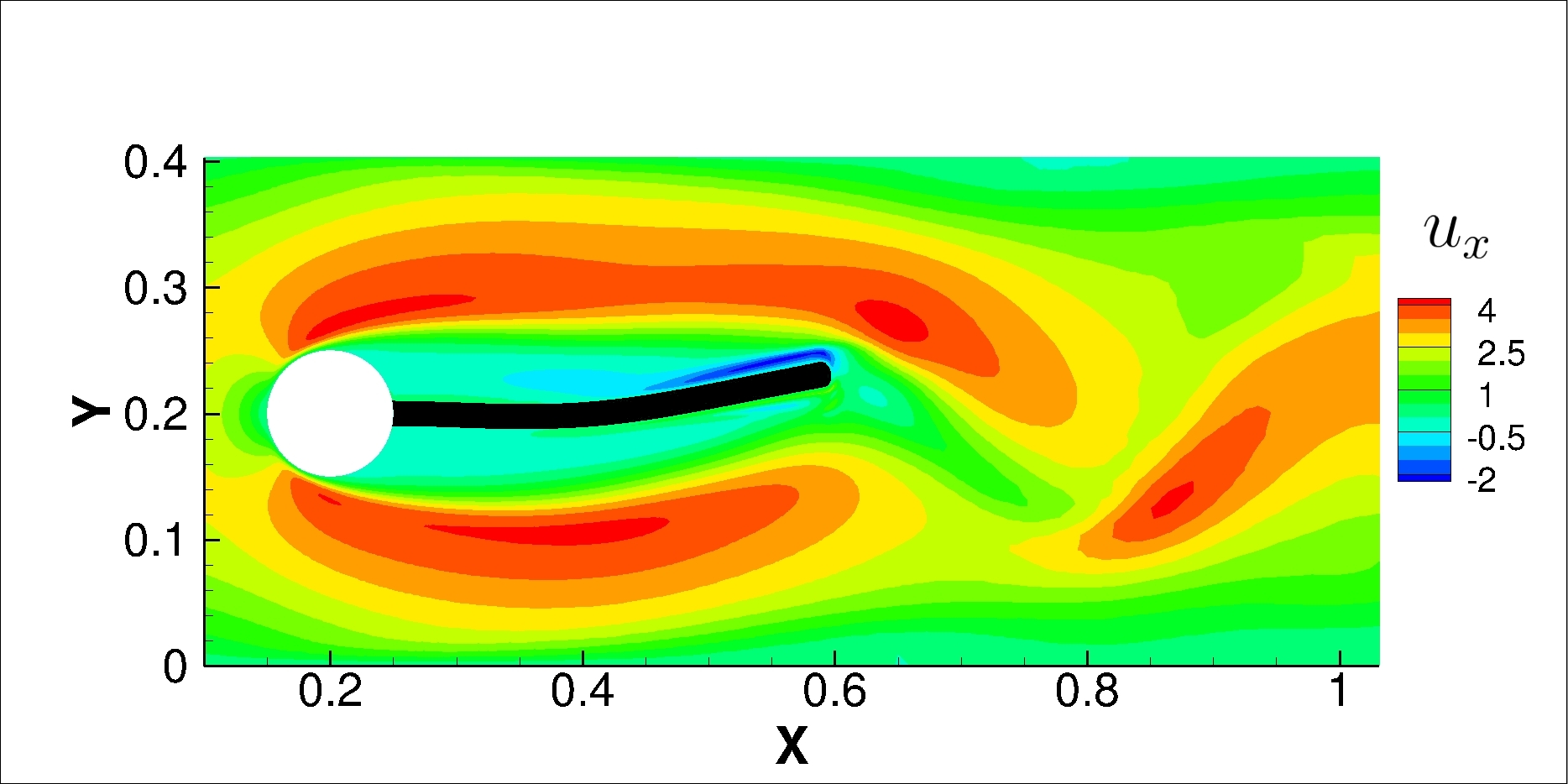}
		\caption*{\hspace{10pt}(a)}
	\end{minipage}
	\hfill
	\begin{minipage}[b]{0.5\textwidth}
		\centering
		\includegraphics[scale=0.5,trim= 2 2 2 2,clip]{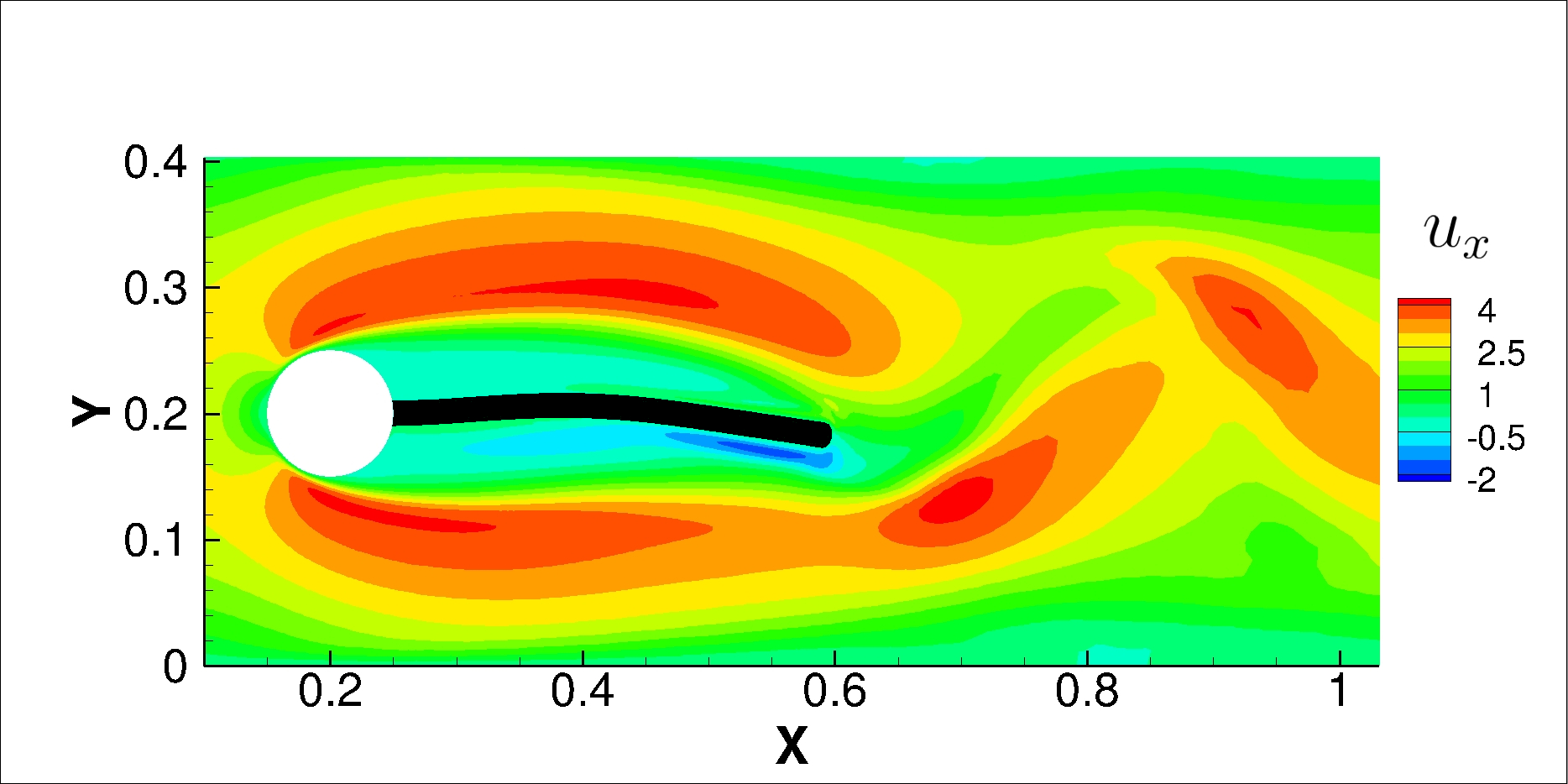}
		\caption*{\hspace{10pt}(b)}
	\end{minipage}

	\caption{Demonstration of the flow field $v_x$  and the deformed solid depicted by $\phi>0$ at (a) $t=5.5$ (b) $t=5.6$.} 
	\label{tkiu}
	\label{}
\end{figure}

Next, we perform a convergence study for the diffuse interface model. Starting from $\varepsilon_0=0.001\times\sqrt{2}$ and $\eta_0=0.1\times\sqrt{2}$, we scale down the diffuse interface parameters by a factor of $\sqrt{2}$ till $\varepsilon_2=0.001/\sqrt{2}$ and $\eta_2=0.1/\sqrt{2}$, where the subscript denotes the number of times of scaling. The mesh is refined accordingly to keep the interface resolution as $h= 4/3 \varepsilon$. The time step is kept as $\Delta t=0.001$ and the vibration is simulated until $t=6$. The  vertical $Y$-direction displacement of the marker point is shown in Fig. \ref{tkcv}. For the most refined case, the solid plate depicted by $\phi>0$ and the contour of $v_x$ at $t=5.5$ and $t=5.6$ are demonstrated in Fig. \ref{tkiu}. By taking the $Y$-displacement between $t=5$ and $t=6$, we calculate the neutral position of the plate by $1/2(\max(u_y)+\min(u_y))$ and the vibration amplitude by $1/2(\max(u_y)-\min(u_y))$, which results in $(1.9\pm32.7\ )\times10^{-3}$. Qualitatively the predicted vibration amplitudes are comparable with the FSI3 case \cite{turek2006proposal}, which uses $\nu=0.4$ while all the rest physical parameters are the same, where the vibration is given by $(1.48\pm34.38\ )\times10^{-3}$.


As being a PDE-based technique, the proposed phase-field-based IGP method can be easily extended to three-dimension with parallel implementation. Currently, the diffusion term required to stabilize the construction of GMV is approximated according to the maximum possible propagation velocity.  Owing to the maximum value criterion, the overestimated diffusion coefficient may induce an increment in the normal velocity gradient. By locally adjusting the diffusion coefficient, the curvature flow can be further minimized. The generalization of the current method to multiphase and flexible multibody problems needs to be further investigated with large deformation and a possibility of topology changes in immersed solid bodies.

\section{Conclusion}
In the present work, we proposed an interface and geometry preserving phase-field method for fully Eulerian FSI problems. The key innovation is to convect the level sets of the order parameter with approximately the same normal velocity in the normal direction such that the convective distortion is reduced. This is achieved by a novel PDE-based construction of an auxiliary gradient-minimizing velocity field. This treatment significantly reduces the time-dependent mobility coefficient associated with free energy minimization in the phase-field formulation. Consequently, there is a lesser volume-conserved mean curvature flow. Furthermore, the GMV is constructed by extending the solid velocity, which excludes the fluid velocity involved in the unified velocity field. Hence the GMV guarantees that the diffuse interface region conforms to the solid bulk. As a result, the description of the fluid-solid interface geometry is substantially improved in both geometric details and global conformity behavior in the newly developed IGP formulation. When tested for fully Eulerian FSI problems namely the flow passing a fixed deformable block in a channel and the cylinder-flexible plate problem, a reduction in the curvature flow at around one to two orders of magnitude is achieved. The proposed IGP method can be employed with other interface-capturing methods such as the level set method where the interface region is resolved with around three to four elements. Since the proposed IGP method relies on the differential equation, it can be easily extended to three-dimension with parallel implementation for complex geometries. 
Finally, the IGP method is of practical usage for fully Eulerian FSI problems involving deformable solid bodies with sharp corners and large aspect ratios.

\section*{Acknowledgement}
The authors would like to acknowledge the Natural Sciences and Engineering Research Council of Canada (NSERC) and Seaspan Shipyards for the funding. This research was supported in part through computational resources and services provided by Advanced Research Computing at the University of British Columbia. 
\appendix
\setcounter{equation}{0} 
\setcounter{figure}{0}

\section{Residual and Jacobian term for the solid stress in 2D}\label{Ap2}
In this appendix, we present the 2D component-wise form of the residual and Jacobian matrix of the solid stress. We start from the convection of the left Cauchy-Green tensor:
\begin{align}
	\frac{\partial B_{11}}{\partial t}=-u\frac{\partial B_{11}}{\partial x}-v\frac{\partial B_{11}}{\partial y}+\left( \frac{\partial u}{\partial  x}B_{11}+ \frac{\partial u}{\partial y} B_{21}\right) +\left(B_{11} \frac{\partial u}{\partial x} +B_{12}\frac{\partial u}{\partial y} \right)=0,\nonumber\\
	\frac{\partial B_{12}}{\partial t}=-u\frac{\partial B_{12}}{\partial x}-v\frac{\partial B_{12}}{\partial y}+\left( \frac{\partial u}{\partial  x}B_{12}+ \frac{\partial u}{\partial y} B_{22}\right) +\left(B_{11} \frac{\partial v}{\partial x} +B_{12}\frac{\partial v}{\partial y} \right)=0,\nonumber\\
	\frac{\partial B_{21}}{\partial t}=-u\frac{\partial B_{21}}{\partial x}-v\frac{\partial B_{21}}{\partial y}+\left( \frac{\partial v}{\partial  x}B_{11}+ \frac{\partial v}{\partial y} B_{21}\right) +\left(B_{21} \frac{\partial u}{\partial x} +B_{22}\frac{\partial u}{\partial y} \right)=0,\nonumber\\
	\frac{\partial B_{22}}{\partial t}=-u\frac{\partial B_{22}}{\partial x}-v\frac{\partial B_{22}}{\partial y}+\left( \frac{\partial v}{\partial  x}B_{12}+ \frac{\partial v}{\partial y} B_{22}\right) +\left(B_{21} \frac{\partial v}{\partial x} +B_{22}\frac{\partial v}{\partial y} \right)=0.\nonumber
\end{align}

To keep the symmetry of the left Cauchy-Green tensor, we let $B_{21}=B_{12}$ in practical implementation. Using the generalized-$\alpha$ method, $\boldsymbol{B}^{n+\alpha}$ can be written as a function of $\boldsymbol{u}^{n+\alpha}$ as follows:
\begin{align}
	&B_{11}^{n+\alpha}=B_{11}^n+\alpha\Delta t \left(1-\frac{\gamma}{\alpha_m}\right)\partial_t B_{11}^n\nonumber\\
	&+\frac{\Delta t\alpha\gamma}{\alpha_m}\left(-u^{n+\alpha}\frac{\partial B_{11}^{n+\alpha}}{\partial x}-v^{n+\alpha}\frac{\partial B_{11}^{n+\alpha}}{\partial y}+2\frac{\partial u^{n+\alpha}}{\partial x}B_{11}^{n+\alpha}+2\frac{\partial u^{n+\alpha}}{\partial y}B_{12}^{n+\alpha}\right),\nonumber\\
	&B_{12}^{n+\alpha}=B_{12}^n+\alpha\Delta t \left(1-\frac{\gamma}{\alpha_m}\right)\partial_t B_{12}^n\nonumber\\
	&+\frac{\Delta t\alpha\gamma}{\alpha_m}\left(-u^{n+\alpha}\frac{\partial B_{12}^{n+\alpha}}{\partial x}-v^{n+\alpha}\frac{\partial B_{12}^{n+\alpha}}{\partial y}+\frac{\partial u^{n+\alpha}}{\partial x}B_{12}^{n+\alpha}+\frac{\partial u^{n+\alpha}}{\partial y}B_{22}^{n+\alpha}+\frac{\partial v^{n+\alpha}}{\partial x}B_{11}^{n+\alpha}+\frac{\partial v^{n+\alpha}}{\partial y}B_{12}^{n+\alpha}\right),\nonumber\\
	&B_{21}^{n+\alpha}=B_{12}^{n+\alpha},\nonumber\\
	&B_{22}^{n+\alpha}=B_{22}^n+\alpha\Delta t \left(1-\frac{\gamma}{\alpha_m}\right)\partial_t B_{22}^n\nonumber\\
	&+\frac{\Delta t\alpha\gamma}{\alpha_m}\left(-u^{n+\alpha}\frac{\partial B_{22}^{n+\alpha}}{\partial x}-v^{n+\alpha}\frac{\partial B_{22}^{n+\alpha}}{\partial y}+2\frac{\partial v^{n+\alpha}}{\partial y}B_{22}^{n+\alpha}+2\frac{\partial v^{n+\alpha}}{\partial x}B_{12}^{n+\alpha}\right).\nonumber
\end{align}
The residual of the stress term can be formed via $\boldsymbol{\sigma}^{n+\alpha}=-p^n\boldsymbol{I}+\mu^\mathrm{s}(\boldsymbol{B}^{n+\alpha}-\boldsymbol{I})$. Finally, the components of the Jacobian matrix of the solid stress are given by:
\begin{align}
	\frac{\delta\left(N_{,x}\sigma^{n+\alpha}_{11}+N_{,y}\sigma^{n+\alpha}_{21}\right)}{\delta u^{n+\alpha}}&=\mu^{\mathrm{s}}\frac{\delta(N_{,x}B^{n+\alpha}_{11}+N_{,y}B^{n+\alpha}_{12})}{\delta u^{n+\alpha}},\nonumber\\
	&=\mu^{\mathrm{s}}N_{,x}\frac{\Delta t\alpha\gamma}{\alpha_m}\left(-N\frac{\partial B_{11}^{n+\alpha}}{\partial x}+2N_{,x}B_{11}^{n+\alpha}+2N_{,y}B_{12}^{n+\alpha}\right)\nonumber\\
	&+\mu^{\mathrm{s}}N_{,y}\frac{\Delta t\alpha\gamma}{\alpha_m}\left(-N\frac{\partial B_{12}^{n+\alpha}}{\partial x}+N_{,x}B_{12}^{n+\alpha}+N_{,y}B_{22}^{n+\alpha}\right),\nonumber
	\end{align}
\begin{align}
	\frac{\delta\left(N_{,x}\sigma^{n+\alpha}_{11}+N_{,y}\sigma^{n+\alpha}_{21}\right)}{\delta v^{n+\alpha}}&=\mu^{\mathrm{s}}\frac{\delta(N_{,x}B^{n+\alpha}_{11}+N_{,y}B^{n+\alpha}_{12})}{\delta v^{n+\alpha}},\nonumber\\
	&=\mu^{\mathrm{s}}N_{,x}\frac{\Delta t\alpha\gamma}{\alpha_m}\left(-N\frac{\partial B_{11}^{n+\alpha}}{\partial y}\right)\nonumber\\
	&+\mu^{\mathrm{s}}N_{,y}\frac{\Delta t\alpha\gamma}{\alpha_m}\left(-N\frac{\partial B_{12}^{n+\alpha}}{\partial y}+N_{,x}B_{11}^{n+\alpha}+N_{,y}B_{12}^{n+\alpha}\right),\nonumber
	\end{align}
	\begin{align}
	\frac{\delta\left(N_{,x}\sigma^{n+\alpha}_{12}+N_{,y}\sigma^{n+\alpha}_{22}\right)}{\delta u^{n+\alpha}}&=\mu^{\mathrm{s}}\frac{\delta(N_{,x}B^{n+\alpha}_{12}+N_{,y}B^{n+\alpha}_{22})}{\delta u^{n+\alpha}},\nonumber\\
	&=\mu^{\mathrm{s}}N_{,x}\frac{\Delta t\alpha\gamma}{\alpha_m}\left(-N\frac{\partial B_{12}^{n+\alpha}}{\partial x}+N_{,x}B_{12}^{n+\alpha}+N_{,y}B_{22}^{n+\alpha}\right)\nonumber\\
	&+\mu^{\mathrm{s}}N_{,y}\frac{\Delta t\alpha\gamma}{\alpha_m}\left(-N\frac{\partial B_{22}^{n+\alpha}}{\partial x}\right),\nonumber
	\end{align}
	\begin{align}
	\frac{\delta\left(N_{,x}\sigma^{n+\alpha}_{12}+N_{,y}\sigma^{n+\alpha}_{22}\right)}{\delta v^{n+\alpha}}&=\mu^{\mathrm{s}}\frac{\delta(N_{,x}B^{n+\alpha}_{12}+N_{,y}B^{n+\alpha}_{22})}{\delta v^{n+\alpha}},\nonumber\\
	&=\mu^{\mathrm{s}}N_{,x}\frac{\Delta t\alpha\gamma}{\alpha_m}\left(-N\frac{\partial B_{12}^{n+\alpha}}{\partial y}+N_{,x} B_{11}^{n+\alpha}+N_{,y}B_{12}^{n+\alpha}\right)\nonumber\\
	&+\mu^{\mathrm{s}}N_{,y}\frac{\Delta t\alpha\gamma}{\alpha_m}\left(-N\frac{\partial B_{22}^{n+\alpha}}{\partial y}+2N_{,y}\partial B_{22}^{n+\alpha}+2N_{,x}B_{12}^{n+\alpha}\right).\nonumber
\end{align}

\section{Deformation of a solid block driven by cavity flow}\label{cavity block}
\begin{figure}[h]
	\centering
	\trimbox{0 70 0 70}{
		\begin{tikzpicture}[scale=2.4,every node/.style={scale=0.5}]
			
			\begin{axis}[ 
				ymin=-1,
				ymax= 3,
				xmax=3,
				xmin=-1,
				xticklabel=\empty,
				yticklabel=\empty,
				ytick style={draw=none},
				xtick style={draw=none},
				axis line style = {draw=none},
				minor tick num=0,
				axis lines = middle,
				label style = {at={(ticklabel cs:1.1)}},
				axis equal
				]
				
				\draw [fill=gray,fill opacity=0.5] (axis cs: 0,0.5)--(axis cs: 0,0)--(axis cs: 2,0)--(axis cs: 2,0.5);
				\draw (axis cs: 0,0.5)--(axis cs: 0,2)--(axis cs: 2,2)--(axis cs: 2,0.5)--(axis cs: 0,0.5);
				
				\draw [-stealth] (axis cs: 0,2)--(axis cs: 0.25,2);
				\draw [-stealth] (axis cs: 0.25,2)--(axis cs: 0.75,2);
				\draw [-stealth] (axis cs: 0.75,2)--(axis cs: 1.25,2);
				\draw [-stealth] (axis cs: 1.25,2)--(axis cs: 1.75,2);
				\draw  (axis cs: 1.75,2)--(axis cs: 2,2);
				\node at (1,1.25) {$\Omega^{\mathrm{f}}$};
				\node at (1,0.25) {$\Omega^{\mathrm{s}}$};
	
				\draw[-stealth] (1,1.75) arc (90:-180:0.5);
			\end{axis}
	\end{tikzpicture}}
	
	\caption{Schematic diagram of the deformation of a solid block driven by cavity flow.}
	\label{bcw}
\end{figure}
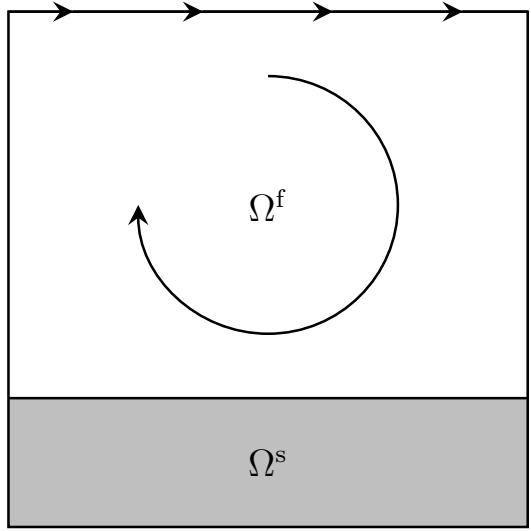

In this appendix, we compare our fully Eulerian FSI solver with the case of deformation of a block of an elastic solid driven by the cavity flow. In this benchmark case, the geometry of the interface is close to a planar interface. The volume-conserved mean curvature flow is almost negligible and the geometry preserving effect is not relatively apparent. Hence, we employ this as the reference study to assess the fully Eulerian solver without the geometry preserving effect. To keep the consistency with the work of \cite{zhao2008fixed}, the convection term in the momentum equations is omitted. A rectangular computational domain $[0,2]\times[0.2]$ is considered in this problem. The order parameter representing the solid block is given by $\phi(x,y,0)=-\tanh\left((y-T)/(\sqrt{2}\varepsilon)\right)$,
where $T=0.5$ is the height of the solid block. While the no-slip boundary condition is applied on the left, right and bottom boundaries, a prescribed velocity in the $X$-direction is applied on the top boundary to drive the formation of the cavity flow which deforms the solid block:
\begin{align*}
	v_{x}(x,2)=0.5\begin{cases}
		\sin^2(\pi x/0.6), &x\in[0,0.3],\\
		1, &x\in[0.3,1.7],\\
		\sin^2(\pi (x-2)/0.6, &x\in[1.7,2.0].\\
	\end{cases}
\end{align*}
The densities of the solid and the fluid are selected as $\rho^{\mathrm{s}}=1$, $\rho^{\mathrm{f}}=1$. The dynamic viscosity of the fluid is selected as $\mu^{\mathrm{f}}=0.2$. The shear modulus of the solid is chosen as $\mu^{\mathrm{s}}=0.2$. The problem setup is illustrated in Fig. (\ref{bcw}).

\subsection{Mesh convergence study}
\begin{figure}[h]
	
	\begin{minipage}[b]{0.5\textwidth}
		\centering
		\includegraphics[scale=0.6,trim= 0 0 0 0,clip]{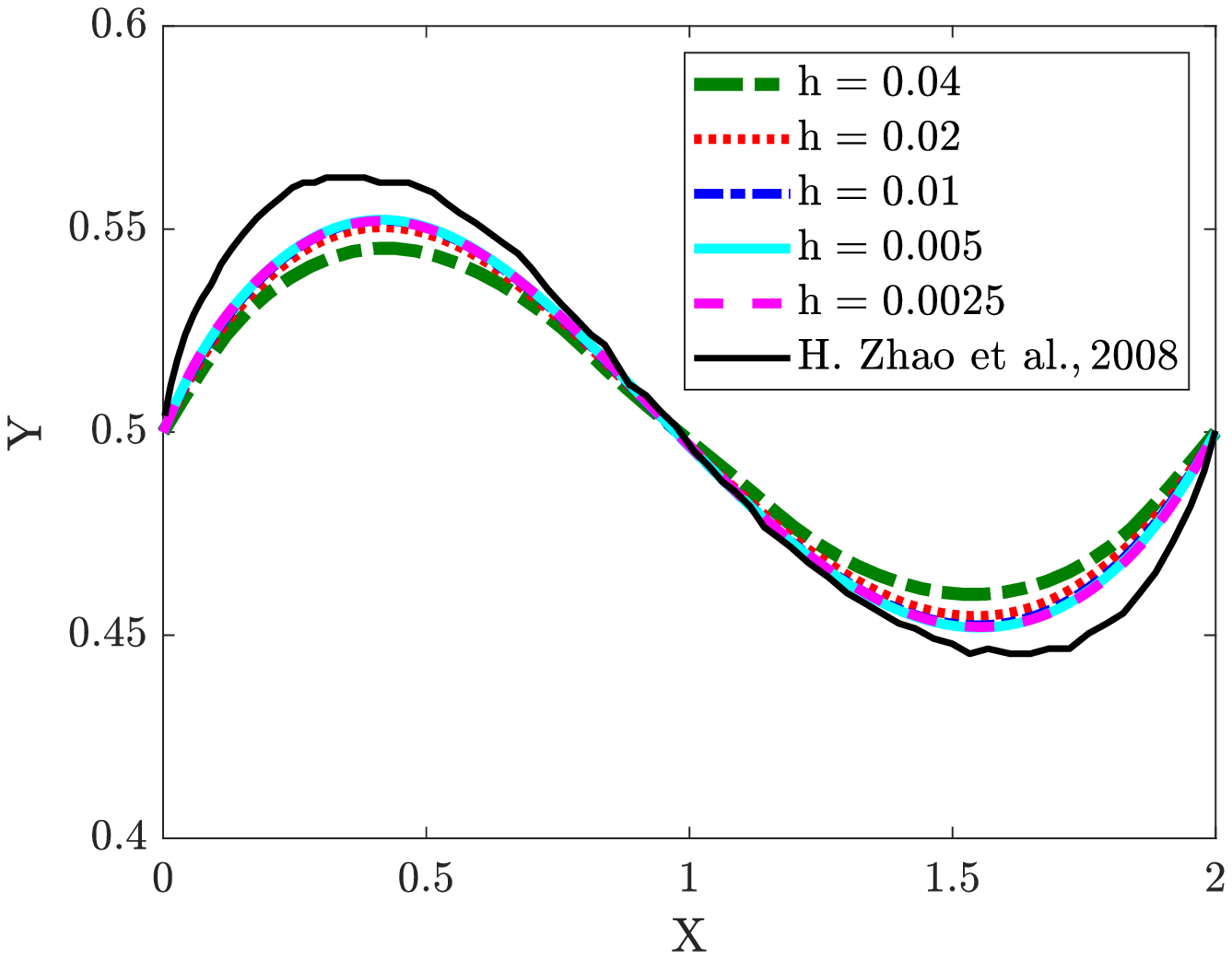}
		\caption*{\hspace{30pt}(a)}
	\end{minipage}
	\hspace{0pt}
	\begin{minipage}[b]{0.5\textwidth}
		\centering
		\includegraphics[scale=0.65,trim= 0 0 0 0,clip]{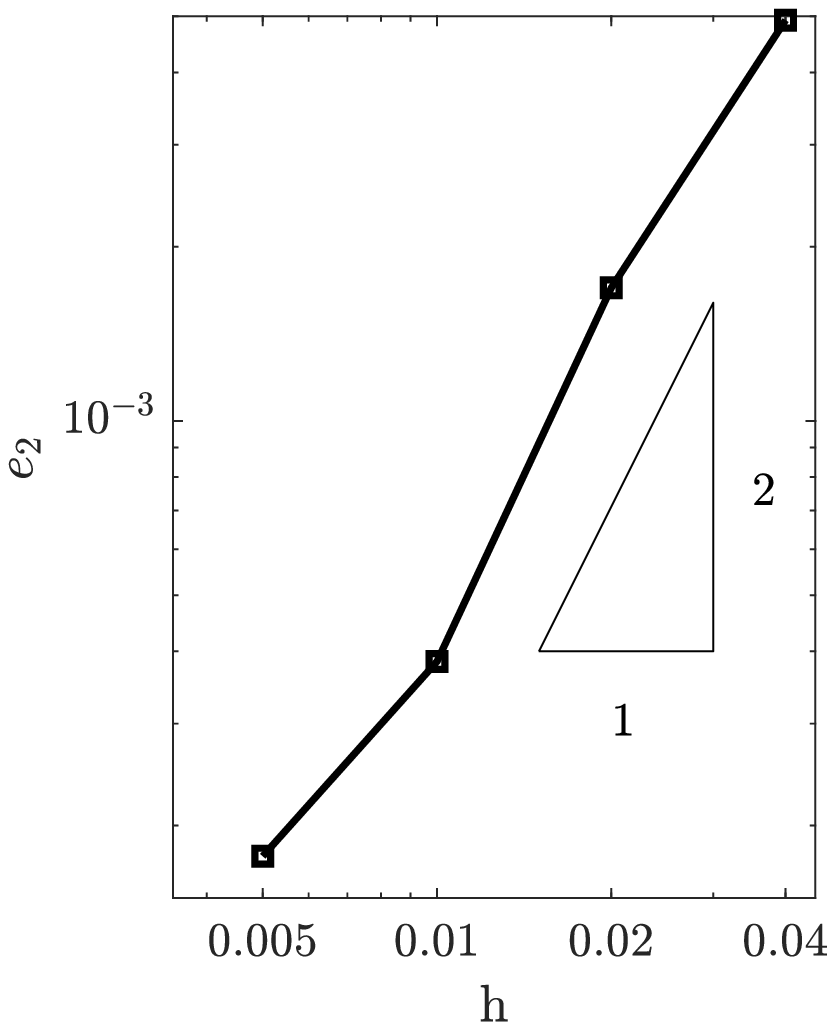}
		\caption*{\hspace{30pt}(b)}
	\end{minipage}
	\caption{Mesh convergence study for the deformation of a solid block driven by cavity flow: (a) interface position, (b) relative $L^2$ error of the order parameter field.} 
	\label{blh}
\end{figure}

We first conduct a systematic mesh convergence study for our fully Eulerian FSI framework employing the IGP method. The interface thickness is selected as $\varepsilon=0.04$. The RMS interface distortion parameter is taken $\eta=0.1$. The time step is chosen as $\Delta t=0.02$ and the solution at $t=20$ is used for the analysis. The computational domain is discretized with a uniform structured triangle mesh. The size of the mesh is bisected from $h=0.04$ to $h=0.005$, while for $h=0.0025$ the structured mesh is only used between $y=0.2$ and $y=0.8$ with unstructured mesh covering the rest of the computational domain. The resulting interface locations are shown in Fig. \ref{blh} (a). To demonstrate the order of convergence of the current framework, we calculate the relative $L^2$ error of the order parameter field, which is defined as $e_2=||\phi-\phi_{\mathrm{ref}}||/||\phi_{\mathrm{ref}}||$, where the solution of $h=0.0025$ is taken as the reference. All the solutions are linearly interpolated to the coarsest mesh, then the $e_2$ errors are evaluated. As shown in Fig. \ref{blh} (b), the current numerical framework is second-order accurate. The difference between the current solution and the reference is due to the lack of convergence in the diffuse interface model.

\subsection{Convergence of the diffuse interface model}
\begin{figure}[h]
	
	\centering
	\includegraphics[scale=0.6,trim= 0 0 0 0,clip]{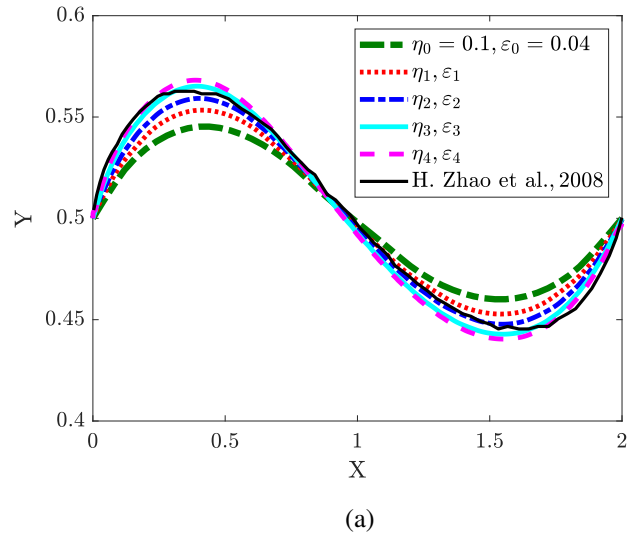}
	\caption*{\hspace{30pt}(a)}
	
	\caption{Convergence of the diffuse interface model. The subscript denotes the number of times of the bisection ($\eta_n=(1/2)^n\eta_0$,$\varepsilon_n=(1/2)^n\varepsilon_0$ ). } 
	\label{blep}
\end{figure}

After the mesh convergence study, we perform a convergence study for the diffuse interface model to compare our solver with the reference data. Similar to the previous cases, in the current convergence study, we bisect $\eta$ and $\varepsilon$ simultaneously from $\eta_0=0.1$, $\varepsilon_0=0.04$ to $\eta_4=0.00625$, $\varepsilon_4=0.0025$, where the subscript denotes the number of times of the bisection. While the time step is selected as $\Delta t=0.02$, the solution at $t=20$ is analyzed. To maintain the proper resolution at the interface, the mesh is refined to $\varepsilon/h=1$. The converged interface position is shown in Fig. \ref{blep}, which shows a good match between the converged solution and the reference data.

\clearpage

\bibliography{bibfile}
\end{document}